\documentclass[aps,onecolumn,notitlepage]{revtex4-1}

\usepackage{amsmath, amsthm, amssymb}
\usepackage{graphicx}
\usepackage{color}
\usepackage{dsfont}
\usepackage{epstopdf}
\usepackage[T1]{fontenc}
\usepackage{textcomp}
\usepackage{array}

\usepackage{natbib}
\usepackage{hyperref}

\pdfoutput=1

\hypersetup{
       colorlinks=true,
       linkcolor=black
}

\begin{document}

%\tableofcontents

\newpage 
\title{Topological Phases: Classification of Topological Insulators and Superconductors of Non-Interacting Fermions, and Beyond}

\author{Andreas  W. W. Ludwig}
\affiliation{Department of Physics, University of California,  Santa Barbara, CA 93106, USA}

\begin{abstract} 
After briefly recalling the quantum entanglement-based view of Topological Phases of Matter in order to outline the  general context,
we give an overview of  different approaches to the classification  problem of  Topological Insulators and Superconductors of non-interacting Fermions.
In particular, we review in some detail  general  symmetry aspects of the "Ten-Fold Way"
which forms the foundation of the classification, and put different approaches to the classification  in relationship with  each other. We end by briefly
mentioning some of the results  obtained on the effect of interactions, mainly in three spatial dimensions.

\end{abstract}

\maketitle

\vskip 1cm

%\tableofcontents

%\newpage 

\section{Introduction} 
Based on  the theoretical and, shortly thereafter,  experimental discovery  of the $Z_2$ Topological Insulators in $d=2$ and $d=3$ spatial dimensions dominated 
by spin-orbit interactions (see \cite{HasanKaneRMP2010,QiZhangRMP2010} for a review), 
the field of Topological Insulators and Superconductors has grown in  the last decade into what is arguably one of the most interesting and stimulating
developments in Condensed Matter Physics. The field  is now  developing at an  ever increasing pace in a number of  directions. While  the understanding of fully interacting phases is still evolving, our understanding of Topological Insulators and Superconductors of non-interacting Fermions is now very well established and complete, and it serves as a stepping stone for further developments. 

\vskip .1cm

Here we will provide
an overview of different  approaches to the classification problem of non-interacting Fermionic Topological Insulators  and Superconductors,  and exhibit  connections
between these approaches which  address  the problem from  very different angles. The exposition is meant to be fairly accessible. 
 To set the stage we begin in section  \ref{sectionEntanglementPerspective} by 
briefly recalling the general quantum entanglement-based  viewpoint of Topological Phases of Matter.  Section \ref{LabelSectionClassificationOfTopInsSupercond}
addresses the classification of Topological Insulators (Superconductors) of non-interacting Fermions.  In particular, in  section \ref{LabelSectionTenFoldWay}
we review in some detail  general symmetry aspects of  the so-called "Ten-Fold Way" which forms the underpinning of the classification problem, stressing its generality as well as  its geometrical interpretation. In section \ref{LabelSubSectionClassificationByTopolgyOfTheBulk} we present  basic ideas of how to bring
topology into the general framework of the "Ten-Fold Way", and explain the result arising from the K-Theory classification of topological band theory for the translationally invariant Topological Insulators (Superconductors) in the bulk. In section \ref{LabelSubSectionClassificatonByLackOfAndersonLocalizationOnSurface}
we review a very different, boundary-based approach to the classification problem which exploits the inability of the boundaries of a non-interacting Topological
Insulator (Superconductor) to support an Anderson-insulating phase even for a strong amount of breaking of translational symmetry  ("disorder") - \lq lack of Anderson Localization\rq.
In section \ref{LabelSubSectionAgreementBetweenBulkAndBoundaryMethods} we address the source of the agreement between the former
bulk-based approach of topological band theory, requiring translational symmetry, and the latter boundary-based approach using  Anderson Localization in which translational
symmetry is intrinsically broken. We review in section \ref{LabelSubSectionPerspectiveOfQuantumAnomalies} a third perspective on the classification problem, that
 is also boundary-based, and exploits the existence of what are  known  as   Quantum Anomalies which prevent the boundary to exist as a consistent quantum theory
on its own, in isolation from the topological quantum state in the bulk.
It was in the  context of the  classification  problem of Topological Insulators (Superconductors) of non-interacting Fermions that
the importance of Quantum Anomalies, well known for  more than three  decades  from  Quantum Field Theory  in Elementary Particle Physics,
was first recognized as a very general tool to characterize these phases.\cite{RyuMooreLudwig2012}
The importance of the characterization in terms of Quantum Anomalies
is that they are universal and  persist beyond the non-interacting regime, and this has recently  been a very active area of study. Finally, in section \ref{LabelSubSectionRecentResults-d-3-PresenceOfInteractions} we mention  results on Fermionic Topological Insulators (Superconductors)
in $d=3$ spatial dimensions
in the presence of interactions. This example indicates  that while there are, expectedly,  in many cases some significant differences between the interacting
and non-interacting Topological Insulators (Superconductors), the fully interacting classification appears to  follow  rather closely  the \lq non-interacting  template\rq.

\vskip .2cm

A number of technical details are delegated to four Appendices.

 \vskip 1cm

\section{Topological Phases - Entanglement Perspective}
\label{sectionEntanglementPerspective}

\noindent Quantum entanglement turns out to provide  an important and very instructive  perspective\cite{ChenGuLiuWenSymmProtectedAndGroupCohomology2013} 
on  topological quantum states of matter. In particular, it is illuminating to  distinguish  two  cases:

\vskip .1cm

\noindent{\it  {\bf Case (1) - No Symmetry Constraints.}}  Let us first consider a situation where  the system under consideration is not subject to
any symmetry constraints.
It   turns out to be useful to distinguish  so-called {\bf (1a)} Short Range Entangled (SRE), 
and {\bf (1b)} Long  Range Entangled (LRE) quantum states $|s\rangle$, 
depending on whether an "initial" state $|s\rangle = |s\rangle_i$  {\bf can}  (SRE), or  {\bf cannot} (LRE)
be continuously transformed into a "final"  direct product state $|s\rangle_f$,
\begin{eqnarray}\nonumber
|s\rangle_f  &=& {\cal T}_g \left [ e^{-i \int_{g_i}^{g_f} \ dg \ {\hat H}(g)} \right ]
|s\rangle_i,\\ \label{AdiabaticTransformation}
|s\rangle_f &=& . . . \otimes |s_1\rangle_{i=1}\otimes |s_2\rangle_{i=2} \otimes |s_3\rangle_{i=3} \otimes ... .
\end{eqnarray}
Here ${\hat H}(g)$ is a local ``Hamiltonian'' (depending on a parameter $g$) on which no symmetry condition is imposed, and ${\cal T}_g$ is the usual
'time-ordering' (in the parameter $g$ acting as \lq time\rq \ for purposes of \lq time-ordering\rq).\footnote{Upon discretizing the $g$-ordered exponential in Eq.(\ref{AdiabaticTransformation}), and  observing that
${\hat H}(g)$ is a sum of terms which are local in space,  one obtains a product (over all ``time''-$g$-steps) of local unitary operators.}
Long Range Entangled (LRE) states turn out to have  what is often  called 
``intrinsic bulk topological order'': that is, they typically possess ground state
degeneracies on topologically non-trivial manifolds, anyonic excitations which may have fractional quantum numbers,  etc.. (These are all properties familiar, e.g.,  from the 2D fractional quantum Hall states, the 2D Toric Code, etc..)
Short Range Entangled (SRE) states, on the other hand, possess no such ``intrinsic bulk topological order''; they can all be continuously  transformed
into each other using Eq. (\ref{AdiabaticTransformation}), i.e.  there is only a single SRE phase in the current situation where no symmetry is imposed.

\vskip .1cm

\noindent{\it {\bf Case (2) - Systems with Symmetry Constraints.}} Let us now consider systems on which  the condition is imposed
that they be invariant under some symmetry group $G$. In this case the class of Short Range Entangled (SRE)
quantum states can be  richer and depends on the particular group $G$. First of all, there are  `standard' Short Range Entangled (SRE) states which simply
 arise from spontaneously breaking of the  symmetry of the system; let us call this
Case  {\bf (2a.1)}. Second, there can be Short Range Entangled (SRE) states in which the symmetry of the system is not broken. These are called Symmetry Protected Topological (SPT) states which form the focus of much of this review; let us call this Case {\bf (2a.2)}.
It turns out that in Case (2a.2)  there can be several distinct phases, all possessing the same symmetry,
such that  in going from one such phase to another a quantum phase transition has to be crossed with the bulk gap closing. On the other hand, if we don't impose the symmetry, the state can be continuously  deformed into a direct product state (since it is as SRE state, case (1a) above).
Well known examples of SPT phases are the spin-1 chain of SU(2) quantum spins in the "Haldane Phase", as well as the non-interacting Fermion topological insulators which form much of the focus of this review. But there are many others, and indeed, SPT phases can be viewed as a generalization of the non-interacting topological insulator phases. Finally, there are Long Range Entangled (LRE) states with  given symmetry constraints. These are called Symmetry Enhanced Topological (SET) phases, Case {\bf (2.b)},  but they will not form the subject of this review.

Let us briefly review key  properties of SPT phases~\cite{ChenGuLiuWenSymmProtectedAndGroupCohomology2013,LuVishwanathClassInteractIntegerTopPhases2D-2012,VishwanathSenthi3DBosTopinsl2013,WangPotterSenthil-Science2013, WangSenthilInteractingFermionic3D-2014,MetlitskiKaneFisherBosTopIns3DWittenEffect2013,FidkowskiChenVishwanathNoAbelianTopOrderSurface3D2013},
case {\bf (2a.2)} above. They have a bulk gap and, as already mentioned,  no intrinsic bulk topological order (i.e. no anyons, no fractional quantum numbers, no ground state degeneracy on topologically non-trivial manifolds, ...) In short,  an SPT phase  
appears to exhibit no interesting bulk properties. However, the distinguishing feature of an SPT phase, that reflects the `topologically non-trivial' nature of the phase, consists in the fact
that the boundary (to vacuum or a topologically different phase)  is always non-trivial in some way: Specifically, the boundary 
\underbar{(i):} must  either spontaneously break the symmetry governing the phase, or \underbar{(ii):} if gapped, must have intrinsic (boundary-, not bulk-) topological order (i.e. anyons etc. ...) , or else, \underbar{(iii)}:  must be gapless. In fact,  the  fundamental and characteristic  property of an SPT phase resides in the nature of its  boundary: {\it The (d-1)-dimensional
boundary of a d-dimensional SPT state cannot exist in isolation as a purely (d-1)-dimensional object. Rather, it  must always be the boundary of
some bulk theory in one dimension higher (i.e. of a d-dimensional SPT state).} We say that the theory on the boundary of an SPT phase  {\it `is anomalous'},
or {\it `has an anomaly'}. This property of SPT phases was first recognized in its general form in the context of non-interacting Fermionic
Topological Insulators in Ref. \cite{RyuMooreLudwig2012}, where it was related to the notion of quantum anomalies familiar from elementary particle
physics.  In the general case,  including interacting theories,  the properties of the boundaries of SPT phases in the various physical spatial dimensions are
today known to be as follows:
\underbar{ 1.)} The (0+1)-dimensional boundary of a (1+1)-dimensional SPT phase is always gapless. (Gaplessness of a quantum theory at a point in space, such as at the (0+1)-dimensional boundary, 
 is understood to be the presence of a zero mode at that point, i.e. a quantum state right at zero energy localized at that point.)
\underbar{ 2.)}  The (1+1)-dimensional boundary of a (2+1)-dimensional SPT phase is either gapless or spontaneously breaks the symmetry defining the phase.
\underbar{3.)}  The (2+1)-dimensional boundary of a (3+1)-dimensional SPT phase either spontaneously breaks the symmetry, carries intrinsic (boundary) topological
order, or is gapless.

As already mentioned briefly, Topological Insulators and Superconductors of non-interacting Fermions provide the simplest examples of SPT phases (they were
also the first examples of such phases that were discovered): A special property  of these non-interacting Fermion SPT phases is that their  boundaries happen to be always gapless [case \underbar{(iii)} above]. As is the case for all SPT phases, they cannot exist in isolation, without being attached to a bulk Topological Insulator (Superconductor) in one dimension higher. Topological Insulators and Superconductors of non-interacting Fermions
can be completely classified in any dimension of space. Here we will review different approaches to this classification, and the interrelation between these approaches. This classification displays extremely interesting and far-reaching general structures.

More general SPT phases are  minimal generalizations of the non-interacting Fermion Topological Insulators (Superconductors) to interacting system. This is currently a very active field of research and a number of interesting results  have  recently emerged. Some will be quoted further below.

It should also be mentioned that there has been much recent progress on Bosonic SPT phases, which are concerned with the physics of topological phases of systems of Bosons, or quantum spins. In particular, Chen, Gu, Liu and Wen\cite{ChenGuLiuWenSymmProtectedAndGroupCohomology2013,ChenGuLiuWenBosonicScience2012}
 pointed out that the notion of Group Cohomology plays an important role as a classifying principle of these (Bosonic) topological systems. Yet,
Group Cohomology in its original form does not appear to exhaustively classify all Bosonic SPT phases, or may require certain extensions. It should also be mentioned that a slightly different direction consisting in the development of approaches focusing directly on physical properties of  SPT phases was initiated in Ref. \cite{LuVishwanathClassInteractIntegerTopPhases2D-2012} 
(and  follow-up work). On the other hand, the full significance of
 a proposed generalization of the  Group Cohomology  approach ("Group Super Cohomology"\cite{GuWenGroupSuperCohomology2014}), aimed at classifying interacting Fermionic SPT phases,  has not been fully understood to-date.
In this review we will not focus on the  Group Cohomology approach, nor on Bosonic SPT phases.

\section{Classification of Topological Insulators and Superconductors of Non-Interacting Fermions, and the Ten-Fold Way}
\label{LabelSectionClassificationOfTopInsSupercond}

Topological Insulators and Superconductors of Non-Interacting Fermions have been completely classified.
Three entirely different approaches to the classification problem   have been used. These are:

\vskip .2cm

\qquad - \underbar{\it Anderson Localization} (Schnyder, Ryu, Furusaki, Ludwig, 2008; 2009; 2010) \cite{SchnyderRyuFurusakiLudwigPRB2008,SchnyderRyuFurusakiLudwigLandau2009,RyuSchnyderFurusakiLudwigNJPhys2010}),

\qquad - \underbar{\it Topology} \underbar{\it (K-Theory)} (Kitaev, 2009)\cite{KitaevLandau2009}),

\vskip .1cm

\noindent  and slightly later  a perspective using 

\qquad - \underbar{\it Quantum Anomalies}  (Ryu, Moore, Ludwig, 2012) \cite{RyuMooreLudwig2012})

\vskip .2cm

\noindent was developed.
%\vskip .1cm
%\noindent 
Below we will review these approaches, and put them in perspective with each other\footnote{A complete classification of all topological insulators and superconductors of non-interacting Fermions in spatial dimensionalities up to $d=3$ was given in \cite{SchnyderRyuFurusakiLudwigPRB2008} in terms of physical symmetries. A systematic regularity (perodicity) of the classification as the dimensionality is varied was discovered by Kitaev in \cite{KitaevLandau2009} for all dimensions through the use of K-theory.
(A systematic pattern as the dimensionality is varied was also discovered for some of the topological insulators by Qi et al. in \cite{QiHughesZhangPRB2008}.)  
An alternative and rather
simple way to understand the regularity (periodicity) of the classification as the dimension is varied was provided in \cite{SchnyderRyuFurusakiLudwigLandau2009,RyuSchnyderFurusakiLudwigNJPhys2010} using the methods of \cite{SchnyderRyuFurusakiLudwigPRB2008}. 
A perspective of the classification using Quantum Anomalies,  which permits a more general point of view that is relevant also in the presence of interactions,
was developed slighty later in \cite{RyuMooreLudwig2012}.}. (For additional recent work  see
\footnote{A very interesting formulation from a more mathematical perspective has also been developed  in \cite{DanFreedGregMooreAnnalesHenriPoincare14-2013-1927}},
\footnote{Quite recently, a somewhat different very interesting approach using topology, which is not based on K-theory, was developed in \cite{KennedyZirnbauer-arXiv-1409.2537,KennedyZirnbauer-Nobel-arXiv-1412.4808}.}.)

%A certain subset of this complete list of Topological Insulators (Superconductors) was obtained by Qi, Hughes and Zhang %(2008)\cite{QiHughesZhang2008}.

\vskip .2cm

In three spatial dimensions for example, only  {\it one} Topological Insulator was known to exist
prior to the appearance of the classification. This was the $Z_2$-Topological 
Insulator\cite{KaneMele2005-1, KaneMele2005-2,FuKanePRB2006, FuKanePRB2007,MooreBalentsPRB2007, RoyPRB2009-1,RoyPRB2009-2,KonigWiedmannBruneRothBuhmannMolenkampScience2007, HsiehQianWrayXiaHorCavaHasanNature2008,HasanKaneRMP2010,QiZhangRMP2010}
characterized by the presence of strong spin-orbit interactions. (This Topological Insulator  belongs to Symmetry Class AII
in the classification scheme discussed in this review.)
As a result of the classification, on the other hand,  it  was found that there exist precisely five distinct types of Topological Insulators (Superconductors)  in every dimension of space. In three spatial dimensions, the four new Insulators (on top of the above-mentioned $Z_2$ Topological Insulator known before), include (i) Topological Superconductors with spin-orbit interactions, and the 3B Helium Superfluid (to be referred to as belonging to Symmetry Class DIII in the classification scheme),  as
well as (ii) Topological Singlet Superconductors (to be referred to as belonging to Symmetry Class CI in the classification scheme). This theoretical prediction, made as a consequence of the classification scheme, has since nucleated experimental activity on both types of systems, three-dimensional topological superconductors and the Helium 3B phase.\footnote{Topological Superconductors were also discussed independently  in \cite{R-Roy-arXiv:0803.2868,QiHughesRaghuZhangPRL2009}.}

\vskip 2cm

\subsection{The Ten-Fold Way: Framework for the Classification of Topological Insulators and Superconductors}
\label{LabelSectionTenFoldWay}

The simplest and most fundamental classification scheme for Topological Insulators and Superconductors of non-interacting Fermions, which is the one we will be reviewing here,
applies to systems for which  {\it no symmetry that is unitarily realized on the first quantized Hamiltonian is required to protect the topological phase}. Unitarily realized symmetries include for example  translational invariance, invariance with respect to some internal symmetry such as e.g. SU(2) spin-rotation symmetry, symmetries of an underlying crystal lattice,  lattice inversion (`parity') symmetry, etc.. It is of course also possible to ask what classification of Topological Insulators and Superconductors
emerges when certain {\it unitarily} realized symmetries are required to protect a topological phase\footnote{see e.g. Ref. \cite{L-Fu-PRL106-2011-106802},\cite{C-K-Chiu-H-Yao-S-Ryu-PRB88-2013-075142,T-Morimoto-A-Furusaki-PRB88-2013-125129,K-Shiozaki-M-Sato-PRB90-2014-165114}.}.  The answer to this question will however clearly not be as universal in that it will depend on the specific  unitarily realized symmetry group required to protect the topological phase. (For Bosonic SPT phases the Group Cohomology approach aims at addressing this question\cite{ChenGuLiuWenSymmProtectedAndGroupCohomology2013}. As already mentioned, the power of Group Cohomology to classify topological phases of Fermions (this is `Group Super Cohomology'\cite{GuWenGroupSuperCohomology2014}) is currently not entirely  understood.) For this reason we ask  here the most universal question possible, i.e. what topological phases of non-interacting Fermions can exist when no unitarily realized symmetry of the first quantized Hamiltonian is required to protect the topological phase. In order to obtain an exhaustive classification that incorporates all possible cases, one first  needs to  have a framework  within which to describe all possible Hamiltonians. This is the framework of the so-called "Ten-Fold Way" that we will now review. After having reviewed this framework we will address the question as to  which of those Hamiltonians are "topological" and which are not.

Of course, even when  a given unitarily  realized  group of symmetries is not required to protect a topological phase, it may still be a symmetry of the  Hamiltonian, i.e. it may commute with it. Given any first quantized  Hamiltonian invariant under an unitarily realized symmetry,  we may always choose a basis in which the Hamiltonian takes on block-diagonal form and the blocks possess no invariance properties with respect to the symmetry. That is,
 there are no  constraints on any of those blocks arising from the unitarily realized symmetry, and thus there will be no linear  operators that commute with 
any of them.
(We will be more explicit below.)  Since we are currently interested in  systems for which no unitarily realized symmetry is required to protect the topological phase, it will  thus  have to be the  properties of these block Hamiltonians that are responsible for the topological properties of the system, and we will now focus on the properties of the blocks. Since 
the block Hamiltonians will not commute with any  linear operator, they must   be very generic Hamiltonians. They  can have very little specific structure\footnote{Since, in particular, translational symmetry is also not required to protect the topological phase, the Hamiltonian and the blocks may lack translational invariance. Such a Hamiltonian is often called ``random'', or ``disordered'', and the block Hamiltonian is an example of a ``random matrix''.}.
 What can we say about the properties of such
very generic first quantized  Hamiltonians? It turns out,  perhaps not unexpectedly, that the only properties they can possess are certain reality conditions, meaning that the Hamiltonian is real or complex in a suitable sense. Such reality conditions have in fact a very transparent physical meaning: They turn out to reflect the properties of the Hamiltonian under time-reversal and charge-conjugation (particle-hole) symmetry  operations. The operations of time-reversal and charge-conjugation are fundamentally  different from
ordinary symmetry operations since they are not realized by unitary, but rather by {\it anti}-unitary operators on the Hilbert space of the first quantized Hamiltonian.
[This is familiar from elementary quantum mechanics  for time-reversal, but is also true for the charge conjugation (particle-hole) operation when acting on the Hilbert space of the first quantized Hamiltonian, as we detail below.] In fact, as we will now briefly review, there are only ten  types of such block Hamiltonians, i.e. 10 ways such a block Hamiltonian can respond to time-reversal and charge conjugation (particle-hole) symmetries. These ten types of generic Hamiltonians were first disovered in foundational work  in the context of random matrix theory by Zirnbauer, and Altland and 
Zirnbauer\cite{ZirnbauerRiemannianSymmetricSpacesJMathPhys1996,AltlandZirnbauerNonstandardSymmClassesPRB1997,HeinznerHuckleberryZirnbauerCMP2005,ZirnbauerOxfordHandbook2010}.

\vskip .1cm

Let us be  more specific. First focus on non-superconducting systems, whose time-evolution is  described in second quantization by the {\it second quantized  Hamiltonian} of the form
\begin{equation}\label{DEFSecondQuantizedHamiltonianNormalSystem}
{\hat H} = \sum_{A,B}   \ \ {\hat \psi}^\dagger_A \  { H}_{A,B} \ {\hat \psi}_B = {\hat \psi}^\dagger H {\hat \psi}
\end{equation}
with Fermion creation and annihilation operators satisfying canonical anti-commutation relations
\begin{equation}\label{CanonicalCommutationRelations}
\{ {\hat \psi}_A, {\hat \psi}_B^\dagger \} = \delta_{A,B}, \quad
\{ {\hat \psi}_A, {\hat \psi}_B \}=\{ {\hat \psi}^\dagger_A, {\hat \psi}_B^\dagger \} =0.
\end{equation}
In the last equality of (\ref{DEFSecondQuantizedHamiltonianNormalSystem})
we have denoted
 by ${\hat \psi}$  the  column vector with components ${\hat \psi}_A$, and  by ${\hat \psi}^\dagger$  the  row vector with components ${\hat \psi}^\dagger_A$.\footnote{As in  (\ref{AppendixAVectorPsiVectorPsiDagger}) of Appendix A, where ${}^t$ denotes the transpose.}
For convenience we have  `regularized' the system on a lattice: The label $A$ denotes the lattice site "$i, j, ...$" \   (i.e. $A, B, ... =i, j, ...$), or may be a combined index, e.g. denoting a lattice site "$i$"  and the orientation of  Pauli-spin "$\sigma=\pm 1$"  at that site (i.e. $A=(i,\sigma)$),  etc.. Hence,  the indices $A,B, ...$ take on $N$ values where $N$ is the number of lattice sites, or twice the number of lattice sites, etc..  %\footnote{${\hat \psi}^\dagger_A$ and ${\hat \psi}_B$ denote Fermion creation and annihilation operators, satisfying 
%canonical commutation relations, 
%$\{{\hat \psi}_A,{\hat \psi}^\dagger_B\} = 
%\delta_{A,B}$, $\{{\hat \psi}_A, {\hat \psi}_B \} =$ $\{{\hat \psi}^\dagger_A, {\hat \psi}^\dagger_B\} =0$.}
The Hamiltonian $H=$ $\{ { H}_{A,B}\}$ in (\ref{DEFSecondQuantizedHamiltonianNormalSystem})
 is then a $N\times N$ matrix of numbers, the {\it first quantized (or single-particle)  Hamiltonian}, and we are interested in the thermodynamic limit $N >> 1$.
%The Hamiltonian ${\hat H}$ in (\ref{DEFSecondQuantizedHamiltonianNormalSystem}) is the  {\it second-quantized   Hamiltonian} which acts on the %(multiparticle) Fermionic Fock Space.
(For a superconducting system, this Hamiltonian is replaced by the Bogoliubov-de-Gennes (BdG) Hamiltonian, and the vector ${\hat \psi}$ is replaced by the {\it Nambu Spinor} - see e.g.  Appendix A for details.  In all cases, superconducting or not,   ${H}_{A,B}$ is a $N\times N$ matrix of numbers, the {\it first quantized Hamiltonian}, and we are interested in the thermodynamic limit $N >> 1$. We continue here
discussing the non-superconducting case.)

\subsubsection{Unitarily Realized Symmetries}
\label{LabelSubSubSectionUnitarily Realized Symmetries}

In the case where the Hamiltonian is invariant under a group $G_0$ of symmetries that are linearly realized on the single particle Hilbert space, there exists a set of unitary $N\times N$ matrices $U$ ($=U_g$ a linear representation of $G_0$ where $g\in G_0$) which commute with the first quantized (\lq single-particle\rq) Hamiltonian,
\begin{equation}\label{DEFUnitarySymmetrySingleParticle}
U H U^\dagger =H.
\end{equation} 
In second quantized language this corresponds to operators ${\hat {\cal U}}$ acting on the Fermion  Fock space via
\begin{equation} \label{DEFUnitarySymmetryFockSpace}
{\hat {\cal U}} \  {\hat \psi}_A \  {\hat {\cal U}}^{-1} \equiv \sum_{B} U^\dagger_{A,B}  \ {\hat \psi}_B,
\qquad \quad 
{\hat {\cal U}} \  {\hat \psi}^\dagger_A \  {\hat {\cal U}}^{-1} \equiv \sum_B \ {\hat \psi}^\dagger_B  \  U_{B, A}
\end{equation}
and commuting with the second quantized Hamiltonian
\begin{equation}\label{UnitarySymmetrySecondQuantizedHamiltonian}
{\hat {\cal U}} \  {\hat H} \  {\hat {\cal U}}^{-1} = {\hat H}.
\end{equation}
In this situation, the first quantized Hamiltonian $H$ (a $N\times N$ matrix)  possesses a block-diagonal structure (as it is well known from basic
quantum mechanics).  Specifically,
the $N$-dimensional vector space ${\cal V}$  spanned by the single-particle states
$|A\rangle$ $={\hat \psi}^\dagger_A|0\rangle$ (where A=1, ..., N, and  $|0\rangle$  denotes the Fock Space vacuum)
decomposes into a direct sum of vector spaces ${\cal V}_\lambda$ associated with certain irreducible representations  (irreps) $\lambda$  of the group $G_0$,
\begin{equation}\label{DecompositionSingleParticleHilbertSpace}
{\cal V} = \bigoplus_\lambda^{certain \ irreps} \  {\cal V}_\lambda.
\end{equation}
In each vector space ${\cal V}_\lambda$ a (orthonormal) basis can be chosen of the form
$|v^{(\lambda)}\rangle_\alpha\otimes|w^{(\lambda)}\rangle_k$ such that the group $G_0$ acts only on $|w^{(\lambda)}\rangle_k$ but not on $|v^{(\lambda)}\rangle_\alpha$,
whereas the single particle  Hamiltonian $H$  acts only $|v^{(\lambda)}\rangle_\alpha$ and not on $|w^{(\lambda)}\rangle_k$.
(Here  $\alpha=1, ..., m_\lambda$ and  $k=1, ..., d_\lambda$
where $d_\lambda$ is the dimension of the  irreducible representation $\lambda$, and
$m_\lambda$ is the multiplicity with which $\lambda$ occurs in the vector space ${\cal V}$.)
Thus each irrep. $\lambda$ defines a 
block
Hamiltonian $H^{(\lambda)}$ which is a $m_\lambda\times m_\lambda$ matrix
with matrix elements $H^{(\lambda)}_{\alpha,\beta} =$ 
$ {}_\alpha\langle v^{(\lambda)}| H | v^{(\lambda)}\rangle_\beta$.\footnote{If the operations arising from the anti-unitary 
symmetries discussed in the following section do not
leave  ${\cal V}_\lambda$ invariant, the block arising from ${\cal V}_\lambda$ may have to be slightly 
enlarged\cite{HeinznerHuckleberryZirnbauerCMP2005,ZirnbauerOxfordHandbook2010}.}

%Let us include into $G_0$ {\it all} unitary operators commuting with the first quantized Hamiltonian $H$. Then, by construction, none of the block %Hamiltonians $H^{(\lambda)}$ is invariant under any unitarily realized symmetries. I.e. there are no unitary operators (matrices) that  commute with
%the matrix $H^{(\lambda)}$. Thus the Hamiltonians $H^{(\lambda)}$ are "symmetry-less" (for any of the  irreps $\lambda$).

We can now ask the following question, which turns  out to have a rather  interesting answer: Fix a symmetry group $G_0$ and consider {\it all possible}
single particle Hamiltonians $H$ which commute with all symmetry operations in  $G_0$ that are unitarily realized on the single
particle Hilbert space.  As we run through the set of all these Hamiltonians,
what sets of matrices does one obtain for the blocks  $H^{(\lambda)}$?
The  answer is  that the resulting set of block Hamiltonians $H^{(\lambda)}$  
is independent of
the symmetry group $G_0$. It is also (essentially)  independent of the irrep. $\lambda$. What is interesting is that there  turn out to be only ten possible such sets of matrices, and  the complete list of  corresponding quantum mechanical time-evolution operators $U^{(\lambda)}(t)=$ $\exp\{ i t H^{(\lambda)}\}$ associated with
these Hamiltonian blocks  is provided in the column  titled "Time evolution operator" of FIG.\ref{fig:TenFoldWay}. Thus, the question we have been asking (and its answer)  
is  very useful  since it has made the problem of listing all block Hamiltonians tractable. It has turned the problem of listing all Hamiltonians  into a finite problem, no matter what the group $G_0$ of unitarily realized symmetries is they are invariant under, and what the irrep. $\lambda$. 
In section \ref{SubSectionTenFoldWay}
below we will elaborate further on this result.

In order to better understand this result, we may ask the question:
What can we say about the block Hamiltonians  $H^{(\lambda)}$?  They can only depend  on very general properties of the quantum mechanical
system under consideration, as they are completely independent of any unitarily realized symmetries. The key to the answer lies in the well known
fact that any symmetry in quantum mechanics must be realized  either by a unitary or  by an anti-unitary operator acting on Hilbert space.
Because  we have already exhausted the properties of a Hamiltionian following from any unitarily realized symmetries,  the only property that the
  block Hamiltonians may depend on is  the  behavior of the underlying quantum mechanics under {\it anti}-unitarily realized symmetries. 
As we will review below, there are only very few anti-unitarily realized symmetry operations  (modulo unitarily realized symmetries),
and this is the reason why the resulting list is finite. Since anti-unitary symmetries involve complex
conjugation, the presence or absence of anti-unitary symmetries  will imply only certain reality conditions on the block Hamiltonians. As we will now review,
there are only ten possibilities for this to happen.

\subsubsection{Anti-Unitarily Realized Symmetries}
\label{LabelSubSubSectionAntiUnitarilyRealizedSymmetries}

We will now briefly list the relevant properties of three symmetry operations  that can appear in connection with anti-unitary symmetries.
At the end of this section we show that there are no others.

\vskip .1cm

{\it Time-Reversal Symmetry:} This is the most familiar anti-unitary symmetry operation. The operator ${\hat {\cal T}}$ implementing time-reversal
symmetry on the Fermion Fock space of second quantization is defined by its action on Fermion creation and annihilation operators by
%\begin{eqnarray}\nonumber
%{\hat {\cal T}} {\hat \psi}_A {\hat {\cal T}}^{-1} &=& \sum_B (U_T)_{A,B} \ {\hat \psi}_B \\
%\nonumber
%{\hat {\cal T}} {\hat \psi}^\dagger_A {\hat {\cal T}}^{-1} &=& \sum_B  \ {\hat \psi}^\dagger_B\ (U^\dagger_T)_{B,A}\\
%\label{DEFTimeReversalSecondQuantizedAlt}
%{\hat {\cal T}} i {\hat {\cal T}}^{-1} &=&  - i \qquad {\rm (anti-unitary)}
%\end{eqnarray}
\begin{equation}\label{DEFTimeReversalSecondQuantized}
{\hat {\cal T}} {\hat \psi}_A {\hat {\cal T}}^{-1} = \sum_B  \  {(U^\dagger_T)}_{A,B} \  {\hat \psi}_B; \quad \  \ 
{\hat {\cal T}} {\hat \psi}^\dagger_A {\hat {\cal T}}^{-1} = \sum_B  \ {\hat \psi}^\dagger_B\ {(U_T)}_{B,A};\quad \ \ 
{\hat {\cal T}} i {\hat {\cal T}}^{-1} = - i  \  {\rm (anti-unitarity)}
\end{equation}
where $U_T$ is a unitary matrix.
The second quantized Hamiltonian ${\hat H}$ is time-reversal invariant if and only if
\begin{equation}\label{DEFSecondQuantizedTimeReversalInvariance}
{\hat {\cal T}} {\hat H} {\hat {\cal T}}^{-1} ={\hat H}.
\end{equation}
Note that when acting on the time-evolution operator ${\hat U}(t) =$ $\exp\{ i t {\hat H}\}$, one has  ${\hat {\cal T}} {\hat U}(t) 
{\hat {\cal T}}^{-1} =$ $ {\hat U}(-t)$, as desired for time-reversal. Now, it is readily seen that  the condition in (\ref{DEFSecondQuantizedTimeReversalInvariance})   implies for the first quantized Hamiltonian the condition
\begin{equation}\label{DEFTimeReversalSingleParticleHamiltonian}
U_T { H}^* U^\dagger_T = + { H}.
\end{equation}
This means that the first quantized Hamiltonian ${H}$ is time-reversal invariant if and only if the Hamiltonian ${H}^*$ (where all matrix elements are complex conjugate) is equal to the original Hamiltonian $H$  up to a `rotation' by a unitary matrix $U_T$.
%$$ {\mathbf T}:
%\qquad
%U_T {\cal H}^* U_T^{-1} = {\cal H}
%$$
It will be convenient to introduce a special notation for the first quantized time-reversal operator (acting on the single-particle space),
\begin{equation}\label{DEFTimeReversalOperatorFirstQuantized}
{\bf T}\equiv {\hat {\cal T}}_{| first \ quantized}.
\end{equation}
Equation (\ref{DEFTimeReversalSingleParticleHamiltonian}) can then be written in the form
\begin{equation}\label{DEFTimeReversalSingleParticleHamiltonianAlt}
{\mathbf T} H {\mathbf T}^{-1}= H, \qquad {\rm where} \qquad {\mathbf T} =  U_T \cdot K. 
\end{equation}
Here  $K$ is the (anti-unitary) operator that implements complex conjugation, $K { H} K^{-1}={ H}^*$, where  $K^2=1$. As is immediately
checked, the square of the time-reversal operator ${\hat {\cal T}}^2$
 is a unitary operator  ${\hat {\cal U}}$  as in (\ref{DEFUnitarySymmetrySingleParticle}, \ref{DEFUnitarySymmetryFockSpace}, 
\ref{UnitarySymmetrySecondQuantizedHamiltonian}), where the associated unitary matrix
 is $U=U_T U^*_T$.
%{\color{blue}{<- check this!}}
 It follows from (\ref{DEFTimeReversalSingleParticleHamiltonian}) that the matrix $U_T U^*_T$ commutes with $H$.
We are interested in the case where $H$ is one of the block Hamiltonians $H^{(\lambda)}$ discussed below (\ref{DecompositionSingleParticleHilbertSpace}). Since these block Hamiltonians turn out to  run over an irreducible space of Hamiltonians (as we will see in
more detail below), the matrix $U_T U_T^*$  must be a multiple of the identity matrix by Schur's Lemma:  $U_T \ U_T^* = e^{i\gamma} {\bf 1}$
(due to the unitarity of that matrix the multiple must be a  phase). Considering the matrix product
%\begin{equation}\label{ThreeUTmatrices}
$U_T\ U_T^* \ U_T =$ $ (U_T \ U_T^*) \ U_T =$ $ U_T \ (U_T^* \ U_T)$,
%\end{equation}
we obtain from the last equality
$e^{i\gamma} U_T= U_T e^{-i\gamma}$ which implies
$e^{i\gamma} = \pm 1$, and thus $U_T \ U_T^* = \pm 1$.

%\vskip 3cm

%It is known from elementary quantum mechanics that the  anti-unitary time-reversal operator ${\bf T} = U_T \cdot K$ 
%(where $K$ denotes complex conjugation, $K {\cal H} K^{-1}={\cal H}^*$, $K^2=1$)
%can square only to plus or minus the identity. Thus
We therefore conclude that there are three ways  a Hamiltonian $H$ can respond to time-reversal symmetry.
Let us denote these three possibilities  by $T=0, +1, -1$:
\begin{equation}\label{ThreePossibilitiesForTimeReversal}
T = \begin{cases}
0, &   {\rm when \ the \ Hamiltonian} \ H  \ {\rm is \ not \ time \ reversal \ invariant}\\
+1, & {\rm when \ the \ Hamiltonian} \ H \  {\rm is  \ time \ reversal \ invariant \ and \ } \  {\mathbf T}^2 = +1 \\
-1, & {\rm when \ the \ Hamiltonian} \  H \ {\rm is  \ time \ reversal \ invariant \ and \ } \  {\mathbf T}^2 = -1.
\end{cases}
\end{equation}
We end by noting that in the case where ${\bf T}^2=-1$, we have ${\hat {\cal T}}^2 {\hat \psi}_A {\hat {\cal T}}^{-2} =$
$ (-1) {\hat \psi}_A$, and  ${\hat {\cal T}}^2 {\hat \psi}^\dagger_A {\hat {\cal T}}^{-2} =$
$ (-1) {\hat \psi}^\dagger_A$. Since a state with  a number of $q$ Fermions is created from the Fock space vacuum by applying $q$ Fermion creation operators, we see
that when ${\bf T}^2=-1$, the second quantized time-reversal operator squares to the Fermion Parity operator,
\begin{equation}\label{TimeReversalFermionParity}
{\hat {\cal T}}^2 = (-1)^{\hat F} = (-1)^{\hat Q}
\end{equation}
where ${\hat Q} = \sum_A {\hat \psi}^\dagger_A {\hat \psi}_A$ is the particle number operator. 
On the other hand,  ${\hat {\cal T}}^2 = {\bf 1}$
when ${\bf T}^2 ={\bf 1}$.

\vskip .5cm

{\it Charge-Conjugation (Particle-Hole) Symmetry:} There turns out to be a  similar way of classifying the behavior of the first quantitzed Hamiltonian
$H$ under charge-conjugation (particle-hole) symmetry. It is most convenient to first recall the definition of the second quantized  operator
${\hat {\cal C}}$
which  implements charge-conjugation (particle-hole) symmetry on the Fermion Fock space:
\begin{equation}\label{DEFChargeConjugationSecondQuantized}
{\hat {\cal C}} {\hat \psi}_A {\hat {\cal C}}^{-1} = \sum_B  \  {(U^{*\dagger}_C)}_{A,B} \  {\hat \psi}^\dagger_B; \quad \  \ 
{\hat {\cal C}} {\hat \psi}^\dagger_A {\hat {\cal C}}^{-1} = \sum_B  \ {\hat \psi}_B\ {(U^{*}_C)}_{B,A};\quad \ \ 
{\hat {\cal T}} i {\hat {\cal T}}^{-1} = + i  \  {\rm (unitarity)}
\end{equation}
where $U_C$ is a unitary matrix. Note that the second quantized  charge-conjugation operator ${\hat {\cal C}}$ is a {\it unitary} operator when
acting on the Fermion Fock space. 
The second quantized Hamiltonian ${\hat H}$ is charge-conjugation (particle-hole)  invariant if and only if
 \begin{equation}\label{DEFSecondQuantizedChargeConjugationInvariance}
{\hat {\cal C}} {\hat H} {\hat {\cal C}}^{-1} ={\hat H}.
\end{equation}
It is easily checked\cite{SchnyderRyuFurusakiLudwigPRB2008} that (\ref{DEFSecondQuantizedChargeConjugationInvariance})
implies for the first quantized Hamiltonian the condition
\begin{equation}\label{DEFChargeConjugationSingleParticleHamiltonian}
U_C { H}^* U^\dagger_C = -  { H}.
\end{equation}
The minus sign arises from Fermi statistics. (Note that because $H$ is Hermitian, we could have also written $H^t$ instead of $H^*$, but the
current notation is most convenient.) In analogy to our discussion of time-reversal, it will be convenient  to introduce a special notation for the first quantized charge-conjugation operator (acting on the single-particle space),
\begin{equation}\label{DEFChargeConjugationHamiltonianAlt}
{ {\mathbf C}} \equiv {\hat {\cal C}}_{first \ quantized}.
\end{equation}
As in the case of time-reversal, equation (\ref{DEFChargeConjugationSingleParticleHamiltonian}) can then be written in the form
\begin{equation}\label{DEFTimeReversalChargeConjugationHamiltonianAlt}
{\mathbf C} H {\mathbf C}^{-1}= -  H, \qquad {\rm where} \qquad {\mathbf C} =   U_C \cdot K.
\end{equation}
(As above, $K$ implements complex conjugation.) Using the same arguments as for time-reversal, 
the square of the charge-conjugation operator ${\hat {\cal C}}^2$
 is a unitary operator  ${\hat {\cal U}}$  as in 
(\ref{DEFUnitarySymmetrySingleParticle}),(\ref{DEFUnitarySymmetryFockSpace}),(\ref{UnitarySymmetrySecondQuantizedHamiltonian}), 
where the associated unitary matrix  $U=U_C U_C^*$ commutes with $H$,
and consequently $U_C U_C^*=\pm 1$.
%{\color{blue} <- check this !!}
We therefore conclude that there are three ways  a Hamiltonian $H$ can respond to charge-conjugation.
\begin{equation}\label{ThreePossibilitiesForChargeConjugation}
C = \begin{cases}
0, &   {\rm when \ the \ Hamiltonian} \ H  \ {\rm is \ not \ charge \ conjugation \ invariant}\\
+1, & {\rm when \ the \ Hamiltonian} \ H \  {\rm is  \  charge \ conjugation \ invariant \ and \ } \  {\mathbf C}^2 = +1 \\
-1, & {\rm when \ the \ Hamiltonian} \  H \ {\rm is  \ charge \ conjugation \ invariant \ and \ } \  {\mathbf C}^2 = -1.
\end{cases}
\end{equation}
By the same arguments as those  used for time-reversal, equation (\ref{TimeReversalFermionParity}),  
we note that in the case where ${\bf C}^2=-1$  the second quantized charge-conjugation operator squares to the Fermion Parity operator,
\begin{equation}\label{ChargeConjugationFermionParity}
{\hat {\cal C}}^2 = (-1)^{\hat F} = (-1)^{\hat Q}.
\end{equation}
On the other hand, ${\hat {\cal C}}^2 = {\bf 1}$ when ${\bf C}^2={\bf 1}$.

\vskip .7cm

{\it Chiral  (Sublattice) Symmetry:}  As we will explain below, we will also need to consider, besides ${\hat {\cal T}}$ and ${\hat {\cal C}}$,
the combined operation
\begin{equation}\label{DEFChiralSymetrySecondQuantized}
{\hat {\cal S}} \equiv {\hat {\cal T}} \cdot {\hat {\cal C}}
\end{equation}
which is  conventionally given the name "chiral", or "sublattice" symmetry (because in certain special cases it reduces to symmetries
that are suitably characterized by these names). ${\hat {\cal S}}$  is an {\it anti-unitary} operator on the Fermion Fock space
whose  action follows from  (\ref{DEFTimeReversalSecondQuantized}),(\ref{DEFChargeConjugationSecondQuantized}):
\begin{equation}\label{DEFChiralSecondQuantized}
{\hat {\cal S}} {\hat \psi}_A {\hat {\cal S}}^{-1} = \sum_B  \  ({U^{* \dagger}_S)}_{A,B} \  {\hat \psi}^\dagger_B; \quad \  \ 
{\hat {\cal S}} {\hat \psi}^\dagger_A {\hat {\cal S}}^{-1} = \sum_B  \ {\hat \psi}_B\ {(U^*_S)}_{B,A};\quad \ \ 
{\hat {\cal S}} i {\hat {\cal S}}^{-1} = - i  \  {\rm (anti-unitarity)}
\end{equation}
where  $U_S \equiv U_T  U_C^*$.
%{\color{blue} <- check this !!}
The second quantized Hamiltonian ${\hat H}$ is  invariant under ${\hat {\cal S}}$  if and only if
 \begin{equation}\label{DEFSecondQuantizedChiralInvariance}
{\hat {\cal S}} {\hat H} {\hat {\cal S}}^{-1} ={\hat H}.
\end{equation}
It is immediately checked  that (\ref{DEFSecondQuantizedChiralInvariance})
implies for the first quantized Hamiltonian the condition
\begin{equation}\label{DEFChiralSymmetrySingleParticleHamiltonian}
U_S{ H} U^\dagger_S= -  { H} \qquad \qquad  (U_S \equiv U_T  U_C^*).
\end{equation}
As for time-reversal and charge-conjugation, it will be convenient  to introduce a special notation for the first quantized chiral  operator (acting on the single-particle space),
\begin{equation}\label{DEFChargeConjugationHamiltonianAlt}
{ {\mathbf S}} \equiv {\hat {\cal S}}_{first \ quantized}= {\mathbf T}\cdot {\mathbf C}   = (U_T  \cdot   K)   \cdot ( U_C \cdot K)  
= U_T \cdot U_C^* = U_S.
\end{equation}
The condition (\ref{DEFChiralSymmetrySingleParticleHamiltonian})  reads
\begin{equation}\label{DEFChiralSymmetrySingleParticleHamiltonianAlt}
{\mathbf S} H {\mathbf S}^{-1} = - H,
\end{equation}
consistent with 
(\ref{DEFChiralSymmetrySingleParticleHamiltonian}), (\ref{DEFTimeReversalSingleParticleHamiltonianAlt}) and (\ref{DEFTimeReversalChargeConjugationHamiltonianAlt}). We see from (\ref{DEFChiralSymmetrySingleParticleHamiltonianAlt})
that   ${\bf S}^2 = (U_S)^2$ commutes with $H$ which implies, by the same argument used above for time-reversal and charge-conjugation,
that it must be a multiple of the identity operator,
\begin{equation}\label{ChiralSquaredSchursLemma}
{\bf S}^2 = (U_S)^2 = e^{i\phi} \ {\bf 1}.
\end{equation}
(The multiple is a phase because the left hand side is unitary.) Recalling that $U_S=U_T U_C^*$, and observing that the phases of both
$U_T$ and $U_C$ are completely arbitary (they will not affect any of the results we have mentioned), we can multiply these two matrices
by suitable phases so  that $e^{i\phi}=1$. Upon making such a  choice of phase we obtain\footnote{See also the discussion in \cite{ChiuTeoSchnyderRyuarXiv1505.03535}.}
\begin{equation}\label{ChiralSquaredSchursLemmaNoPhase}
{\bf S}^2 = (U_S)^2 = {\bf 1}.
\end{equation}

We note that in (\ref{DEFChiralSymetrySecondQuantized}) we could  have equally well chosen the other order, obtaining a slightly different version of the chiral symmetry operation
\begin{equation}\label{DEFChiralSymmetryPrimedSecondQuantized}
{\hat {{\cal S}'}} \equiv
 {\hat {\cal C}}\cdot  {\hat {\cal T}}.
\end{equation}
	One can immediately check that the corresponding unitary matrix is now $U_{S'}= U_C U_T^*$. As explained in Appendix \ref{LabelSectionAppendixAlternativeChiralSymmetry}, that  yields precisely the same results as the original chiral symmetry operation
${\hat {\cal S}}$, but corresponds simply to writing the first quantized Hamiltonian $H$ in another basis.

Thus, in general there are two 
 possibilities  denoted by  $S=0, 1$ by  which a Hamiltonian $H$ can respond to chiral (sublattice) symmetry ${\bf S}$:
\begin{equation}\label{TwoPossibilitiesForChiralSymmetry}
S= \begin{cases}
0, &   {\rm when \ the \ Hamiltonian} \ H  \ {\rm does \  not \ anti-commute \ with \ chiral \ symmetry} \ {\bf S} \\
1, &  {\rm when \ the \ Hamiltonian} \ H  \ {\rm anti-commutes \ with \ chiral \ symmetry} \ {\bf S}.
\end{cases}
\end{equation}

\vskip .2cm

To summarize: Considering operators on the Fermionic Fock space of second quantization, the operators ${\hat {\cal T}}$ and ${\hat {\cal S}}$ are anti-unitary operators, whereas the composition
${\hat {\cal C}}=$
${\hat {\cal T}}^{-1} \cdot {\hat {\cal S}}$ is a unitary operator. All these operators {\it commute} with the second quantized Hamiltonian ${\hat H}$
when the corresponding operators are symmetries.
On the  other hand, considering operators acting on the single-particle Hilbert space of first quantization, the corresponding operators
${\bf T}$ and ${\bf C}$ are anti-unitary, whereas ${\bf S}$ is unitary. While ${\bf T}$ commutes with the first quantized Hamiltonian $H$,
the operators 
${\bf C}$ and ${\bf S}$ {\it ant}i-commute with $H$, when the corresponding operations are symmetries.

\vskip .2cm
\noindent Let us now come back to two items that we had briefly mentioned, but not yet   addressed further.
\vskip .1cm

(i): First, we observe that we need to consider {\it only  one} time-reversal and  {\it only one}  charge-conjugation operation. We will work in the first quantized formulation where these operators are both  anti-unitary.  Assume there were, e.g., two time-reversal operators ${\bf T}_1=U_{T_1} \cdot K$
and  ${\bf T}_2 = U_{T_2} \cdot K$. Then, the composition ${\bf T}_1 \cdot {\bf T}_2 =$ $U_{T_{12}}\equiv  U_{T_1} U_{T_2}^*$  is a unitary operator. This means that ${\bf T}_2 =$ ${\bf T}_1^{-1} \cdot U_{T_{12}}=$
$K \cdot U_{T_1}^\dagger \cdot U_{T_{12}}=$ $ U_{T_1}^{* \dagger} \cdot U^*_{T_{12}}\cdot K$. Therefore, ${\bf T}_2$
is equal to ${\bf T}_1$ modulo (left-) multiplication by a unitary operation,
${\bf T}_2=$
$ U_{12} \cdot {\bf T}_1$ (where $U_{12}=$ $U_{T_1}^{* \dagger} \cdot U^*_{T_{12}}\cdot U_{T_1}^\dagger$). Hence we can go back to
section \ref{LabelSubSubSectionUnitarily Realized Symmetries} and enlarge the existing group $G_0$ of symmetries that
are unitarily realized on the single particle Hilbert space by the group element $U_{12}$ to a new group ${G'}_0$. Upon repeating all the previous
analysis using  ${G'}_0$ instead of $G_0$, there will  be no difference in extending the unitarily realized
symmetries by the anti-unitary symmetry elements
${\bf T}_1$ or ${\bf T}_2$.  (Recall from section \ref{LabelSubSubSectionUnitarily Realized Symmetries} that the  structure of the block Hamiltonians
which is the focus of our interest does in fact not depend on the group $G_0$.) So we do not get a new result when we include a second anti-unitary time-reversal
operator ${\bf T}_2$ in addition to ${\bf T}_1$. A completely analogous argument can be made in  case there
are two (anti-unitary) charge-conjugation operators ${\bf C}_1$ and ${\bf C}_2$.  In summary,  we only need to consider
{\it one}  time-reversal operator ${\bf T}$ and
 {\it one} charge-conjugation operator ${\bf C}$.

Note however that it is not possible to dispose of  the anti-unitary
charge-conjugation operation ${\bf C}$ while keeping only one ${\bf T}$, or vice versa. Indeed, while it is true that the composition
${\bf S}=$  ${\bf T} \cdot {\bf C}$ is a unitary operator,  invariance under the chiral symmetry, equation (\ref{DEFSecondQuantizedChiralInvariance}), implies that ${\bf S}$ {\it anti-}commutes with the first quantized Hamiltonian $H$,
equation (\ref{DEFChiralSymmetrySingleParticleHamiltonianAlt}).  ${\bf S}$ can never {\it commute} with $H$. For this reason
it is not possible to dispose of,  say ${\bf C}$,  in favor of ${\bf T}$  (or vice versa) by augmenting $G_0$ by the element ${\bf S}$.
We must keep ${\bf S}$ explicitly  in our analysis, and we will see below the effect this has.

\vskip .1cm

(ii): Second, the above discussion can be directly extended to include Bogoliubov-de-Gennes (BdG)  Hamiltonians for Fermionic quasiparticles
in superconductors. All that is necessary is  to replace the column vector ${\hat \psi}$ in (\ref{DEFSecondQuantizedHamiltonianNormalSystem})
by the column vector ${\hat \chi}$, the Nambu Spinor (see e.g. (\ref{DEFNambuSpinorAlternative}) of Appendix (\ref{LabelSectionAppendixSuperconductor})). The entire discussion in the previous subsection (\ref{LabelSubSubSectionUnitarily Realized Symmetries})  as well as in the current Subsection (\ref{LabelSubSubSectionAntiUnitarilyRealizedSymmetries}) goes through analogously. The main
difference is that the Nambu Spinor  ${\hat \chi}$  is not independent from its conjugate,
\begin{equation}\label{NambuSpinorRealityCondition}
\left
({\hat \chi}^\dagger\right )^t 
= \tau_1 {\hat \chi},
\end{equation}
where $\tau_1$ is a Pauli matrix in \lq particle-hole space\rq (see (\ref{AppendixAVectorChiVectorChiDagger}) and (\ref{DEFTauParticleHoleSpace})
of Appendix (\ref{LabelSectionAppendixSuperconductor})). This leads to the fact that the first quantized BdG Hamiltonian $H$ automatically satisfies
the
charge-conjugation symmetry condition (\ref{DEFChargeConjugationSingleParticleHamiltonian})
\begin{equation}\label{BdGChargeConjugationSymmetryMainText}
\tau_1H^*\tau_1=\tau_1 H^t \tau_1= -  H.
\end{equation}
(See (\ref{BdGChargeConjugationSymmetry}) of Appendix (\ref{LabelSectionAppendixSuperconductor}).)

More formally, we  can describe both systems, normal and superconducting, within the same language if we replace
 the column vector ${\hat \psi}$ in (\ref{DEFSecondQuantizedHamiltonianNormalSystem}), (\ref{CanonicalCommutationRelations})
and all subsequent equations in sections (\ref{LabelSubSubSectionUnitarily Realized Symmetries}) and (\ref{LabelSubSubSectionAntiUnitarilyRealizedSymmetries})
by the general symbol ${\hat \Psi}$ which can denote normal systems (in which case ${\hat \Psi} \rightarrow {\hat \psi}$ as in  (\ref{DEFSecondQuantizedHamiltonianNormalSystem})) or  superconducting sytems  (in which case  ${\hat \Psi} \rightarrow {\hat \chi}$, the Nambu Spinor (\ref {DEFNambuSpinorAlternative})).

\vskip .5cm

\subsubsection{The Ten-Fold Way}
\label{SubSectionTenFoldWay}

We now return to the problem of classifying all first quantized Hamiltonians $H$  which are invariant under some symmetry group $G_0$ of
symmetries that are unitarily realized on the singe-particle Hilbert space. (Note that this also includes the case where  the symmetry group $G_0$ is trivial, 
$G_0=\{ {\bf 1}\}$.) 
As reviewed in section \ref{LabelSubSubSectionUnitarily Realized Symmetries}, these Hamiltonians are characterized by "symmetry-less" block Hamiltonians $H^{(\lambda)}$. Since the only quantum mechanical
symmetries not yet accounted for by invariance under $G_0$  are symmetries that are  {\it anti-unitarily} realized on the single-particle Hilbert space, it must be that these blocks can be classified by their behavior under these so-far not-yet-accounted-for {\it anti-unitary} symmetries. As we have seen in the previous subsection, there can only be {\it two}  such {\it anti-unitary} symmetries, namely  time-reversal ${\mathbf T}$ and charge-conjugation ${\mathbf C}$; as also  mentioned above,
we need to consider in addition  their  product, the chiral (sublattice) symmetry ${\mathbf S}=$ ${\mathbf T} \cdot {\mathbf C}$, since this operation can never be included into the group $G_0$ of unitarily realized symmetries (because it can never commute with $H$). Therefore, the problem of classifying all block Hamiltonians $H^{(\lambda)}$, which we for brevity simply again denote by $H$,  has been reduced to the   problem of classifying all possible ways  in which $H$ can respond to time-reversal, charge-conjugation and chiral (sublattice) symmetries\footnote{We can also say that the problem has ben reduced to the finite problem of extending the unitarily realized symmetry group to the most
general symmetry group of a quantum mechanical Hamiltonian which also includes all possible {\it anti}-unitary symmetry operations. - See also Ref.s \cite{HeinznerHuckleberryZirnbauerCMP2005,ZirnbauerOxfordHandbook2010}.}.
- Note that we have {\it not} just  picked
"at random" three arbitrary  possible symmetries of quantum mechanics which happen to be time-reversal (${\bf T}$), charge-conjugation (${\bf C}$) and chiral symmetry (${\bf S}$) and used them
to classify quantum mechanical Hamiltonians $H$. Rather, after  systematically eliminating {\it all} {\it unitarily realized} symmetries, there are only three additional   symmetries left
that a quantum mechanical system can possibly possess: These are ${\bf T}$, ${\bf C}$ and ${\bf S}$. It is for this reason that  the current scheme
provides a {\it complete} classification of {\it all} single-particle Hamiltonians $H$.

\begin{figure}[!hbt]
  \includegraphics[width=\columnwidth]{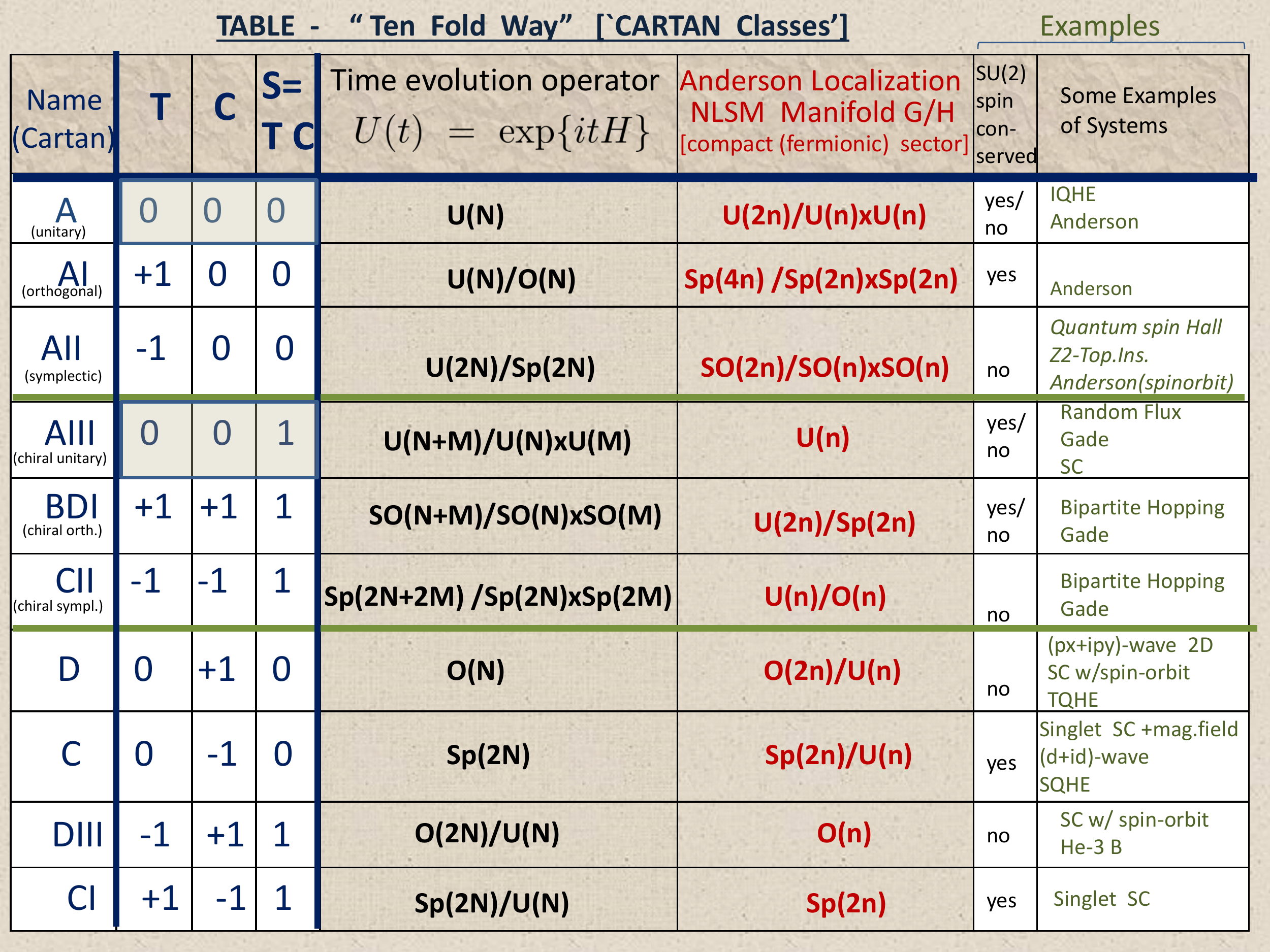}
  \caption{
  The ``Ten-Fold Way''. Listed are the ten generic symmetry classes of first quantized Hamiltonians. This is a complete list of Hamiltonians which possess no unitarily realized symmetries. As reviewed in the text, the different symmetry classes can be characterized by their  different behaviors of the first quantized Hamiltonian under time-reversal ($\bf T$), charge-conjugation ($\bf C$), and chiral symmetry (${\bf S}={\bf T} \cdot{\bf C }$).  The column \lq Name (Cartan)\rq \ lists the name given to the symmetric space appearing
in the column \lq Time evolution operator\rq \ by \'Elie Cartan in his classification scheme of symmetric spaces. The column \lq Anderson Localization NLSM Manifold G/H\rq 
\ lists the (compact sectors of the) target space of the Non-Linear Sigma Model describing Anderson Localization physics in the corresponding symmetry class.
} 
\label{fig:TenFoldWay} 
\end{figure}

The classification goal is now  easily achieved as follows:
Note that it would  at first appear that there are $3 \times 3$ $=9$ ways a first quantized Hamiltonian ${ H}$ can respond to time-reversal and particle-hole operations. This is not quite (but almost) true: As discussed, one needs to consider also the product ${\bf S} \equiv {\bf T} \cdot {\bf C}$.
It turns out that for  eight of the 9 choices the value of $S$ is uniquely fixed by the transformation property of the Hamiltonian under ${\bf T}$
and ${\bf C}$. These are the eight choices where the value of $T$ or $C$, or of both,  is not zero. There is however one of the nine cases, namely the case where the Hamiltonian is not invariant under time-reversal nor under  particle-hole operations, $T=C=0$,
where the value of $S$ is not fixed by the bebavior of ${H}$ under ${\bf T}$ and ${\bf C}$: It can be either $S=0$ or $S=1$. Therefore we obtain $(3\times 3 -1) +2$ $=10$ possibilities. 
Each of these 10 possibilities is called a "symmetry class". The 10 symmetry classes
are listed in FIG. (\ref{fig:TenFoldWay}).

\vskip .1cm

The column ``Time evolution operator'' of FIG.  (\ref{fig:TenFoldWay}) shows what type of $N\times N$ matrix the first quantized time-evolution
${U}(t)=$ $\exp\{ i t { H} \}$ is. For example, the first row ("Cartan Symmetry Class A") 
lists systems with a Hamiltonian ${H}$ that possesses neither time-reversal (${T}=0$),
nor charge-conjugation (${ C}=0$), nor chiral symmetry (${ S}=0$). There is no constraint on such a Hamiltonian,  so it is a general
hermitian $N\times N$ matrix (apart from conditions of  locality). Therefore, the time-evolution operator is a general a unitary $N\times N$ matrix which is the meaning of the entry $U(N)$
in the first row of the Figure. To illustrate how to obtain the other entries in this column  of the Figure, let us discuss the case of a Hamiltonian invariant under a (anti-unitary) time-reversal symmetry ${\bf T}$ which squares to plus the identity, ${\bf T}^2 =+ {\bf 1}$. This case is labeled by the "Cartan symbol" AI.  We know that in this case there exists a basis in which the Hamiltonian is represented by a real symmetric matrix ${H}_s$. To understand the nature of the time-evolution operator in this case, let us first choose an arbitrary hermitean Hamiltonian ${ H}$, and decompose it into symmetric and antisymmetric pieces, ${ H} = {1\over 2} ( {H} + { H}^t) + {1\over 2} ( { H} - { H}^t)$ $={ H}_s + { H}_a$, where the superscript ${}^t$ denotes the transposed matrix. Hence,  in a suitable basis, we can write the time-reversal symmetric Hamiltonian as ${H}_s = {H} - { H}_a$; upon exponentiation,
$\exp\{ i t { H} \} $ $= {\rm unitary}$ $\in U(N)$, and
$\exp\{ i t { H}_a \} =$ ${\rm orthogonal}$ $\in O(N)$. 
Therefore, the time-evolution operator of the time-reversal invariant system with ${\bf T}^2 = +1$  is an element of the 
coset space $\exp\{ i t {H}_s \}$ $\in U(N)/O(N)$. This is the meaning of the entry in the  4th row of the Figure (with heading: "Time evolution operator"). The form of the time-evolution operator in all remaining eight  cases  can be determined analogously. What is interesting (and surprising) is that the result  obtained for the list of ten 
time-evolution operators has a geometrical meaning. In the early parts of the last century, the mathematician \'Elie Cartan asked himself the following, seemingly completely unrelated  question: What are all possible generalizations of spheres? (A sphere is an example of a space that has a constant curvature everywhere.)  More precisely, can one write down a list of all possible Riemannian spaces (i.e. those that have a Riemannian metric) which have the same curvature everywhere (technically, where the Riemann curvature tensor is (covariantly) constant), and which have only a single curvature scale? Cartan found the answer in the year 1926: The list of `constant curvature spaces'  turns out to be precisely  the set of ten (coset) spaces listed under the column ``Time evolution operator'' of FIG. (\ref{fig:TenFoldWay}) !\footnote{ To be precise, the set of ten (coset) spaces listed in FIG. (\ref{fig:TenFoldWay}) coincide with Cartan's list of `{\it large}' symmetric spaces. Cartan found actually more than those ten spaces, the additional ones arising when the classical groups $U(N)$, $O(2N)$ and $Sp(2N)$ are replaced by  exceptional Lie groups. These additional symmetric spaces are however not of interest for the physics of topological insulators, since there  one is actually interested in the thermodynamic limit of the system, where $N$ goes to infinity. The additional, {\it exceptional}  spaces,  appearing  in Cartan's  list, do not possess a parameter like $N$ that can be taken to infinity.}

%\vskip .2cm

Let us now turn attention to the
6th column of  FIG. (\ref{fig:TenFoldWay}) 
 (with the heading 
 ``Anderson Localization NLSM Manifold G/H [compact (fermionic) sector]"). Note that the same set of 10 symmetric spaces  appears as in the previous column (Time evolution operator), except that their order is  permuted. In our context, this column refers to the $\overline{d}=$ $(d-1)$-dimensional {\it boundary}
of the  topological insulator in $d$ spatial  bulk dimensions.
A characteristic property of a non-interacting Fermion Topological Insulator is that while the bulk is insulating, there must always exist extended degees of freedom which are confined to the boundary\footnote{I.e. there must always exist eigenfunctions of the Hamiltonian with support on (near) the boundary which are extended along the boundary (i.e. are not square integrable).}, whose presence is protected by the topological nature of the bulk of the system. Physically, 
the presence of these extended degrees of freedom at the boundary implies that in contrast to the bulk, the boundaries conduct, depending on the case, electrical current or heat like a metal.
(The existence of such extended (gapless) degrees of freedom on the boundary of the fully gapped bulk state may be viewed as an operational definition of the topological insulator.)  These  extended  boundary degrees of freedom remain to be present when translational symmetry is broken, even if translational  symmetry breaking is arbitrarily strong (but the corresponding energy scale is still well  below the gap of the bulk state), because their existence is a consequence of the topological properties of the bulk, and translational symmetry (a unitarily implemented symmetry) is not necessary to protect the topological nature of the  quantum state.
The situation where translational invariance is broken (in practice often by the presence of impurities placed randomly within the sample, or due the presence of `random' potentials, which typically  result from the presence of such impurities) is 
commonly referred to as `disordered' or `random'. The fact that these boundary degrees of freedom remain extended in the presence of translational symmetry breaking  is very unusual because of the phenomenon of {\it Anderson localization}\cite{Anderson-P-WPhysRev109-1958-1492,Lee-P-A-Ramakrishnan-T-V-1985-RevModPhys57-1985-287}, which says that  in ordinary systems (which are not boundaries of Topological Insulators or Superconductors) spatially extended eigenstates of the Hamiltonian become localized (i.e. exponentially decaying in space),
 at least for sufficiently strong breaking of translational symmetry (e.g. due the presence of `disorder' potentials breaking translational symmetry).  In practice the presence of extended eigenstates
at the boundary of a topological insulator, even if translational symmetry is broken, implies that the boundary conducts electrical current or heat (similar to a metal), whereas the presence of only localized eigenstates  would mean that the boundary is an electrical or thermal insulator. The boundaries of Topological Insulators (Superconductors) therefore always evade  (``by definition") the phenomenon of Anderson localization.
The theoretical description of Anderson localization phenomena is known to be very systematic and geometrical.  For a Hamiltonian in one of the ten symmetry  classes, the system (in the current situation  the system in question is that at the boundary) is known to be described at length scales much larger than the `mean free path' (which decribes
the microscopic scale at which translational invariance is violated)  by a Non-Linear Sigma Model (NLSM). A NLSM is a system like that describing the classical statistical mechanics of a Heisenberg 
ferromagnet. The only difference is that while for the Heisenberg ferromagnet a unit vector `spin' is assigned to every point in space, which lives  on a  2-dimensional unit sphere,  in a general NLSM the unit vector `spin' is replaced by an element of one of the ten Cartan  symmetric spaces which, as we have mentioned, are all possible  generalizations of spheres.  These are listed in the 
6th  column of FIG.  (\ref{fig:TenFoldWay})  with heading
 ``Anderson Localization NLSM Manifold G/H [compact (fermionic) sector]".
\footnote{Note that the symmetric space G/H is a different space from the space of  which the time-evolution operator is an element.}
The general NLSM is decribed by a field theory whose (Boltzmann-type)  weight is $\exp \{ - S \}$ where
\begin{equation}
\label{NLSMaction}
S =
{1\over g}
\int 
%\int_{{\rm \tiny{boundary}}}
d^{\overline{d}} r \  \sum_{\mu=1}^{\overline{d}} \ 
Tr \left
( \partial_\mu \Phi(r) \partial_\mu \Phi(r) \right )
\end{equation}
and  the integral is over the $\overline{d}=$ $(d-1)$-dimensional boundary of the Topological Insulator in $d$ spatial dimensions.
Here $\Phi(r)$ is a matrix field\footnote{ For example, for symmetry class A, $\Phi(r) \in$ $G/H=$ $U(2n)/[U(n)\times U(n)]$.}
which is an element of the symmetric space G/H listed in the NLSM column of FIG. (\ref{fig:TenFoldWay}). [(The space G/H of which the `generalized Heisenberg spin' $\Phi(r)$ is an element, is usually referred to as the `target space',  or the `target manifold' of the NLSM.
\footnote{So far we have not yet specified the value the index $n$ in the column of 
 FIG. (\ref{fig:TenFoldWay}) is to take on.
 This is related to the technical details used when describing `disordered' or `random' systems. There are two
ways to do this. The first is the so-called replica trick, where one should take the limit $n\to 0$. This  way is less intuitive and we will not comment on it here. The other, equivalent way, is  to extend the  NLSM  target space in which the degrees of freedom
$\Phi(r)$ live, to contain fermionic degrees of freedom and an additional non-compact version of the space G/H; the resulting space then possesses (graded) `supersymmetry'. None of these details are important for the discussion here. This is because ultimately we will  only be interested in certain `topological terms' that can be added to the $S$ in Eq.(\ref{NLSMaction}), and it is known that those can only arise from the compact Bosonic part of the target space, which is the space listed in the ``NSLM'' column of FIG.  (\ref{fig:TenFoldWay}); here $n$ has to be a positive integer, and the theory is independent of the choice of $n$, as long as it is large enough (this is a consequence of the mentioned underlying supersymmetry). It turns out that the  non-compact Bosonic  part of the target space (not listed in the Figure), as well as the Fermionic part, can both not contribute to these topological terms.}.)

We end by noting, as already mentioned at the beginning of section \ref{LabelSectionClassificationOfTopInsSupercond}, that a very interesting formulation of the Ten-Fold Way from a more mathematical perspective was recently developed in  \cite{DanFreedGregMooreAnnalesHenriPoincare14-2013-1927}.

\vskip 2cm

\subsection{Classification by Topology of the Bulk: Translationally Invariant Case (K-Theory)}
\label{LabelSubSectionClassificationByTopolgyOfTheBulk}

\begin{figure}[!hbt]
\includegraphics[width=10cm]{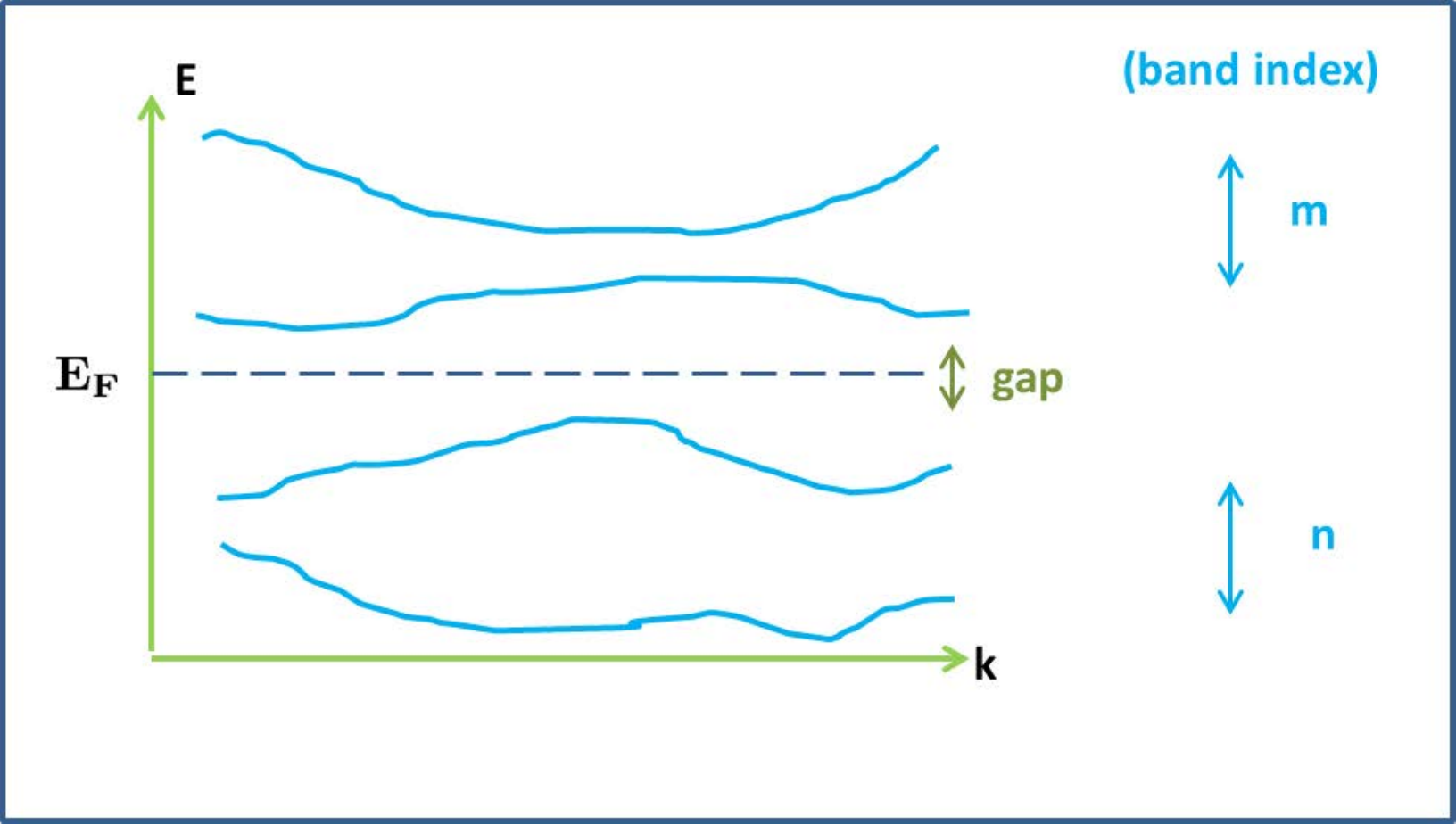}
 \caption{Typical band structure of  a gapped band insulator. There are $n$ filled bands, and $m$ empty  bands.
$E_F$ denotes the Fermi energy.
 } 
\label{fig:BandStructure} 
\end{figure}

In this section we review how topology can be implemented in the general framework of the Ten-Fold-Way classification of Hamiltonians if
we impose translational symmetry (so we can label all states by a momentum eigenvalue), and if
we impose the condition that there is an excitation  gap that separates all filled  from all empty bands, so that the bulk is a "band insulator". In particular, in the presence of translational symmetry we can  write  the single-particle Hamiltonian $H$ in momentum space in the form
\begin{equation}\label{BandStructure}
H(\vec k) \ |u_a (\vec k)\rangle \ = \ E_a(\vec k) \  |u_a(\vec k)\rangle 
\end{equation}
where ${\vec k}$ is the $d$-dimensional wavevector which is an element of  the Brillouin zone (a torus).   Here
$a$ denotes the band index; we consider the case of $n$ filled and $m$ empty bands - see FIG. (\ref{fig:BandStructure}).

Since we are interested in the topological properties of the system, we may continuously deform the Hamiltonian $H(\vec k)$
to bring it into a simplified form where all filled bands have energy $E=-1$ and all empty band have energy $E=+1$ ("spectral flattening"). 
By definition, 
any topological properties
will remain unchanged by such continuous deformations. We thereby obtain the "simplified Hamiltonian"
\begin{equation}\label{SimplifiedHamiltonianFlattenedSpectrum}
{\mathbf {\cal \bf Q}}(\vec k) = {\rm Hamiltonian \ where} \ \ \ E_a(\vec k) = 
 \begin{cases}
+1,  &   {\rm empty \ bands}  \\
-1,  &   {\rm filled \ bands}.
\end{cases}
\end{equation}

\vskip 1cm

\subsubsection{Basic Ideas underlying the Classification and Simplest Example for Classification in the Bulk}
\label{LabelBasicIdeasAndSimplestExampleOfTopologicalCluassificationInTheBulk}
In order to illustrate the idea of how to input information about topology, consider the simplest case of a Hamiltonian in a symmetry class that
has no symmetry conditions at all. This is symmetry class $A$ in FIG.  \ref{fig:TenFoldWay}. Since the
 simplified Hamiltonian ${\mathbf {\cal \bf Q}}(\vec k) $ has $n$ eigenvalues $E=-1$ and $m$ eigenvalues $E=+1$, it can be written in the form
\begin{equation}\label{SimplifiedHamiltonianClassAGrassmannian}
{\mathbf {\cal \bf Q}}(\vec k)  = {\cal U}(\vec k) \ \Lambda \ {\cal U}^\dagger(\vec k), \qquad {\rm where} \quad 
\Lambda=
\left ( 
\begin{matrix}
{\bf 1}_m & {\bf 0} \\
{\bf 0} & {\bf - 1}_n
\end{matrix}
\right ),
\quad {\rm and} \quad {\cal U}(\vec k) \in U(m+n).
\end{equation}
$\Lambda$ describes the eigenvalues, and
${\cal U}(\vec k) \in U(m+n)$ describes the collection of all eigenvectors of ${\mathbf {\cal \bf Q}}(\vec k)$.
Now, when ${\cal U}(\vec k) $ is of the form
\begin{equation}\label{LabelCosetsGrassmannian}
{\cal U}(\vec k) =
\left ( 
\begin{matrix}
U_1(\vec k) & 0 \\
0 &  U_2(\vec k)
\end{matrix}
\right ), \qquad {\rm where} \quad U_1(\vec k) \in U(m) \quad {\rm and} \quad  U_2(\vec k) \in U(n),
\end{equation}
${\mathbf {\cal \bf Q}}(\vec k) $ remains unchanged from the value $=\Lambda$ that it takes when   ${\cal U}(\vec k)={\bf 1}_{m+n}$ . Therefore ${\mathbf {\cal \bf Q}}(\vec k) $ only changes when  ${\cal U}(\vec k)$ is a non-trivial  element 
of the coset space $G_{m,m+n}({\mathbf C})=$ $U(m+n)/[U(m)\times U(n)]$ which is conventionally called the (complex)  "Grassmannian"
$G_{m,m+n}({\mathbf C})$,  i.e. when 
${\cal U}(\vec k) \in U(m+n)/[U(m)\times U(n)]$.
Therefore we have established that every ground state of the simplified Hamiltonian  ${\mathbf {\cal \bf  Q}}(\vec k) $ (where all $n$ bands are filled,
and all $m$ bands are empty) is described by a  map from the Brillouin zone (BZ)  into the (complex) Grassmannian,
\begin{eqnarray}\label{MapBrillouinZoneComplexGrassmannian}
{\mathbf {\cal \bf Q}}:  BZ &\rightarrow& G_{m,m+n}({\mathbf C})= U(m+n)/[U(m)\times U(n)]
\\
{\vec k} &\rightarrow& {\mathbf {\cal \bf Q}}(\vec k).
\end{eqnarray}
Each such map describes a ground state, i.e. a filled Fermi sea of occupied states. Now, we want to know the answer to the question: 
How many different ground states of this kind,
i.e. how many such different  maps are there that cannot be continuously deformed into each other? For simplicity, let us assume that the Brillouin
zone is a $d$-dimensional sphere (and not a $d$-dimensional torus, which it really is - we come back to the  torus case shortly).
In this case (i.e. for a $d$-dimensional "spherical" Brillouin zone) the answer is well known: It  is given by the 
\begin{equation}\label{HomotopyGroupsOfComplexGrassmannian}
{Homotopy  \ Group} \quad
\pi_d\left (U(m+n)/[U(m)\times U(n)]\right )
\end{equation}
 which is known.

For example, when the spatial dimension is $d=2$, then it is known that
$\pi_{d=2}\left (U(m+n)/[U(m)\times U(n)]\right ) =$  ${\bf Z}$, the set of all integers. This means that for every integer there is a ground state, and  ground
states which are  assigned different integers  cannot be continuously deformed into each other (without closing the gap of the underlying bulk Hamiltonian  $H(\vec k)$).  This  particular result
is of course  not new, since the topological quantum state in $d=2$ spatial dimensions that appears for a  single-particle Hamiltonian on which
no symmetry conditions are imposed is the 2D integer quantum Hall state. The mentioned integer just counts the number of branches
of chiral edge states, which tells us exactly which integer quantum Hall plateau the state describes. 

Let us try the case of spatial dimension $d=3$, where $\pi_{d=3}\left ( U(m+n)/[U(m)\times U(n)]\right ) =$ $ \{\bf 1\}$, i.e. the Homotopy Group is trivial and consists only of a single element. Therefore, our calculation predicts that in the present  symmetry class (class A) there is only one state: All other states can be continously
deformed into it. We have recovered the well known result that there are no integer quantum Hall states in $d=3$ spatial dimension.

\vskip 1cm

\subsubsection{General Classification in the Bulk - Results and K-Theory}
\label{LabelGeneralCluassificationInTheBulkResultsAndKTheory}

Conceptually, the classification of Topological Insulators and Superconductors employing the current approach of determining  the topology of the bulk states in the presence of translational invariance, proceeds analogously for all symmetry classes. In particular, for each of the 10 symmetry classes
listed in FIG. \ref{fig:TenFoldWay}, we need to find the corresponding "simplified Hamiltonian" ${\mathbf {\cal \bf Q}}(\vec k)$. This task
however turns out to be 
more  complicated than that in (\ref{MapBrillouinZoneComplexGrassmannian}) for symmetry class A, because 
the simplified Hamiltonian ${\cal \bf Q}(\vec k)$  now has to reflect the
invariance under the symmetries (time-reversal, charge-conjugation, chiral) which define the symmetry class. The corresponding list
is displayed  in {TABLE \ref{fig:ListOfSimplifiedHamiltoniansProjectors}}. 
%\begin{figure}[!hbt]
%%\includegraphics[width=\columnwidth]{TableListOfProjectors.pdf}
%\includegraphics[width=11cm]{TableListOfProjectors.pdf}
 %\caption{List of "simplified Hamiltonians" ${\mathbf {\cal \bf Q}}(\vec k)$ for all ten symmetry classes. (The "simplified Hamiltonians"
%are also called "projectors",  since they are related to projection operators onto the filled Fermi sea\cite{FootnoteProjectors}.
%%footnote{Here ${\cal Q}(\vec k)=$
%%${\bf 1} - 2 P(\vec k)$ where $P(\vec k)$ is the projection opertor onto the filled Bloch states.}
%As mentioned in the text, the (complex) Grassmannian is the coset space $G_{m,n}({\mathbf C})= U(m+n)/[U(m)\times U(n)]$.
%The meaning of the matrices $q(k)$, which appear in all symmetry classes possessing chiral symmetry, is briefly explained in Appendix %\ref{LabelSectionAppendixChiralClassesFIGSimplifiedHamiltonians}.)
 %- From Ref. \cite{SchnyderRyuFurusakiLudwigPRB2008}.
 %} 
%label{fig:ListOfSimplifiedHamiltoniansProjectors} 
%\end{figure}
\begin{table}[!hbt]
\begin{center}
\resizebox{0.65\textwidth}{!}{
%\begin{table*}[t]
%\begin{center}
{
\renewcommand{\arraystretch}{1.5}
\begin{tabular}{c c }\hline \hline
& \\
Cartan Class 
& Simplified Hamiltonian\\ 
& \\ \hline
A &
$\{\, {\cal \bf Q}(k)\in G_{m,m+n}(\mathbb{C})\, \}$
\\ 
AI &
$\{\, {\cal \bf Q}(k)\in G_{m,m+n}(\mathbb{C})\, |\, {\cal \bf Q}(k)^*={\cal \bf Q}(-k)\, \}$ 
\\ 
AII &
$\{\, {\cal \bf Q}(k)\in G_{2m,2(m+n)}(\mathbb{C})\,|\,({i}\sigma_y)
 {\cal \bf Q}(k)^*(-{i}\sigma_y) ={\cal \bf Q}(-k)\, \}$
\\ 
AIII
& $\{\, {\bf q}(k)\in \mathrm{U}(m)\, \}$
\\ 
BDI
&
$\{\, {\bf q}(k)\in \mathrm{U}(m)| {\bf q}(k)^*={\bf q}(-k)\, \}$
\\ 
CII
&
$\{\, {\bf q}(k)\in \mathrm{U}(2m)\, |\,({i}\sigma_y) {\bf q}(k)^*(-{i}\sigma_y)
 ={\bf q}(-k)\, \}$
\\ 
D &
$\{\, {\cal \bf Q}(k)\in G_{m,2m}(\mathbb{C})\, |\, \tau_x {\cal \bf Q}(k)^* \tau_x = -{\cal \bf Q}(-k)\,\}$ 
\\ 
C &
$\{\, {\cal \bf Q}(k)\in G_{m,2m}(\mathbb{C})\,|\,\tau_y {\cal \bf Q}(k)^* \tau_y = -{\cal \bf Q}(-k)\,\}$
\\ 
DIII &
$\{\, {\bf q}(k)\in \mathrm{U}(2m)\,|\, {\bf q}(k)^T = -{\bf q}(-k)\,\}$
\\
CI
  &$\{\, {\bf q}(k)\in \mathrm{U}(m) \, |\, {\bf q}(k)^T ={\bf q}(-k) \, \}$
\\ 
 \hline \hline
\end{tabular}
}
}
\end{center}
\caption{List of "simplified Hamiltonians" ${\mathbf {\cal \bf Q}}(\vec k)$ for all ten symmetry classes\cite{SchnyderRyuFurusakiLudwigPRB2008}.
%footnote{Here ${\cal Q}(\vec k)=$
%${\bf 1} - 2 P(\vec k)$ where $P(\vec k)$ is the projection opertor onto the filled Bloch states.}
As mentioned in the text, the (complex) Grassmannian is the coset space $G_{m,m+n}({\mathbf C})= U(m+n)/[U(m)\times U(n)]$.
The meaning of the matrices ${\bf q}(k)$, which appear in all symmetry classes possessing chiral symmetry, is briefly explained in Appendix \ref{LabelSectionAppendixChiralClassesFIGSimplifiedHamiltonians}.
}
\label{fig:ListOfSimplifiedHamiltoniansProjectors}
\end{table}

The list of 10 "simplified Hamiltonians" displayed in TABLE \ref{fig:ListOfSimplifiedHamiltoniansProjectors} exhibits an interesting structure which is most easily revealed by 
looking at this list either
in  the special case of zero spatial dimensions, $d=0$, or in general spatial dimensions  $d$ at special points ${\vec k}=$ ${\vec k}_0$ in the Brillouin
zone which satisfy the property that $+{\vec k}_0$ and $-{\vec k}_0$ differ by a reciprocal lattice vector.
The list of the corresponding "simplified Hamiltonian" matrices ${\mathbf {\cal \bf  Q}}(\vec k)$ is displayed in the rightmost column 
TABLE \ref{fig:ListOfClassifyingSpacesForAllSYmmetryClasses}.  Before commenting on  this list, let us first elaborate on the subdivision of the 10 Cartan symmetry classes into
"complex" and "real" ones. A look at FIG. \ref{fig:TenFoldWay} reveals that there are exactly two  symmetry classes (classes A and AIII) which
don't possess any  invariance under either of the two anti-unitary symmetries ${\bf T}$ or ${\bf C}$. All remaining eight symmetry classes
are invariant under one or both of these anti-unitary symmetries: In particular,  recall from (\ref{DEFTimeReversalSingleParticleHamiltonian})
and
(\ref{DEFChargeConjugationSingleParticleHamiltonian})
that each of the two anti-unitary symmetries imposes a {\it reality condition} on the Hamiltonian $H$. Thus we see that classes A and AIII
are the only two classes in which  no reality condition whatsoever is imposed on the Hamiltonian. It is for this reason that they are called "complex". On the other hand, 
in the remaining eight symmetry classes the Hamiltonian satisfies at least one reality condition. For this reason these symmetry classes
are called "real".
Now we come to the main point of TABLE \ref{fig:ListOfClassifyingSpacesForAllSYmmetryClasses}: It   reveals the remarkable observation
that  the  simplified Hamiltonians
${\bf {\cal \bf Q}}(\vec k)$ run  again over the 10 symmetric spaces. But now these spaces  appear yet  in a different order as compared to the order in the 
column "Time evolution operator", and to the column  "Anderson localization" in FIG. \ref{fig:TenFoldWay} .

%\begin{figure}[!hbt]
%%\includegraphics[width=\columnwidth]{CartanClassesClassifyingSpaces.pdf}
%\includegraphics[width=16cm]{CartanClassesClassifyingSpaces.pdf}
% \caption{List of time-evolution operators and "simplified Hamiltonians" ${\mathbf {\cal \bf  Q}}(\vec k)$ in either zero spatial dimension, or in %arbitrary spatial dimensions at special points ${\vec k}=$ ${\vec k}_0$ in the Brillouin
%zone which satisfy the property that $+{\vec k}_0$ asnd $-{\vec k}_0$ differ by a reciprocal lattice vector. The sets of "simplified Hamiltonians" %${\mathbf {\cal \bf  Q}}(\vec k)$ listed in the rightmost column are the so-called "Classifying Spaces" of K-Theory. They are called $C_0$ and %$C_1$
%for the complex symmetry classes, and $R_q$ with $q=0, 1, 2, ..., 7$ (modulo 8) in the real symmetry classes. The Classifying Spaces run again %over all symmetric spaces, but in an order that is different from that enountered in the column "Time evolution operator". Note that in the left %column  "Time evolution operator" the numbers $N$ and $M$ physically denote numbers of lattice sites. In the right column "Classifying space" the %numbers
%$N$ and $M$ denote the numbers of filled and empty bands, generalizing the setup in FIG. \ref{fig:BandStructure}.
%- From Ref. \cite{RyuSchnyderFurusakiLudwigNJPhys2010}.
%\label{fig:ListOfClassifyingSpacesForAllSYmmetryClasses} 
%\end{figure}

\begin{table}[!hbt]
\begin{center}
\resizebox{0.99\textwidth}{!}{
%\begin{small}
{
\renewcommand{\arraystretch}{1.3}
\begin{tabular}{cccccc}
\hline
Cartan
&Time evolution operator\ 
& 
& & ${\cal \bf Q}({\vec k})$ \\
label
& 
$\exp\{ \mathrm{i} t {\cal H}\}$
& 
& & Classifying  Space
           \\ \hline \hline 
A &
$\mathrm{U}(N)\times \mathrm{U}(N)/\mathrm{U}(N)$
& \qquad \qquad   ("complex") \qquad \qquad  
&
&
$\mathrm{U}(N+M)/\mathrm{U}(N)\times \mathrm{U}(M)$&$=C_0$ 
\\ 
{\bf AIII }& 
$\mathrm{U}(N+M)/\mathrm{U}(N)\times \mathrm{U}(M)$
& 
& &  $\mathrm{U}(N)\times \mathrm{U}(N)/ \mathrm{U}(N)$ & $=C_1$
\\  \hline 
AI &
$\mathrm{U}(N)/\mathrm{O}(N)$
&
& &
$\mathrm{O}(N+M)/\mathrm{O}(N)\times \mathrm{O}(M)$ & $=R_0$
\\ 
{\bf BDI}
&
$\mathrm{O}(N+M)/\mathrm{O}(N)\times \mathrm{O}(M)$
&
& & 
$\mathrm{O}(N)\times \mathrm{O}(N)/\mathrm{O}(N)$ &$=R_1$
\\  
D &
$\mathrm{O}(N)\times \mathrm{O}(N)/\mathrm{O}(N)$
& 
&& $\mathrm{O}(2N)/ \mathrm{U}(N)$ & $=R_2$
\\
{\bf DIII} &
$\mathrm{SO}(2N)/\mathrm{U}(N)$
&  \qquad \qquad  ("real") \qquad  \qquad 
& & $\mathrm{U}(2N)/ \mathrm{Sp}(2N)$ & $=R_3$ 
\\ 
AII &
$\mathrm{U}(2N)/\mathrm{Sp}(2N)$
&
& &
$\mathrm{Sp}(N+M)/\mathrm{Sp}(N)\times \mathrm{Sp}(M)$ &$=R_4$ 
\\ 
{\bf CII} &
$\mathrm{Sp}(N+M)/\mathrm{Sp}(N)\times \mathrm{Sp}(M)$
& 
& & $\mathrm{Sp}(N)\times\mathrm{Sp}(N)/\mathrm{Sp}(N)$ &$=R_5$
\\ 
C &
$\mathrm{Sp}(2N)\times \mathrm{Sp}(2N)/\mathrm{Sp}(2N)$
& 
& & $\mathrm{Sp}(2N)/ \mathrm{U}(N)$ &$=R_6$ 
\\ 
{\bf CI} &
$\mathrm{Sp}(2N)/\mathrm{U}(N)$
& 
& & $\mathrm{U}(N)/ \mathrm{O}(N)$ &$=R_7$ 
\\ \hline
\end{tabular}
%\end{small}
}
}
\end{center}
\caption{
List\cite{RyuSchnyderFurusakiLudwigNJPhys2010} of time-evolution operators and "simplified Hamiltonians" ${\mathbf {\cal \bf  Q}}(\vec k)$ in either zero spatial dimension, or in arbitrary spatial dimensions at special points ${\vec k}=$ ${\vec k}_0$ in the Brillouin
zone which satisfy the property that $+{\vec k}_0$ asnd $-{\vec k}_0$ differ by a reciprocal lattice vector. The sets of "simplified Hamiltonians" ${\mathbf {\cal \bf  Q}}(\vec k)$ listed in the rightmost column are the so-called "Classifying Spaces" of K-Theory. They are called $C_0$ and $C_1$
for the complex symmetry classes, and $R_q$ with $q=0, 1, 2, ..., 7$ (modulo 8) in the real symmetry classes. The Classifying Spaces run again over all symmetric spaces, but in an order that is different from that enountered in the column "Time evolution operator". Note that in the left column  "Time evolution operator" the numbers $N$ and $M$ physically denote numbers of lattice sites. In the right column "Classifying space" the %numbers
$N$ and $M$ denote the numbers of filled and empty bands, generalizing the setup in FIG. \ref{fig:BandStructure}.
%\label{RMT, NLsM}
%\label{fig:TableThreeOccurrencesOfTenCartanSpaces}
\label{fig:ListOfClassifyingSpacesForAllSYmmetryClasses} 
}
\end{table}

At this point is it basically clear conceptually  how to go about the classification in the bulk. We have to repeat the analysis of section \ref{LabelBasicIdeasAndSimplestExampleOfTopologicalCluassificationInTheBulk} for all the remaining nine  "simplified Hamiltonians"
${\bf {\cal  \bf Q}}(\vec k)$ listed in TABLE \ref{fig:ListOfSimplifiedHamiltoniansProjectors}. It turns out that this is a rather involved task,
due to constrains arising from the three symmetry operations defining the ten symmetry classes. However, it turns out that K-Theory
is the tool that precisely answers this question. While we will not review here  the apparatus of K-Theory, it is easy
to state the result and explain the physical features appearing in it. Let us list the result for the real symmetry classes. They are denoted by the symbols $R_q$ which are defined in the rightmost column of TABLE  \ref{fig:ListOfClassifyingSpacesForAllSYmmetryClasses}. 
The result due to Kitaev  reads\cite{KitaevLandau2009}
\begin{equation}\label{BulkClassificationKTheoryResult}
\pi \left (  \overline{{\rm T}^d}, R_q \right) \ \ =  \ \ \pi_0 \left( R_{q-d}\right ) \ \  \oplus \ \  \bigoplus_{s=0}^{d-1} \  \ 
\left ( 
\begin{matrix}
d \\
s
\end{matrix}
\right )
 \ \  
\pi_0\left ( R_{q-s} \right ).
\end{equation}
The symbol $\pi \left (  \overline{{\rm T}^d}, R_q \right)$ appearing on the left hand side of the above equation
generalizes  the Homotopy Groups appearing in (\ref{HomotopyGroupsOfComplexGrassmannian}). The generalization is twofold.
The above symbol generalizes (i)  the Homotopy Group to maps from the actual
d-dimensional Brillouin zone torus (hence the appearance of $T^d$)  and not from 
the  d-dimensional "spherical  Brillouin zone",
and (ii) to all the eight real symmetry classes which are labeled by the symbold $R_q$, defined
in TABLE  \ref{fig:ListOfClassifyingSpacesForAllSYmmetryClasses}, 
taking into account all the constraints listed in TABLE \ref{fig:ListOfSimplifiedHamiltoniansProjectors} (hence the "bar" over $T^d$).
 The result on the right hand side of (\ref{BulkClassificationKTheoryResult})
contains nothing but the zeroth Homotopy Groups of the various Classifying spaces $R_q$. (These are known and will be listed explicitly   in subsequent  Tables or Figures.) The result 
on the right hand side of (\ref{BulkClassificationKTheoryResult}) consists of two pieces, the first piece "$\pi_0 \left( R_{q-d}\right )$"
and remaining sum "$\bigoplus_{s=0}^{d-1} \left ( 
\begin{matrix}
d \\
s
\end{matrix}
\right )
 \ \  
\pi_0\left ( R_{q-s} \right )$"
. The first piece turns out to be the universal piece of interest to us, which is not tied to the presence of translational
invariance.\footnote{Clearly, since we are here discussing Topological Phases that are not protected by any unitarily realized symmetry, the presence
of translational invariance - one of the unitarily realized symmetries - should be of no relevance here. In particular, destroying translational
invariance by e.g. adding randomly placed impurities in the system, or by destroying the crystal lattice altogether by allowing for a finite
density of lattice defects, should have no effect on the topological properties we are discussing.} If this  first piece 
is non-trivial we have what is called a "strong Topological Insulator or Superconductor", whose topological properties would persist even if translational symmetry was broken. On the other hand it is clearly seen already from the structure of
the second piece that it relies on the presence of translational order: The various $d$ terms in the summand of the second piece
describe the presence or
absence of a Topological Insulator or Superconductor phase on submanifolds of dimension $s=$ $(d-1), (d-2), ..., 0$. In particular, the term
with $s=(d-1)$ describes the presence or absence of a Topological Insulator (Superconductor) in hyperplanes of the $d$-dimensional
lattice (of codimension $=1$), $s=(d-2)$ describes submanifolds of the lattice of condimension $=2$ (e.g. one lines in a $d=3$ dimensional lattice), etc.. The term $s=0$ describes such toplogical properties associated with points. The factor $\left ( {}^d_s \right )$ counts the number
of ways the $s$-dimensional objects can be placed into the $d$-dimensional lattice.  None of these lower dimensional objects
would be defined in the absence of the lattice. For this reason the second piece characterizes  "weak Topological Insulators (Superconductors)" which  can exist in the system.\footnote{These weak Topological Insulators (Superconductors) were also identified  in \cite{RyuSchnyderFurusakiLudwigNJPhys2010} using a physical argument.}
Finally, there is a corresponding expression for the two complex symmetry classes A and AIII, and the corresponding
complex Classifying Spaces called  $C_q$, $q=0,1$. By keeping only the first, universal piece on the right hand side of (\ref{BulkClassificationKTheoryResult}) as well as the corresponding "complex" version, one arrives at the Table of Topological Insulators and Superconductors displayed in 
TABLE \ref{fig:TableOfTopInsSupercond}. 
%\footnote{It should be noted that a somewhat  different approach to topological band theory has recently been developed in
%\cite{KennedyZirnbauer-arXiv-1409.2537}; see also the contribution \cite{KennedyZirnbauer-Nobel-arXiv-1412.4808} to the present  Nobel Symposium. The reader is %urged to consult these papers for further details.}

We end by noting that, as already  mentioned at the beginning of section \ref{LabelSectionClassificationOfTopInsSupercond}, a somewhat different
approach to topological band theory, which does not use K-Theory, was very recently developed in
\cite{KennedyZirnbauer-arXiv-1409.2537} (see also the contribution \cite{KennedyZirnbauer-Nobel-arXiv-1412.4808} to the present  Nobel Symposium).

%\begin{figure}[!hbt]
%%\includegraphics[width=\columnwidth]{TableOfTopInsSupercond.pdf}
%\includegraphics[width=16cm]{TableOfTopInsSupercond.pdf}
 %\caption{Table of Topological Insulators and Superconductors
%- From Ref. %\cite{SchnyderRyuFurusakiLudwigPRB2008,SchnyderRyuFurusakiLudwigLandau2009,KitaevLandau2009,RyuSchnyderFurusakiLudwigNJPhys2010}.} 
%\label{fig:TableOfTopInsSupercond} 
%\end{figure}

%%%%%%%%%%%%%%%%%%%%%%%%%%%%%%%%%%%%
%\begin{table}[t!]
\begin{table}[!hbt]
%\begin{table}[h]
%\paragraph{complex case:}
\begin{center}
\resizebox{0.80\textwidth}{!}{
%\begin{small}
{
\renewcommand{\arraystretch}{1.1}
\begin{tabular}{lccccccccc}\hline
Cartan$\backslash d$ & \quad  0 \quad   & \quad  1 \quad  & \quad  2 \quad  & \quad  3 \quad  & \quad  4 \quad  & \quad  5 \quad  & \quad  6 \quad  
& \quad  7 \quad  & \quad  8 \quad  \\ \hline\hline
{\footnotesize  \it Complex   case:} &  &  &  &  &  &  &  &  &   \\ 
A   &  $\mathbb{Z}$   &   0   &   $\mathbb{Z}$   &   0  
    &  $\mathbb{Z}$   &   0   &  $\mathbb{Z}$   &  0  
    & $\mathbb{Z}$ \quad   $\cdots$
\\
%\hline
AIII  & 0 & $\mathbb{Z}$ & 0 & $\mathbb{Z}$
      & 0 & $\mathbb{Z}$ & 0 & $\mathbb{Z}$
      & 0  \quad  $\cdots$
\\
%\hline
%\end{tabular}
%\end{center}
%\paragraph{real case:}
%\begin{center}
%\begin{tabular}{cccccccccccccc}\hline
%Cartan$\backslash d$ & \quad  0 \quad  & \quad  1 \quad  & \quad  2 \quad  & \quad  3 \quad  & \quad  4 \quad  &  \quad 5 \quad  & \quad  6 \quad  &
%\quad  7 \quad  & 8 & 9 & 10 & 11 & 12\\ \hline\hline
{\footnotesize \it  Real   case:} &  &  &  &  &  &  &  &  &   \\ 
AI  & $\mathbb{Z}$ & 0 & 0 
    & 0 & $2\mathbb{Z}$ & 0 
    & $\mathbb{Z}_2$ & $\mathbb{Z}_2$ & $\mathbb{Z}$ 
     \quad $\cdots$ \\ 
%\hline
BDI & $\mathbb{Z}_2$ & $\mathbb{Z}$ & 0 & 0 
    & 0 & $2\mathbb{Z}$ & 0 
    & $\mathbb{Z}_2$ & $\mathbb{Z}_2$  \quad  $\cdots$ \\ 
%\hline
D   & $\mathbb{Z}_2$ & $\mathbb{Z}_2$ & $\mathbb{Z}$ 
    & 0 & 0 & 0 
    & $2\mathbb{Z}$  & 0 & $\mathbb{Z}_2$ 
     \quad $\cdots$ \\ 
%\hline
DIII& 0 & $\mathbb{Z}_2$ & $\mathbb{Z}_2$ & $\mathbb{Z}$ 
    & 0 & 0 & 0 
    & $2\mathbb{Z}$  & 0 \quad    $\cdots$ \\ 
%\hline
AII & $2\mathbb{Z}$  & 0 & $\mathbb{Z}_2$ 
    & $\mathbb{Z}_2$ & $\mathbb{Z}$ & 0 
    & 0 & 0 & $2\mathbb{Z}$ 
      \quad $\cdots$\\ 
%\hline
CII & 0 & $2\mathbb{Z}$  & 0 & $\mathbb{Z}_2$ 
    & $\mathbb{Z}_2$ & $\mathbb{Z}$ & 0 
    & 0 & 0  \quad $\cdots$\\ 
%\hline
C   & 0  & 0 & $2\mathbb{Z}$  
    & 0 & $\mathbb{Z}_2$  & $\mathbb{Z}_2$ 
    & $\mathbb{Z}$ & 0 & 0 
     \quad $\cdots$ \\ 
%\hline
CI  & 0 & 0  & 0 & $2\mathbb{Z}$  
    & 0 & $\mathbb{Z}_2$  & $\mathbb{Z}_2$ 
    & $\mathbb{Z}$ & 0  \quad  $\cdots$ \\ \hline
\end{tabular}
}
}
\end{center}
\caption{
Table of Topological Insulators and Superconductors\cite{SchnyderRyuFurusakiLudwigPRB2008,SchnyderRyuFurusakiLudwigLandau2009,KitaevLandau2009,RyuSchnyderFurusakiLudwigNJPhys2010} .
}
\label{fig:TableOfTopInsSupercond} 
\end{table}

%%%%%%%%%%%%%%%%%%%%%%%%%%%%%%%%%%%%%%%%%%%%%%%%%

%Classification of topological insulators and  superconductors as a function of spatial dimension $d$
%and symmetry class, indicated by the ``Cartan label'' (first column). The definition of the ten  generic
%symmetry classes of single particle Hamiltonians (due to Altland and Zirnbauer\cite{Zirnbauer96,Altland97}%)
%is given in  table~ \ref{TableSymmetryClassesTwo}. The symmetry classes are grouped in two separate lists,
%complex and real cases, depending on whether the Hamiltonian is complex, or whether one
%(or more) reality conditions (arising from time-reversal or charge-conjugation symmetries)
%are imposed on it; the symmetry classes are ordered in such a way
%that a periodic pattern in dimensionality becomes visible \cite{KitaevLandau100Proceedings}.
%(See also the discussion in subsection \ref{Review-Classification-Hamiltonian} and table \ref{homotopy}.)
%The symbols $\mathbb{Z}$ and $\mathbb{Z}_2$ indicate that the topologically distinct phases within a given 
%symmetry class of topological insulators (superconductors) are characterized by an integer invariant ($\mathbb{Z}$),
%or a $\mathbb{Z}_2$ quantity, respectively.  The symbol ``$0$'' denotes the case when
%there exists no topological insulator (superconductor), i.e., when all quantum ground states are topologically equivalent to the trivial state.
%}
%\label{periodic table}
%%%%%%%%%%%%%%%%%%%%%%%%%%%%%%%%%%%%%%%%%%%%%%

\vskip 2cm

\subsection{Classification by Lack of Anderson Localization on the Boundary (Non-Linear Sigma Models)}
\label{LabelSubSectionClassificatonByLackOfAndersonLocalizationOnSurface}

The technically simplest way to obtain the Table of Topological Insulators and Superconductors, TABLE  \ref{fig:TableOfTopInsSupercond},
is to focus on the classification of {\it boundaries} of the system, as opposed to the approach in the previous section \ref{LabelSubSectionClassificationByTopolgyOfTheBulk} which focused on the bulk. The characterization  of  Topological Phases in terms of  their boundaries
has proven more generally to be one of the most successful tools in this subject area. This is related to the fact that these boundaries must 
exhibit an "anomaly" of some kind, a fact that was first recognized in its general form in the context of non-interacting Fermionic
Topological Insulators and Superconductors in \cite{RyuMooreLudwig2012}, and this is  discussed in section \ref{LabelSubSectionPerspectiveOfQuantumAnomalies} below. The
characterization of more general Topological Phases, including interacting systems, by anomalies at their boundaries has become
 a key tool in this area. (See also the comments in sections \ref{sectionEntanglementPerspective} and \ref{LabelSubSectionPerspectiveOfQuantumAnomalies}.)
In the case of non-interacting Fermionic Topological Insulators (Superconductors) the anomalous properties of the sample boundaries
manifest themselves through the fact that  the boundaries must always possess extended states. This means that the boundaries
always conduct electrical current or heat similar to a metal. Because of the well known phenomenon of Anderson localization this provides
a very convenient and quick way to classify Toplogical Insulators (Superconductors), as we will now review.

As already mentioned above, 
Anderson localization is the phenomenon that, at least for sufficiently  strong breaking of translational invariance (typically introduced
by potentials that arise from randomly located impurities - "disorder"),  spatially extended eigenstates of the Hamiltonian $H$  
 tend to become
localized in space (typically exponentially decaying in space about a point)\cite{Anderson-P-WPhysRev109-1958-1492, Lee-P-A-Ramakrishnan-T-V-1985-RevModPhys57-1985-287}. The resulting phase is the so-called {\it Anderson Insulator} which, hence, typically
always occurs at least for sufficiently strong breaking of translational invariance. Now, the point is that the Anderson Insulator is a phase
that does not conduct electrical current nor heat, and therefore is not allowed to occur at the boundaries of a Topological Insulator or
Superconductor. Thus, the boundary of any Topological Insulator or Superconductor must entirely evade the phenomeon of
Anderson Localization. As we will now  see, this  condition can be fairly easily exploited at the technical level, and this directly yields
the classification of Topological Insulators (Superconductors), TABLE \ref{fig:TableOfTopInsSupercond}.

\vskip .1cm

In summary, the current method of classification amounts to the following: {\it We reduce the problem of classifying Topological Insulators (Superconductors) in $d$ spatial dimensions to a problem of Anderson Localization in $\overline{d}=$ $(d-1)$ dimensions (i.e.
"at the boundary"). By studying the lack of Anderson Localization in $\overline{d}=$ $(d-1)$  spatial
dimensions, we solve the classification problem  of Topological Insulator (Superconductors) in $d$ spatial dimensions.}

\vskip .1cm

\begin{figure}[!hbt]
\includegraphics[width=18.5cm]{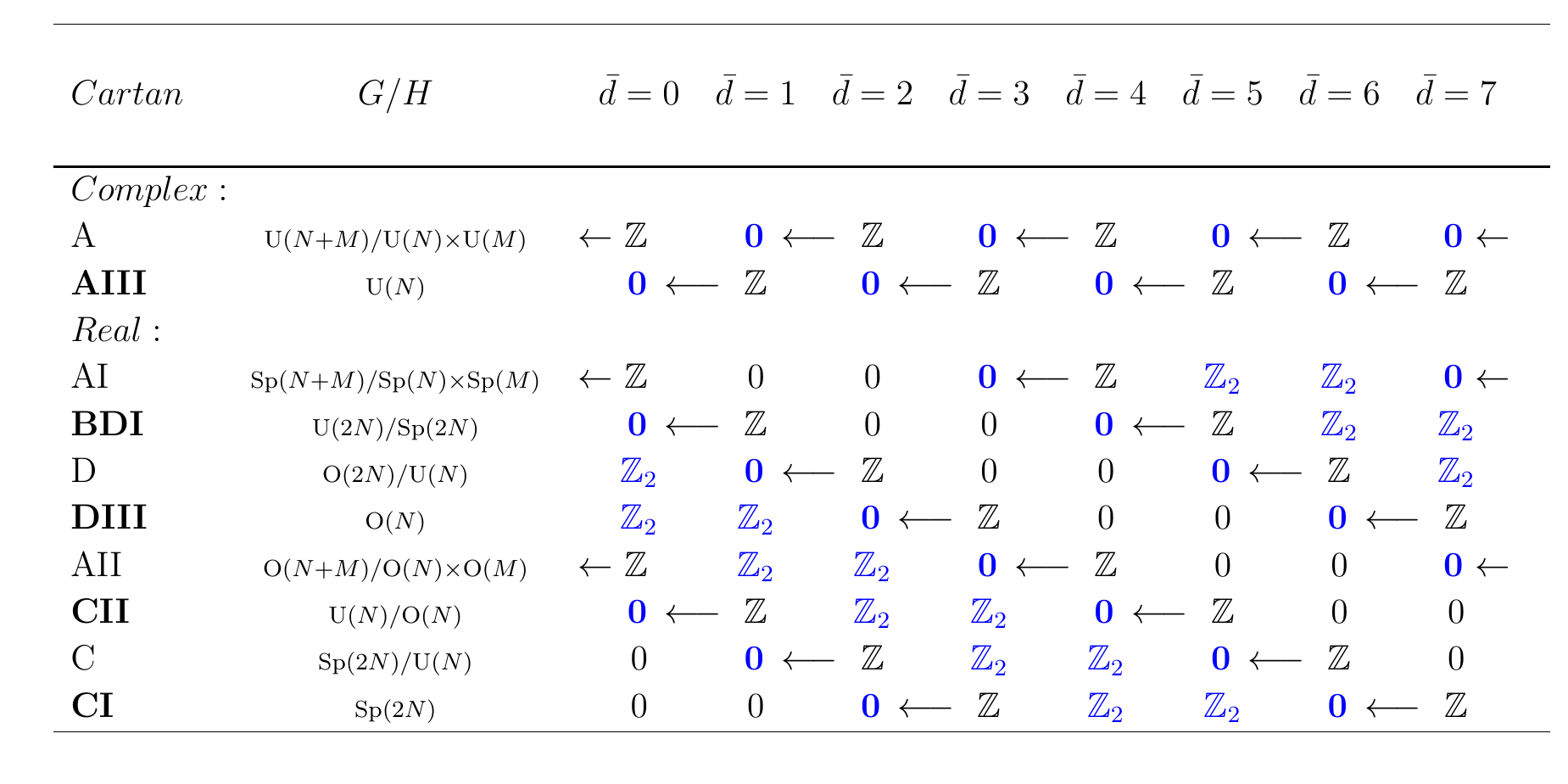}
 \caption{Table of Homotopy Groups $\pi_s \left ( G/H\right )$  of the ten symmetric spaces appearing in the 
Ten-Fold-Way.\cite{ LundellLectNotesMath1992,RyuSchnyderFurusakiLudwigNJPhys2010}.
%Lundell-A-T-1992-Lect.-Notes-Math.-1509-2509}\cite{RyuSchnyderFurusakiLudwigNJPhys2010}  
Here, according to (\ref{ExistenceOfZ2AndWZWTerms-a}),
$s= \overline{d}$  when   $\pi_s \left ( G/H\right )=$ ${\bf Z_2}$, whereas  according to  (\ref{ExistenceOfZ2AndWZWTerms-b}),
$s=\overline{d}+1$ when   $\pi_s \left ( G/H\right )=$ ${\bf Z}$.
%The left-arrows indicate that due  to (\ref{ExistenceOfZ2AndWZWTerms-b}) the boundary dimension $\overline{d}$ at which
%the Topological Insulator (Superconductor) with the corresponding ${\bf Z}$ classification  is located   at the {\it end} of the arrow. 
 The left-arrows indicate that due  to (\ref{ExistenceOfZ2AndWZWTerms-b}) the boundary dimension $\overline{d}$ of  the Topological Insulator (Superconductor) with the corresponding ${\bf Z}$ classification  is located  at the {\it end} of the arrow. After moving all entries ${\bf Z}$ to the end of the corresponding arrows, and after shifting all columns of the  Table in
FIG. \ref{fig:TableOfHomotopyGroupsForAllTenSymmetricSpaces}  - this implements the rules specified in (\ref{ExistenceOfZ2AndWZWTerms-a},
\ref{ExistenceOfZ2AndWZWTerms-b}) - one obtains from FIG. \ref{fig:TableOfHomotopyGroupsForAllTenSymmetricSpaces} directly
the Table of Topological Insulators and Superconductors, TABLEs \ref{fig:TableOfTopInsSupercond} and \ref{fig:TableOfTopInsSupercondCOPY}.
%- From \cite{RyuSchnyderFurusakiLudwigNJPhys2010, LundellLectNotesMath1992}. 
 } 
\label{fig:TableOfHomotopyGroupsForAllTenSymmetricSpaces} 
\end{figure}

In order to understand how to implement this program in practice, all one needs to appreciate
is that the answer to the question about the lack of Anderson Localization is a question in Field Theory. It is well known
that the  theoretical description of problems of 
Anderson Localization is very systematic and geometrical (See e.g. \cite{EfetovBook1997,Wegner-F-J-1979-Z-PhysB38-1979-113,ZirnbauerRiemannianSymmetricSpacesJMathPhys1996}). Microscopically, the underlying first quantized Hamiltonian $H$ is  a member of one of the ten
symmetry classes, FIG. \ref{fig:TenFoldWay}, {\it  in every realization of disorder}. Physically, this is in our current discussion the Hamiltonian
describing the $\overline{d}=$ $(d-1)$ dimensional boundary of the $d$-dimensional Topological Insulator (Superconductor) in question which will  lie in the same symmetry class as the bulk.
It turns out that  at long length scales  (i.e. at scales much larger than the "mean free path", the corresponding small length scale
in this kind of problem) a description in terms of a Non-Linear Sigma Model (NLSM) field theory emerges. As already briefly mentioned above, 
a  general NLSM is a simple
generalization of the theory we use in standard Statistical Mechanics  to describe the classical Heisenberg Ferromagnet. The Ferromagnet
can be formulated as a model of  a unit vector spin, at every point in space,
that points to the surface of a two-dimensional sphere (in three dimensional spin space).
The energy is simply a gradient square of such a spin configuation. A general NLSM of interest in the present context  is the same type 
of  theory, except that the
field, the unit vector spin that used to represent an element of a sphere in the Ferromagnet, is now replaced by an element of one of the
ten symmetric spaces listed in the 6th column of FIG.\ref{fig:TenFoldWay} (with the heading "Anderson Localization, NLSM Manifold G/H").
As mentioned towards the end of Section \ref{SubSectionTenFoldWay}, the ten symmetric spaces  $G/H$ appearing in FIG.\ref{fig:TenFoldWay}
are generalizations of spheres (according to the seminal work by the mathematician
\'Elie Cartan). The general  NLSM is therefore described by a  "Boltzmann"-type weight (\lq action\rq) of the form already displayed in (\ref{NLSMaction}),
where the coupling constant $g$ changes with the amount of translational symmetry breaking ("disorder"). NLSMs of the form (\ref{NLSMaction})
will  always possess an Anderson Insulator phase, at least for sufficiently large values of the coupling constant $g$. In the Anderson Insulator
phase the correlation length\footnote{The correlation length is called in this context  usually localization length} is finite. At the boundary of a Topological Insulator (Superconductor),
 the corresponding NLSM must however always  describe a (electrical or heat) conductor, which means that the correlation length must be infinite.
Now, the NLSM on the  $\overline{d}$ dimensional boundary  completely evades the phenomenon of Anderson Localization if a certain {\it extra term of topological origin
with no adjustable parameter} can be added to the action  (\ref{NLSMaction}) of the NLSM\footnote{See  Ref.s \cite{RyuSchnyderFurusakiLudwigNJPhys2010} and 
footnote 22 in Ref. \cite{SchnyderRyuFurusakiLudwigLandau2009}}:
\begin{equation}
\label{NLSMactionWithTopTerm}
S =
{1\over g}
\int 
%\int_{{\rm \tiny{boundary}}}
d^{\overline{d}} r \  \sum_{\mu=1}^{\overline{d}} \ 
Tr \left
( \partial_\mu \Phi(r) \partial_\mu \Phi(r) \right ) +  S_{top}\{\Phi(r)\}, \qquad \Phi(r) \in G/H={\rm "target \ space"}.
\end{equation}
The question whether a suitable such 
{\it term of topological origin} $S_{top}$   exists or not,  depends (i) on the "target space" $G/H$ of the NLSM which is determined
by the symmetry class of the underlying Topological Insulator (Superconductor) in $d$ spatial dimension, as determined by
 FIG.\ref{fig:TenFoldWay}, and (ii) on the
dimension $\overline{d}=$ $(d-1)$ of the boundary on which the NLSM is defined.
There is however a simple answer to this question: It is the Homotopy Group of the target space $G/H$ which determines whether a suitable term exists. Namely, the target space $G/H$ of the NLSM allows for
\begin{eqnarray}\label{ExistenceOfZ2AndWZWTerms-a}
&(a)&:   \quad  {\rm a} \ Z_2 \ {\rm topological \ term}  \qquad  \qquad  \qquad   \ \ \ 
{\rm \it  if \ and \ only \ if} \qquad 
\pi_{\overline{d}}\left ( G/H\right ) = 
\pi_{d-1}\left (G/H \right )
 = {\mathbf Z_2} 
\\\label{ExistenceOfZ2AndWZWTerms-b}
&(b)&: \quad  {\rm a \ {Wess-Zumino-Witten}  \ term} \  \qquad {\rm \it if \ and \ only \ if}  \qquad 
\pi_{\overline{d}+1}\left ( G/H\right ) =
\pi_d\left ( G/H \right )
= {\bf Z}.
\end{eqnarray}
We will now implement this rule to determine the list of Topological Insulators (Superconductors). To this end, we display in
FIG. \ref{fig:TableOfHomotopyGroupsForAllTenSymmetricSpaces} a Table containing the  list of the Homotopy Groups $\pi_s \left ( G/H\right )$
for all 10 symmetric spaces $G/H$. The left-arrows indicate that due  to (\ref{ExistenceOfZ2AndWZWTerms-b}) the boundary dimension $\overline{d}$ of  the Topological Insulator (Superconductor) with the corresponding ${\bf Z}$ classification  is located
at  the position to which
 the {\it end} of the arrow points. After moving all entries ${\bf Z}$ to the end of the corresponding arrows, and after shifting all columns of  the Table in FIG.  \ref{fig:TableOfHomotopyGroupsForAllTenSymmetricSpaces} (i.e. $\overline{d}\to$ $d = \overline{d}+1$) - this implements the rules specified in (\ref{ExistenceOfZ2AndWZWTerms-a},
\ref{ExistenceOfZ2AndWZWTerms-b}) - one obtains from FIG.  \ref{fig:TableOfHomotopyGroupsForAllTenSymmetricSpaces} directly
the Table of Topological Insulators and Superconductors, TABLE \ref{fig:TableOfTopInsSupercondCOPY} (which is  a copy of TABLE \ref{fig:TableOfTopInsSupercond}, reproduced again for the convenience of the reader).

%{TableOfHomotopyGroups.pdf}
%After moving all entries ${\bf Z}$ to  the end of the corresponding arrows,  and after shifting all columns by one ($d=\overline{d}+1$) -  this  %implements the rule specified in (\ref{ExistenceOfZ2AndWZWTerms-a}, \ref{ExistenceOfZ2AndWZWTerms-b}) - 
%one obtains  from (\ref{fig:TableOfHomotopyGroupsForAllTenSymmetricSpaces}) directly  the Table of 
%Topological Insulators and Superconductors, FIG. \ref{fig:TableOfTopInsSupercond}.
%After moving all entries ${\bf Z}$ to  the end of the corresponding arrows,  and after shifting all columns by one ($d=\overline{d}+1$) -  this  %implements the rule specified in (\ref{ExistenceOfZ2AndWZWTerms-a}, \ref{ExistenceOfZ2AndWZWTerms-b}) - 
%one obtains  from (\ref{fig:TableOfHomotopyGroupsForAllTenSymmetricSpaces}) directly  the Table of
 %Topological Insulators and Superconductors, FIG. \ref{fig:TableOfTopInsSupercond}.

%\begin{figure}[!hbt]
%%\includegraphics[width=\columnwidth]{TableOfTopInsSupercond.pdf}
%\includegraphics[width=16cm]{TableOfTopInsSupercond.pdf}
% \caption{Table of Topological Insulators and Superconductors
%- From Ref. \cite{SchnyderRyuFurusakiLudwigPRB2008,SchnyderRyuFurusakiLudwigLandau2009,KitaevLandau2009,RyuSchnyderFurusakiLudwigNJPhys2010}.} 
%\label{fig:TableOfTopInsSupercondCOPY} 
%\end{figure}

%%%%%%%%%%%%%%%%%%%%%%%%%%%%%%%%%%%%
%\begin{table}[t!]
\begin{table}[!hbt]
%\begin{table}[h]
%\paragraph{complex case:}
\begin{center}
\resizebox{0.80\textwidth}{!}{
%\begin{small}
{
\renewcommand{\arraystretch}{1.1}
\begin{tabular}{lccccccccc}\hline
Cartan$\backslash d$ & \quad  0 \quad   & \quad  1 \quad  & \quad  2 \quad  & \quad  3 \quad  & \quad  4 \quad  & \quad  5 \quad  & \quad  6 \quad  
& \quad  7 \quad  & \quad  8 \quad  \\ \hline\hline
{\footnotesize  \it Complex   case:} &  &  &  &  &  &  &  &  &   \\ 
A   &  $\mathbb{Z}$   &   0   &   $\mathbb{Z}$   &   0  
    &  $\mathbb{Z}$   &   0   &  $\mathbb{Z}$   &  0  
    & $\mathbb{Z}$ \quad   $\cdots$
\\
%\hline
AIII  & 0 & $\mathbb{Z}$ & 0 & $\mathbb{Z}$
      & 0 & $\mathbb{Z}$ & 0 & $\mathbb{Z}$
      & 0  \quad  $\cdots$
\\
%\hline
%\end{tabular}
%\end{center}
%\paragraph{real case:}
%\begin{center}
%\begin{tabular}{cccccccccccccc}\hline
%Cartan$\backslash d$ & \quad  0 \quad  & \quad  1 \quad  & \quad  2 \quad  & \quad  3 \quad  & \quad  4 \quad  &  \quad 5 \quad  & \quad  6 \quad  &
%\quad  7 \quad  & 8 & 9 & 10 & 11 & 12\\ \hline\hline
{\footnotesize \it  Real   case:} &  &  &  &  &  &  &  &  &   \\ 
AI  & $\mathbb{Z}$ & 0 & 0 
    & 0 & $2\mathbb{Z}$ & 0 
    & $\mathbb{Z}_2$ & $\mathbb{Z}_2$ & $\mathbb{Z}$ 
     \quad $\cdots$ \\ 
%\hline
BDI & $\mathbb{Z}_2$ & $\mathbb{Z}$ & 0 & 0 
    & 0 & $2\mathbb{Z}$ & 0 
    & $\mathbb{Z}_2$ & $\mathbb{Z}_2$  \quad  $\cdots$ \\ 
%\hline
D   & $\mathbb{Z}_2$ & $\mathbb{Z}_2$ & $\mathbb{Z}$ 
    & 0 & 0 & 0 
    & $2\mathbb{Z}$  & 0 & $\mathbb{Z}_2$ 
     \quad $\cdots$ \\ 
%\hline
DIII& 0 & $\mathbb{Z}_2$ & $\mathbb{Z}_2$ & $\mathbb{Z}$ 
    & 0 & 0 & 0 
    & $2\mathbb{Z}$  & 0 \quad    $\cdots$ \\ 
%\hline
AII & $2\mathbb{Z}$  & 0 & $\mathbb{Z}_2$ 
    & $\mathbb{Z}_2$ & $\mathbb{Z}$ & 0 
    & 0 & 0 & $2\mathbb{Z}$ 
      \quad $\cdots$\\ 
%\hline
CII & 0 & $2\mathbb{Z}$  & 0 & $\mathbb{Z}_2$ 
    & $\mathbb{Z}_2$ & $\mathbb{Z}$ & 0 
    & 0 & 0  \quad $\cdots$\\ 
%\hline
C   & 0  & 0 & $2\mathbb{Z}$  
    & 0 & $\mathbb{Z}_2$  & $\mathbb{Z}_2$ 
    & $\mathbb{Z}$ & 0 & 0 
     \quad $\cdots$ \\ 
%\hline
CI  & 0 & 0  & 0 & $2\mathbb{Z}$  
    & 0 & $\mathbb{Z}_2$  & $\mathbb{Z}_2$ 
    & $\mathbb{Z}$ & 0  \quad  $\cdots$ \\ \hline
\end{tabular}
}
}
\end{center}
\caption{
Table of Topological Insulators and Superconductors\cite{SchnyderRyuFurusakiLudwigPRB2008,SchnyderRyuFurusakiLudwigLandau2009,KitaevLandau2009,RyuSchnyderFurusakiLudwigNJPhys2010} .
}
%\label{fig:TableOfTopInsSupercond} 
\label{fig:TableOfTopInsSupercondCOPY} 
\end{table}

%%%%%%%%%%%%%%%%%%%%%%%%%%%%%%%%%%%%%%%%%%%%%%%%%

\vskip 2cm

\subsection{Mathematical Reason for Agreement between Bulk-  and Boundary- based Classifications}
\label{LabelSubSectionAgreementBetweenBulkAndBoundaryMethods}

%\subsection{Mathematical Reason for agreement between Bulk and Boundary Methods: Symmetry Property of Table of Homotopy Groups}
%\label{LabelSubSectionAgreementBetweenBulkAndBoundaryMethods}

While the bulk-based  classification method from  section 
\ref{LabelSubSectionClassificationByTopolgyOfTheBulk}, based on an analysis of topological band theory, 
and the  boundary-based classification from section \ref{LabelSubSectionClassificatonByLackOfAndersonLocalizationOnSurface},
 based on the lack of Anderson Localization on the boundary, must yield the same result on physical grounds (as they do),
it is a priori not clear what the technical (i.e. mathematical) reason for this agreement is, since the two methods appear to be based on 
 conditions that look mathematically very different. By explicitly  comparing these two conditions, we will  now exhibit a certain "symmetry" property  inherent in  Table of Homotopy Groups,
FIG. \ref{fig:TableOfHomotopyGroupsForAllTenSymmetricSpaces}, that is responsible for the agreement of the two methods.

%\begin{figure}[!hbt]
%%\includegraphics[width=\columnwidth]{TableOfHomotopyGroups.pdf}
%\includegraphics[width=16cm]{ThreeOccurrencesOfCartanSpaces}
 %\caption{Comparison of the order of the three occurrences of the 10 Cartan Symmetric Spaces. For the eight real cases,
%the last column ("Classifying space") lists the spaces   $R_p$ where
%$p=0, 1, ..., 7$ $({\rm mod} \  8)$ in ascending order downwards,  whereas the corresponding symmetric
% space in the middle column ("Fermionic replica NL$\sigma$M target space")
%is $R_{4-p}$. As seen from the Figure, the corresponding symmetric space in the column "Time evolution operator"
%is $R_{p-1}$. - The corresponding shifts for the two complex symmetric spaces are also clearly visible. - Adapted from %\cite{RyuSchnyderFurusakiLudwigNJPhys2010}.
 %} 
%\label{fig:TableThreeOccurrencesOfTenCartanSpaces} 
%\end{figure}

\begin{table}[!hbt]
\begin{center}
\resizebox{0.99\textwidth}{!}{
%\begin{small}
{
\renewcommand{\arraystretch}{1.3}
\begin{tabular}{cccccc}
&&&$R_{4-p}$& &$R_p$ \\
&&&$\Downarrow$& &$\Downarrow$ \\
\hline
Cartan
&Time evolution operator\ 
& Fermionic replica
& & Classifying \\
label
& 
$\exp\{ \mathrm{i} t {\cal H}\}$
& NLSM target space
& & space
           \\ \hline \hline 
A &
$\mathrm{U}(N)\times \mathrm{U}(N)/\mathrm{U}(N)$
&
$\mathrm{U}(2n)/\mathrm{U}(n)\times \mathrm{U}(n)$ &$=C_0$
&
$\mathrm{U}(N+M)/\mathrm{U}(N)\times \mathrm{U}(M)$&$=C_0$ 
\\ 
{\bf AIII }& 
$\mathrm{U}(N+M)/\mathrm{U}(N)\times \mathrm{U}(M)$
& $\mathrm{U}(n)\times \mathrm{U}(n)/ \mathrm{U}(n)$
& $=C_1$&  $\mathrm{U}(N)\times \mathrm{U}(N)/ \mathrm{U}(N)$ & $=C_1$
\\  \hline 
AI &
$\mathrm{U}(N)/\mathrm{O}(N)$
&
$\mathrm{Sp}(2n)/\mathrm{Sp}(n)\times \mathrm{Sp}(n)$ 
& $=R_4$ &
$\mathrm{O}(N+M)/\mathrm{O}(N)\times \mathrm{O}(M)$ & $=R_0$
\\ 
{\bf BDI}
&
$\mathrm{O}(N+M)/\mathrm{O}(N)\times \mathrm{O}(M)$
& $\mathrm{U}(2n)/ \mathrm{Sp}(2n)$ 
& $=R_3$& 
$\mathrm{O}(N)\times \mathrm{O}(N)/\mathrm{O}(N)$ &$=R_1$
\\  
D &
$\mathrm{O}(N)\times \mathrm{O}(N)/\mathrm{O}(N)$
& $\mathrm{O}(2n)/ \mathrm{U}(n)$
& $=R_2$& $\mathrm{O}(2N)/ \mathrm{U}(N)$ & $=R_2$
\\
{\bf DIII} &
$\mathrm{SO}(2N)/\mathrm{U}(N)$
& $\mathrm{O}(n)\times \mathrm{O}(n)/\mathrm{O}(n)$ 
& $=R_1$ & $\mathrm{U}(2N)/ \mathrm{Sp}(2N)$ & $=R_3$ 
\\ 
AII &
$\mathrm{U}(2N)/\mathrm{Sp}(2N)$
& $\mathrm{O}(2n)/\mathrm{O}(n)\times \mathrm{O}(n)$
& $=R_0$ &
$\mathrm{Sp}(N+M)/\mathrm{Sp}(N)\times \mathrm{Sp}(M)$ &$=R_4$ 
\\ 
{\bf CII} &
$\mathrm{Sp}(N+M)/\mathrm{Sp}(N)\times \mathrm{Sp}(M)$
& $\mathrm{U}(n)/ \mathrm{O}(n)$ 
&$=R_7$ & $\mathrm{Sp}(N)\times\mathrm{Sp}(N)/\mathrm{Sp}(N)$ &$=R_5$
\\ 
C &
$\mathrm{Sp}(2N)\times \mathrm{Sp}(2N)/\mathrm{Sp}(2N)$
& $\mathrm{Sp}(2n)/ \mathrm{U}(n)$ 
& $=R_6$ & $\mathrm{Sp}(2N)/ \mathrm{U}(N)$ &$=R_6$ 
\\ 
{\bf CI} &
$\mathrm{Sp}(2N)/\mathrm{U}(N)$
& $\mathrm{Sp}(2n)\times \mathrm{Sp}(2n)/\mathrm{Sp}(2n)$
& $=R_5$ & $\mathrm{U}(N)/ \mathrm{O}(N)$ &$=R_7$ 
\\ \hline
\end{tabular}
%\end{small}
}
}
\end{center}
\caption{Comparison\cite{RyuSchnyderFurusakiLudwigNJPhys2010}
 of the order of the three occurrences of the 10 Cartan Symmetric Spaces. (Symmetry Classes with Chiral Symmetry are marked  in bold face). For the eight real cases,
the last column ("Classifying space") lists the spaces   $R_p$ where
$p=0, 1, ..., 7$ $({\rm mod} \  8)$ in ascending order downwards,  whereas the corresponding symmetric
 space in the middle column ("Fermionic replica NL$\sigma$M target space")
is  seen to be $R_{4-p}$. -  As also seen from the TABLE, the corresponding symmetric space in the column "Time evolution operator"
is $R_{p-1}$ as compared to the last column.  The corresponding shift  $C_0 \leftrightarrow$ $C_1$ 
for the two complex symmetric spaces $C_0$ and $C_1$  is  also clearly visible.
%\label{RMT, NLsM}
\label{fig:TableThreeOccurrencesOfTenCartanSpaces}
}
\end{table}

To this end, it is useful to compare the three different occurrences of the 10 Cartan Symmetric Spaces in the classification. This can be seen from 
TABLE  \ref{fig:TableThreeOccurrencesOfTenCartanSpaces}:  Let us first focus on the eight  real classes, which are listed in descending order in
the 4th (last) column of this TABLE, and are denoted by  $R_p$ $(p=0, 1, ..., 7$ (mod $8$). The 3rd column  of the TABLE, with the heading "Fermionic replica NLSM
target spaces", lists the NLSM target spaces $G/H$ of Anderson Localization, which  appear in the permutation $R_{4-p}$ $\Leftrightarrow$ $R_p$ as compared to the 4th column, 
as also  indicated in the TABLE. (We also see that the corresponding symmetric space in the 2nd column  with the heading "Time evolution operator"
is $R_{p-1}$.) Now, the conditions (\ref{ExistenceOfZ2AndWZWTerms-a},
\ref{ExistenceOfZ2AndWZWTerms-b}) for obtaining a Topological Insulator (Superconductor) based on the Boundary method (Anderson Localization - NLSM) are displayed on the right hand sides of (\ref{ComparisonBulkBoundaryCondition}). Here we used the property
$\ \pi_d\left ( R_p\right ) = \pi_0\left (R_{p+d}\right )$ which can immediately be read off from the  Table of Homotopy Groups, FIG. \ref{fig:TableOfHomotopyGroupsForAllTenSymmetricSpaces}. On the other hand, the conditions (\ref{BulkClassificationKTheoryResult})
based on the Bulk method (Topology of Bulk Band Structure) are displayed on the left hand sides of  (\ref{ComparisonBulkBoundaryCondition}).
As mentioned above, it is not immediately obvious that the conditions arising from the "Bulk" and "Boundary" methods are the same. However, it is not difficult to check using FIG.\ref{fig:TableOfHomotopyGroupsForAllTenSymmetricSpaces} that these conditions are indeed the same, which is a certain built-in "symmetry"  property of the Table of Homotopy Groups,  FIG.\ref{fig:TableOfHomotopyGroupsForAllTenSymmetricSpaces}.
 (Details are provided in  Appendix
(\ref{LabelSectionAppendixSymmetryPropertiesHomotopyGroups}).)
%\vskip 1cm 
\begin{equation}\label{ComparisonBulkBoundaryCondition}
\left \{
\begin{matrix}
{\rm \bf\underbar{\bf Bulk:} \ Topological \ Band \ Theory}&& {\rm \bf \underbar{\bf Boundary:} \ Anderson \ Localization (NLSM)} \\
&&\\
(a):\qquad \qquad \textcolor{blue}{\pi_0\left ( R_{q-d}\right)= {\bf Z_2}} \qquad &{\rm \it  if \ and \ only \ if}&    \left [ \pi_{d-1}\left ( R_{4-q}\right) = \right ] \ \ 
\textcolor{blue}{\pi_0\left ( R_{3-(q-d)}\right ) 
= {\bf Z_2}}\\ 
(b): \qquad  \qquad \textcolor{red}{ \pi_0\left (R_{q-d}\right )={\bf Z}} \  \qquad    &{\rm \it  if \ and \ only \ if}&  
 \   \ \left [  \pi_{d}\left ( R_{4-q}\right) =\right ] \ \ \textcolor{red}{\pi_0 \left( R_{4-(q-d)}\right)
 ={\bf Z}}  \\
&&\\
                                                                                       & {\rm using:} \ \pi_d\left ( R_p\right ) = \pi_0\left (R_{p+d}\right ) &
\end{matrix}
\right \}
\end{equation}

\vskip 2cm

\subsection{Perspective of Quantum Anomalies}
\label{LabelSubSectionPerspectiveOfQuantumAnomalies}

As already mentioned at the beginning of  section \ref{LabelSectionClassificationOfTopInsSupercond}, a third classifying principle has emerged, 
besides those discussed in sections \ref{LabelSubSectionClassificationByTopolgyOfTheBulk} and \ref{LabelSubSectionClassificatonByLackOfAndersonLocalizationOnSurface}. This new classifying principle, which is based on  Quantum Anomalies
that  are  forced
to occur at the boundary of Topological Insulators and Superconductors, is probably the most general such principle as it will
extend also to interacting theories; it has been the topic of much recent discussions. (For a  very short (and incomplete)  list of references to recent discussions see 
e.g. \cite{SuleChenRyuPRB2013,RyuNobelProc2015,WangSenthilInteractingFermionic3D-2014,KapustinThorngrenTurzilloWang-arXiv2014,Witten-arXiv2015}.)
This  principle was first recognized in Ref. \cite{RyuMooreLudwig2012}
in its general form in the context of non-interacting Fermionic Topological Insulators and Superconductors.
The perspective of Quantum Anomalies of Topological Insulators (Superconductors) 
can be viewed as  a generalization of the boundary-based classification principle invoking Anderson Localization  that was reviewed in section \ref{LabelSubSectionClassificatonByLackOfAndersonLocalizationOnSurface}. It relies on the notion  that the {\it boundary
of a Topological Insulator (Superconductor) cannot exist as an isolated system in its own dimensionality. Rather it must always be attached
to a higher dimensional bulk.} Technically, the inability of the boundary of the system to exist in isolation is rooted  in certain  ill-defined properties
 the boundary would possess  in isolation.  It was in Ref. \cite{RyuMooreLudwig2012} that such  ill-defined properties of the boundary
of non-interacting Fermionic Topological Insulators (Superconductors) were
in general related to the notion
of Quantum Anomalies known from work  in the 1980ies
 in relativistic Quantum Field Field Theory and  Elementary Particle Physics (see e.g. \cite{AlvarezGaumeWittenNPB1983,AlvarezGaumeGinspargAnnPhys1985}).
We will  review the basic ideas of \cite{RyuMooreLudwig2012} in this section.

In general, a quantum system possesses a Quantum Anomaly if the corresponding  classical system is invariant under some (global or gauge)
 symmetry, and  this symmetry gets lost in the process of quantization. For technical reasons it is useful to take advantage of the following
result that emerged from the classification of Topological Insulators and Superconductors: It turns out  that every 
Topological Insulator
(Superconductor) phase of non-interacting Fermions, in any dimension, has a massive Dirac Hamiltonian representative in the same topological class\cite{RyuSchnyderFurusakiLudwigNJPhys2010,KitaevLandau2009,TeoKanePRB82-2010-115120}.
Since
we are only interested in the {\it topological properties}
of the phase, we are free to consider the Dirac Hamiltonian representative. Next, we couple the Dirac Hamiltonian representative
to suitable space and time dependent classical background fields. These could be a $U(1)$-gauge field if the Topological Insulator
 (Superconductor)
has a conserved $U(1)$ charge, an $SU(2)$-gauge field if it has $SU(2)$ symmetry, etc.. If, on the other hand, we have a Topological Superconductor which is not invariant under any continuous symmetries (examples are known in Cartan Symmetry Classes D and DIII),
then we can still couple the Dirac Hamiltonian to a background gravitational field, i.e. we put it in a curved background\footnote{Coupling
to weak gravitational fields can physically be viewed, in the condensed matter physics context,
as technical trick to write a Kubo-formula for thermal conductivities\cite{LuttingerPaperThermalKubo-PhysRev1964,RyuMooreLudwig2012}.}. In either case we end up, after integrating out the gapped Fermions, with an effective action for the classical space and time dependent background fields (in the gravitational
case, the effective action depends on the background metric).  For example in the case of a 
$U(1)$ background gauge field $A_\mu(x)$ we obtain an effective action
\begin{eqnarray}\label{Effective-U-1-GaugeField-Action}
e^{-W_{eff}[A_\mu] } 
\ &\equiv& \ 
\int{\cal D}[{\bar \psi},\psi] \ e^{-S[{\bar \psi}, \psi; A_\mu]}. \\ \nonumber
\end{eqnarray}
We work in imaginary time where $x$ denotes the $D=d+1$ space-time  coordinates $x=(\tau, {\vec x})$, and ${\vec x}$ the spatial
coordinates. As indicated in the second line of the above equation, the effective action for the gauge field,  which
physically describes the responses of the system to the external classical perturbation $A_\mu$,
can have, besides a possible "standard" or "usual" term (in $D=4$ space-time dimensions and 
in the case of a $U(1)$ gauge field this will be Maxwell action for electromagnetism in the medium
of the Topological Insulator (Superconductor)),  an unusual or  anomalous  response arising from a topological term $W_{top}$.
That latter (anomalous) response would never exist for an "ordinary" system describing a boundary that is allowed to exist in isolation without
an attached bulk. 
It only occurs for boundaries of Topological Insulators (Superconductors)\footnote{Such anomalous responses occur also for more general Symmetry Protected Topological (SPT) phases (see the discussion in section \ref{sectionEntanglementPerspective}).}.

\vskip .1cm
It turns out that 
in general we need to distinguish two types of such anomalous responses which will be defined in more detail below: {\bf (i) Chern-Simons type responses}, occuring in odd space-time dimensions $D=d+1=$
$2n-1$, and
{\bf  (ii) Theta term type responses}, occuring in even space-time dimensions $D=d+1=$ $2n$. In the following section we will describe  two well-known special
cases of these anomalous responses in $D=3$ and $D=4$ space-time dimensions.
 In the subsequent section, we will generalize them and show how to use  the  so-obtained generalized responses to predict various  Topological Insulators and Superconductors occuring in TABLE \ref{fig:TableOfTopInsSupercond} .

\vskip 1cm

\subsubsection{Examples: Chern-Simons term in Integer Quantum Hall effect in $D=3$ space-time dimensions, and Theta Term in standard ${\bf Z_2}$ Topological Insulator in $D=4$ space-time dimensions.}
\label{LabelSubSubSectionExamples2DIntegerQuantumHall3DAxion}

\begin{figure}[!hbt]
\includegraphics[width=5cm]{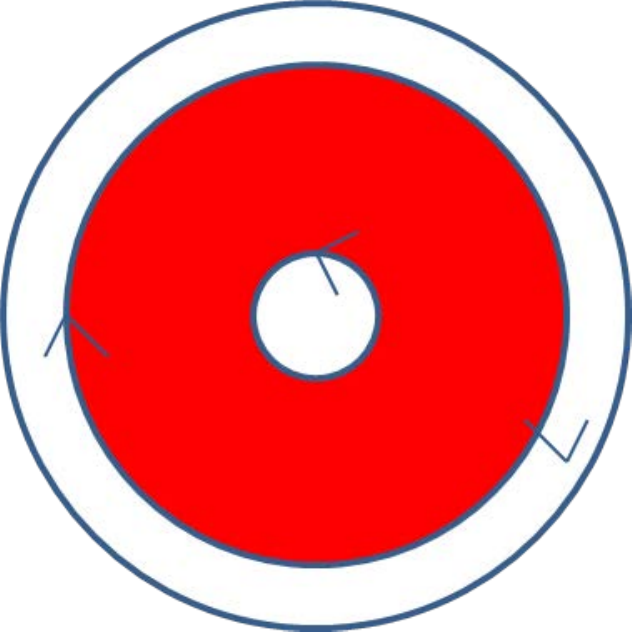}
 \caption{Annulus Geometry for the Quantum Hall Effect. In the present discussion we consider only the case
of the Integer Quantum Hall effect.
 } 
\label{fig:IntegerQuantumHallAnnulus} 
\end{figure}

{\bf (i) Chern-Simons type response:} {\it $d=2$ ($D=3$)  Integer Quantum Hall effect.}  Let us consider an (spatial)  annulus filled with the Integer Quantum Hall state  - marked red in  FIG. 
\ref{fig:IntegerQuantumHallAnnulus}. There are two counterpropagating chiral edge modes at the boundaries of the annulus. Let us
consider the theory describing one of these  boundaries in $D=2=(1+1)$ dimensional space-time. We know that charge conservation at such a boundary
is spoiled by quantum effects: charge will \lq leak\rq  \ from the boundary into the bulk (the annulus) and eventually  to the other boundary. The
$d=2$  Integer
Quantum Hall state is in Cartan Symmetry Class $A$, and is thus  one of the Topological Insulators in the 10-fold Table. The \lq leaking\rq \ of charge  indicates that the  bounday cannot exist in isolation  but
must be the boundary of a Topological Insulator in one dimension higher. 
How is this seen  from the perspective of the bulk? As is well known, the topological part of the action defined in (\ref{Effective-U-1-GaugeField-Action}) is in the present case a  (Abelian) Chern-Simons
term
\begin{equation}\label{TwoPlusOne-U-1-ChernSimonsTerm}
W_{top}
=
i \sigma_{xy}  
\int d^2{\vec x} d \tau \ {(-1)\over 4 \pi} \epsilon^{\mu\nu\lambda} A_\mu\partial_\nu A_\lambda,
\qquad {\rm where} \quad \sigma_{xy} = {\rm integer}.
\end{equation}
(Here $\sigma_{xy}$ is the Hall conductivity measured in units of $e^2/h$.) As we will discuss in more detail for the general cases,
a Chern-Simons term defined on a space-time manifold with boundaries is not invariant under gauge transformations performed on the gauge field
$A_\mu$. It is this lack of
gauge invariance of the bulk that will have to be compensated for by the theory describing  the boundary, so that the combined system is gauge
invariant (as it is). So  the  appearance of the Chern-Simons term $W_{top}$  of the $D=3$ dimensional bulk indicates that there is anomaly
at the boundary, and it allows us  to identify and quantify this anomaly of  the
theory describing the boundary. As we will discuss in the next section, various generalized versions of Chern-Simons terms will occur in all odd space-time dimensions.

\vskip 1cm
{\bf (ii) Theta-term type response:} {\it $d=3$ ($D=4$)  ${\bf Z_2}$ Topological Insulator.} Let us consider the "standard" ${\bf Z_2}$ Topological Insulator
in Cartan Symmetry Class AII
in $d=3$ spatial dimensions (see e.g. \cite{HasanKaneRMP2010}). It was first observed in \cite{QiHughesZhangPRB2008}
that after minimally coupling to a $U(1)$ gauge field $A_\mu$ 
the effective action for the gauge field contains, besides the usual Maxwell term, a topological term which is
an example of a class of topological terms that are  in general called  {\it $\theta$-terms},
\begin{eqnarray}\label{ThreeDimensionsATwoThetaTerm}
W_{top}&=&
i \theta \int_{{\cal M}_4}
{(-1)\over 32 \pi^2}
\epsilon^{\mu\nu\kappa\lambda} F_{\mu\nu}F_{\kappa\lambda}  \ d^3{\vec x}d \tau
\equiv \\ \nonumber
&\equiv& i \theta \int_{{\cal M}_4} \ \ \Omega_4,
\qquad \qquad \qquad \left (  {\rm where:} \ \  \Omega_4 \equiv 
{(-1)\over 32 \pi^2}
\epsilon^{\mu\nu\kappa\lambda} F_{\mu\nu}F_{\kappa\lambda}  \ d^4 x \right )
\end{eqnarray}
($F_{\mu\nu}=\partial_\mu A_\nu - \partial_\nu A_\mu$).
Here  ${\cal M}_4$ is a $D=4$ dimensional space-time manifold, and
we have denoted the integrand by the differential 4-form $\Omega_4$ defined by the above equation. When ${\cal M}_4$ is
a manifold without boundaries
then the integral is an integer $n$  and thus $W_{top}=i \theta n$, which implies that $\exp \{- W_{top} \}=$
$\exp\{- i \theta n\}$ is periodic under $\theta \to \theta+2\pi$. Moreover, since one can write
$W_{top}\propto $ $i \theta \int_{{\cal M}_4} d^3{\vec x}d\tau \ {\vec E}\cdot {\vec B}$, one sees that  $\theta\to$ $-\theta$ under time-reversal
so that $\theta=0, {\rm or} \ \pi$ (mod \ $2 \pi$). The value $\theta=0 \  ({\rm mod} \ 2 \pi)$ applies to the trivial Insulator,
and $\theta=\pi  \  ({\rm mod} \ 2\pi)$ to the Topological Insulator.

Now, the integrand in (\ref{ThreeDimensionsATwoThetaTerm}) is a total derivative
\begin{equation}\label{U-1-Theta-Term-Total-Derivative}
\epsilon^{\mu\nu\kappa\lambda} F_{\mu \nu} F_{\kappa\lambda} = 4 \partial_\sigma
\left ( \epsilon^{\mu\nu\rho\sigma} A_\mu\partial_\nu A_\rho\right ).
\end{equation}
Therefore, if the space-time manifold ${\cal M}_4$ has a boundary $\partial {\cal M}_4$, one obtains an integral over the boundary,
\begin{eqnarray}\label{ThreeDimensionsATwoThetaTermIntegrateToBoundary}
W_{top}
&=&
 i \sigma_{xy} \int_{\partial {\cal M}_4}
{(-1)\over 4 \pi} \epsilon^{\mu\nu\lambda} A_\lambda \partial_\nu A_\lambda \ d^2{\vec x}d\tau \equiv \\ \nonumber
&\equiv & i \sigma_{xy} \int_{\partial {\cal M}_4} \ 2\pi \ \ \Omega^{(0)}_3, \qquad \qquad
\left ( {\rm where:} \ \Omega^{(0)}_3 = {(-1)\over 8 \pi} \epsilon^{\mu\nu\lambda} A_\lambda \partial_\nu A_\lambda \ d^3 x \right )\\ \nonumber
\qquad \qquad  {\rm and:}  \ \ \sigma_{xy}&=&\pm {\theta\over 2 \pi}.
\end{eqnarray}
Writing for short
\begin{eqnarray}\label{OmegaFour=dOmegaZeroThree}
\Omega_4 &=& d \Omega^{(0)}_3 \\ \nonumber
W_{top}&=&  i \theta \int_{{\cal M}_4} \Omega_4 = 2 \pi i {\theta\over 2 \pi} \int_{\partial {\cal M}_4} \Omega^{(0)}_3,
\end{eqnarray}
we now see that
\begin{equation}\label{ThetaEqualsPiSigmaxyEqualsOneHalf}
{\rm when} \quad \theta=\pi \qquad \Rightarrow \qquad \sigma_{xy}=\pm {1\over 2}.
\end{equation}
This is the main conclusion. To interpret it, we need to understand that the sign is picked by applying an infinitesimally small
time-reversal breaking perturbation to the boundary. This induces  a gapped Quantum Hall state on the surface 
characterized by a Chern-Simons term as in (\ref{TwoPlusOne-U-1-ChernSimonsTerm}) and a non-vanishing value of
$\sigma_{xy}$. However, equation (\ref{ThetaEqualsPiSigmaxyEqualsOneHalf}) says that the resulting $\sigma_{xy}$ is
not an integer. This is impossible to get in a Quantum Hall effect of non-interacting electrons. Therefore, the resulting boundary
with a value of $\sigma_{xy}=\pm {1\over 2}$ is not allowed to occur in a purely two-dimensional system. The two-dimensional
boundary must be attached to a three-dimensional bulk. Therefore, the result in (\ref{ThetaEqualsPiSigmaxyEqualsOneHalf})
tells us that the boundary possesses an "anomaly".

\vskip 1cm

\subsubsection{General Anomalous Responses and Descent Relations}
\label{LabelSubSubSectionGeneralAnomalousResponsesDescentRelations}

It is known how to construct
such  actions $W_{top}$ for anomalous responses
 in great generality and in general space-time dimensions $D$ 
(see e.g. \cite{NakaharaBook,FujiwaraSuzukiBook}). 

\vskip .1cm

{\it General $\theta$-terms.} We will first explain the general construction of  the $\theta$-term.
For a gauge field (Abelian or non-Abelian, e.g. $U(1)$ or $SU(2)$) let 
\begin{equation}\label{DEFCalFMuNu}
{\cal F}_{\mu\nu} \equiv {1\over 2} F_{\mu\nu} dx^\mu \wedge dx^\nu
\end{equation}
where $F_{\mu\nu}$ is the field strength (for non-Abelian symmetry groups where the gauge field transforms in an $r$-dimensional
representation, this is an $r\times r$ matrix - in order to accomodate the non-Abelian case, we write in the next equation traces over this
$r$-dimensional space, $r=1$ for Abelian symmetries). Now define the so-called {\it Chern-Character} (see e.g. \cite{NakaharaBook})
\begin{equation}\label{DEFChernCharacter}
ch \left ( {\cal F} \right )
\equiv
r + \left ( {i \over 2 \pi} \right ) tr {\cal F}
+{1\over 2!} \left  ( {i \over 2 \pi} \right )^2 tr {\cal F}^2  + ...
\qquad \qquad ({\rm Chern-Character}).
\end{equation}
Similarly, for a gravitational background field, we construct using the definition
\begin{equation}\label{DEFCalRMuNu}
{{\cal R}_\mu}^\nu \equiv
{1\over 2} {R_{\alpha\beta\mu}}^\nu \ dx^\alpha\wedge dx^\beta,
\end{equation}
where 
${R_{\alpha\beta\mu}}^\nu$ is the curvature tensor,
the so-called {\it Dirac Genus} (see e.g. \cite{FujiwaraSuzukiBook})
\begin{equation}\label{DEFDiracGenus}
{\hat A}\left ( {\cal R} \right )
\equiv
1 + {1\over (2 \pi)^2} {1\over 12}   tr {\cal R}^2
+{1\over (2 \pi)^2 } \left  [ {1\over 288} tr {\cal R}^2  + {1\over 360} tr {\cal R}^4\right ] + ...
\qquad \qquad ({\rm Dirac \ Genus}).
\end{equation}
%Similarly, for a gravitational background field, we construct using the definition
%\begin{equation}\label{DEFCalRMuNu}
%{{\cal R}_\mu}^\nu \equiv
%{1\over 2} {R_{\alpha\beta\mu}}^\nu \ dx^\alpha\wedge dx^\beta
%\end{equation}
Note that Chern-Character and Dirac Genus are "power series" in \lq differentials\rq \   $dx^\mu$. If we are in $D$ space-time dimensions we have only
a number $D$  of different differentials $d x^\mu$, and if we multiply them all together (using the wedge  product) we obtain the (oriented)  space-time volume element $d^D x=$
$d x^1\wedge d x^2 \wedge ... \wedge d x^D$.
Therefore, if we
pick out of any of the "power series"  $ch \left ( {\cal F} \right )$, or
${\hat A}\left ( {\cal R} \right )$, or the product $ch \left ( {\cal F} \right ){\hat A}\left ( {\cal R} \right )$
the term with precisely $D$ differentials, we obtain a term
$\Omega_D \equiv {\cal A}_D(x) d^D x$ where ${\cal A}_D(x)$ is just a {\it function}.
This is the generalization of the expression $\Omega_4$ in (\ref{ThreeDimensionsATwoThetaTerm}). We may sumarize  this process
by writing
for the three cases of  so-called gauge-, gravitational-, and mixed- anomaly terms:
\begin{equation}\label{DEFGaugeGravitationsMixedAnomaly}
\begin{matrix}
\Omega_D \equiv ch \left ( {\cal F}\right )_{| D};  \qquad \qquad   
 &  \Omega_D\equiv {\hat A}\left ( {\cal R} \right )_{| D};  \qquad  \qquad  
& \Omega_D \equiv 
\left (
ch \left ( {\cal F} \right ){\hat A}\left ( {\cal R} \right )
\right )_{| D};
& \qquad \qquad  \left ( {\rm where:}  \ \Omega_D = {\cal A}_D \ d^D x\right )   \\ 
{\rm \bf (a):   gauge-}  \qquad & \textcolor{red}{{\rm \bf  (b):  gravitational-}} &   \textcolor{blue}{ {\rm \bf  (c): mixed-}} & \qquad  \qquad {\rm anomalous \ response}
\end{matrix}
\end{equation}

As already mentioned, the  so-obtained terms $\Omega_D$ are  generalizations of the  $\theta$-term 
$W_{top}$ which appeared in the  special  example of equation  (\ref{ThreeDimensionsATwoThetaTerm}). Note that   all
terms $\Omega_D$ are  gauge invariant, since they are constructed out of the gauge invariant objects $F_{\mu\nu}$ and
${R_{\alpha\beta\mu}}^\nu$.\footnote{In the gravitational case this amounts to invariance under  infinitesimal $SO(D)$ transformations in the
tangent space of the Euclidean space-time manifold (see \cite{RyuMooreLudwig2012} for more details).}
 Moreover,  since the Chern-Character and Dirac Genus "generating functions" contain only terms with an {\it even} number of  differentials, we clearly
see that $\Omega_D=0$ for {\it odd} space-time dimension $D$.

\vskip .5cm

{\it Descent Relations.} It turns out that $\Omega_D$ is always a total derivative\footnote{in the sense of differential forms},
\begin{equation}\label{DEFFirstDescentEquation}
\Omega_D = d \Omega^{(0)}_{D-1}
\qquad \qquad \left ( {\rm First \ Descent \ Relation}, \ \ D={\rm even} \right )
\end{equation}
which is a generalization of (\ref{OmegaFour=dOmegaZeroThree}). Therefore, we obtain for
a $D$-dimensional space-time manifold ${\cal M}_D$ with boundary $\partial {\cal M}_D$
\begin{equation}\label{GeneralIntegralThetaTermsBoundaryChernSimons}
\int_{ {\cal M}_D}
{\cal A}_D  \ d^D x \ = \ \int_{ {\cal M}_D} \Omega_D \ = \ \int_{\partial {\cal M}_D} \Omega^{(0)}_{D-1}, 
\qquad \qquad D = {\rm even}.
\end{equation}
In analogy with  (\ref{OmegaFour=dOmegaZeroThree}), the term $\Omega^{(0)}_{D-1}$ is a generalization of the Chern-Simons term.

Let us now focus on the generalized Chern-Simons term $\Omega^{(0)}_{D-1}$. In complete analogy  with the special case of the Chern-Simons
term in
(\ref{TwoPlusOne-U-1-ChernSimonsTerm}), and the discussion in the paragraph below that equation, the general Chern-Simons
term $\Omega^{(0)}_{D-1}$ is not invariant under a gauge transformation $A_\mu \to$ $A_\mu + v_\mu$. Rather, denoting by $\delta_v$ the change of this
term by an infinitesmal gauge transformation (linear in $v_\mu$), it is know that\footnote{By definition, $\Omega^{(1)}_{D-2}$ is linear in $v_\mu$.}
\begin{equation}\label{SecondDescentRelation}
\delta_v \Omega^{(0)}_{D-1} = d \Omega^{(1)}_{D-2}, \qquad \qquad ({\rm Second \ Descent \ Relation}, \ \  D={\rm even}).
\end{equation}

\vskip .4cm
We are now ready to write the general forms of (\ref{TwoPlusOne-U-1-ChernSimonsTerm}) and (\ref{OmegaFour=dOmegaZeroThree}),(\ref{ThetaEqualsPiSigmaxyEqualsOneHalf}):

\vskip .2cm

{\bf (i) Chern-Simons type response:} {\it $D=2n-1$ odd.} This generalizes (\ref{TwoPlusOne-U-1-ChernSimonsTerm}). If ${\cal M}_{2n-1}$ is an odd-dimensional
space-time manifold with boundary ${\partial {\cal M}_{2n-1}}$, then the Second Descent Relation (\ref{SecondDescentRelation}) implies
\begin{equation}\label{IntegralOfSecondDescentRelation}
\delta_v \int_{{\cal M}_{2n-1}} \ \Omega^{(0)}_{2n-1} \ = \int_{{\cal M}_{2n-1}} \Omega^{(1)}_{2n-2}
=
\int_{\partial {\cal M}_{2n-1}} \Omega^{(1)}_{2n-2}.
\end{equation}
After normalizing
$\Omega^{(0)}_{2n-1}$ by a suitable multiplicative factor,  
the topological part of the effective action  is
\begin{equation}\label{GeneralWtopforChernSimonsTerm}
W_{top}=  i \sigma \int_{{\cal M}_{2n-1}} \ \Omega^{(0)}_{2n-1},
\end{equation}
where $\sigma$ generalizes the constant $\sigma_{xy}$
in (\ref{TwoPlusOne-U-1-ChernSimonsTerm}).
It then follows from (\ref{IntegralOfSecondDescentRelation}) which is a consequence of the
Second Descent Relation (\ref{SecondDescentRelation})
that, as in the previous special case,  the bulk theory is not invariant under gauge transformation and a term violating gauge
invariance appears at the boundary $\partial {\cal M}_{2n-1}$ of the  manifold ${\cal M}_{2n-1}$. Therefore, the boundary theory must generate a compensating term so that the total system
is invariant. Hence, the boundary on its own is not gauge invariant and thus not consistent - it cannot exist in isolation without the bulk.

\vskip .2cm

{\bf (ii) Theta-term type response:} {\it  $D=2n$ even.} This generalizes (\ref{OmegaFour=dOmegaZeroThree},\ref{ThetaEqualsPiSigmaxyEqualsOneHalf}). Upon using a suitable normalization of $\Omega_{2n}$ the integral
$\int_{{\cal M}_{2n}} \Omega_{2n}$  over a $D=2n$-dimensional
manifold ${\cal M}_{2n}$ without boundary  is an integer, and 
\begin{equation}\label{GeneralWtopForThetaTerm}
W_{top}=i \theta \int_{{\cal M}_{2n}} \Omega_{2n}
\end{equation}
leads  to periodicity of the theory under
$\theta\to$ $\theta+2 \pi$  for the same reason as discussed in the paragraph below (\ref{ThreeDimensionsATwoThetaTerm}).
The Topological Insulators (Superconductors) we will be applying this to will have a symmetry (time-reversal or chiral)  which will fix
the value of $\theta$ again to $\theta = 0,$  ${\rm or} \ \pi$ (mod $2\pi$) in the same way as below 
 (\ref{ThreeDimensionsATwoThetaTerm}). On  a manifold ${\cal M}_{2n}$ with boundary,
$W_{top}$ is, due to
(\ref{GeneralIntegralThetaTermsBoundaryChernSimons}),
% the First Descent Relation  (\ref{DEFFirstDescentEquation}), 
a Chern-Simons term on the boundary ${\partial {\cal M}_{2n}}$;
as in (\ref{OmegaFour=dOmegaZeroThree},\ref{ThetaEqualsPiSigmaxyEqualsOneHalf})  its coefficient
does not carry the value allowed for a  theory defined in isolation in $2n-1$ space-time dimensions.  For that reason, again, such a boundary cannot exist in isolation without being attached to a higher dimensional bulk.

\vskip 1cm

\subsubsection{Applications to Topological Insulators and Superconductors}
\label{LabelSubSubSectionApplicationsToTopologicalInsulatorsAndSuperconductors}

In this section we review the method employed in Ref. \cite{RyuMooreLudwig2012} to predict the existence of  Topological Insulators or Superconductors by locating the spatial dimensions in which anomalous responses $W_{top}$ of the kind discussed above exist. These
are associated with what are called "perturbative anomalies".
%We will restrict ourselves entirely to the cases of Topological Insulators (Superconductors) with an integer classification. 
As we will
see, the presence of anomalous responses arising in this way allows us to predict the occurrence of all Topological Insulators (Superconductors)
which are characterized by  an integer classification in TABLEs \ref{fig:TableOfTopInsSupercond} and  \ref{fig:TableOfTopInsSupercondCOPY},  to which we  mostly  restrict attention here. - At the end of this
section,  we also review the conjecture made in  Ref. \cite{RyuMooreLudwig2012}
that the remaining cases,  i.e. those with a ${\bf Z}_2$ classification, as well as a full picture based on anomalies, valid also
in the presence of interactions, will
 arise from the consideration of global anomalies (different from the above
"perturbative anomalies"). To support this conjecture we give several examples in the classification which  may follow from
certain global gravitational anomalies. (Recent work on global anomalies in Topological Insulators includes e.g. \cite{RyuNobelProc2015,HsiehChoRyu-arXiv2015,Witten-arXiv2015}, which confirms the validity of this conjecture.)

\vskip .2cm

{\bf (a): Complex Symmetry Classes.} First consider the complex symmetry classes A and AIII in TABLE \ref{fig:periodic table with gravity} which
naturally
%Since  charge-conjugation symmetry is absent in these symmetry classes (FIG. \ref{fig:TenFoldWay}), they
 possess $U(1)$ symmetry.
Upon minimally coupling a $U(1)$ gauge field, we read off from the Chern Character (using case $(a)$ of (\ref{DEFGaugeGravitationsMixedAnomaly})) that a $\theta$-term $\Omega_D$ can exist in every  space-time
dimension $D=2k$ ($k=$ integer). The resulting Chern-Simons term $\Omega^{(0)}_{D-1}$,  from  (\ref{GeneralWtopforChernSimonsTerm})  and the first descent relation (\ref{DEFFirstDescentEquation}), predicts that a Topological Insulator can exist
in every odd space-time dimension $D=2k-1$, i.e. in all  even space dimensions $d=2k$. These are marked with a black  bold face ${\bf Z}$ in 
TABLE \ref{fig:periodic table with gravity}.  - Moving on,  we make use of the $\theta$-term $\Omega_{2k}$  which,  as discussed below
(\ref{GeneralWtopForThetaTerm}),
 provides an anomalous response in
every even space-time dimension $D=2k$, i.e. in all  odd spatial dimensions $d=2k-1$, and hence predicts that  a Topological Insulator can exist
in those dimensions. These are marked with  a black $Z$ (not bold face)
 in TABLE \ref{fig:periodic table with gravity}. Note that the current argument for the AIII cases only shows that  a Topological Insulator characterized by an odd  integer (thus including the case where the integer $=1$) can exist in these dimensions,
and does not demonstrate the existence of the entire  integer classification\footnote{Though, in the non-interacting case one can obtain the general integer from the  winding of the phase of the  partition function as detailed explicitly  in Ref. \cite{RyuMooreLudwig2012} for the case of symmetry class DIII in $d=3$. - See also the discussion in Ref. \cite{QiWittenZhangAxionTopSuperconductors-PRB2013} obtaining the general
integer from an alternative point of view for symmetry class DIII in $d=3$}.
Nevertheless,  the argument shows that there are non-trivial entries in the corresponding positions in the Table.
(The non-interacting $Z$-classification of symmetry class AIII does in general collapse in the presence of interactions to a smaller group, as has been explicitly
demonstrated in $d=3$  spatial dimensions in Ref. \cite{WangSenthilInteractingFermionic3D-2014} - see also section \ref{LabelSubSectionRecentResults-d-3-PresenceOfInteractions}. A full description of the AIII cases in the presence of interactions, and more general, of all Topological Insulators (superconductors)  in the presence of interactions not fully  characterized by perturbative anomalies,  was conjectured\cite{RyuMooreLudwig2012} to be provided by a global anomaly; this
 has recently been confirmed - see 
\cite{RyuNobelProc2015,HsiehChoRyu-arXiv2015,Witten-arXiv2015} for a recent discussion.)

\vskip .3cm 

{\bf Real Symmetry Classes.} It is useful to divide the eight real symmetry classes into two groups. The first group includes classes AI,
AII, CII which are naturally realized as normal (as opposed to superconducting) electronic systems, and thus
carry a natural notion of $U(1)$ symmetry. Moreover, one realization of symmetry class BDI can also be  considered to have a conserved $U(1)$ symmetry.
On the other hand, the
natural realizations of symmetry classes D and DIII possess no conserved $U(1)$ quantity; while classes C and CI can be realized as singlet superconductors
with conserved $SU(2)$ spin - and so there is a conserved $U(1)$ charge - it is natural to treat all symmetry classes that have a standard realization as superconducting systems on an equal footing.
Therefore we subdivide the eight real  symmetry classes in the two groups, the first group consisting of  classes  (D, DIII, C, CI) and the second group consisting of  classes (AI, BDI, AII, CII).

\vskip .1cm

\textcolor{red}{{\bf (b): Purely gravitational anomalous responses} (D, DIII, C, CI).} We consider coupling the members of the  first group   (D, DIII, C, CI) to a purely gravitational background. We read off from the Dirac Genus  (using case $(b)$ of (\ref{DEFGaugeGravitationsMixedAnomaly})) that  a $\theta$-term $\Omega_D$  can exist in every space-time
dimension $D=4k$ ($k=$ integer). The resulting  (gravitational) Chern-Simons term $\Omega^{(0)}_{D-1}$,  from  (\ref{GeneralWtopforChernSimonsTerm}) and the first descent relation (\ref{DEFFirstDescentEquation}),
predicts  that a Topological Insulator (Superconductor)  can exist 
in every space-time dimension $D=4k-1$,  corresponding to spatial dimensions $d=4k-2$, i.e. to 
spatial dimensions $d=2$ (mod 8) and $d=6$ (mod 8).
The intersections of these dimensions $d$ with the group (D, DIII, C, CI) of symmetry classes are marked
with a  red bold face \textcolor{red}{${\bf Z}$}  in 
TABLE \ref{fig:periodic table with gravity}.  Note that this predicts precisely the spatial dimensions of the Topological Superconductors in these classes
with an integer classification. 
In analogy with the Quantum Hall case discussed below (\ref{TwoPlusOne-U-1-ChernSimonsTerm}), in the present gravitational case the conservation law
for  energy and momentum
is lost at the boundary; energy and momentum  \lq leaks\rq \ from the boundary into the bulk. Thus, 
the presence of the corresponding Chern-Simons term, and hence of  the integer classification, 
is expected to be robust to adding interactions in these cases. (While this has been well known to be the case for classes D and C in $d=2$, relatively
recent work\cite{KapustinThorngrenTurzilloWang-arXiv2014}, aimed at the interacting cases using the cobordism approach, appears to
be in
agreement with our prediction
for $d=6$ in symmetry class D.) - Moving on to make use of the (purely gravitational version of) the  $\theta$-term $\Omega_{D}$  which,  as discussed below
(\ref{GeneralWtopForThetaTerm}),
can exist in every space-time dimension $D=4k$
and can thus provide 
an anomalous response in those space-time dimensions
corresponding to  spatial dimensions $d=4k-1$, i.e. to spatial dimensions $d=3$ (mod 8) and $d=7$ (mod 8). These are marked with  a red \textcolor{red}{$Z$} (not bold face)
 in TABLE \ref{fig:periodic table with gravity}. Similar comments as those for AIII above are in order here. The present argument only shows that in these dimensions  Topological Superconductors in classes DIII and CI can exist which are  characterized by an odd integer in the integer classification and does not demonstrate the existence of the entire integer classification\footnote{Though, in the non-interacting case one can obtain the general integer from the  winding of the phase of the  partition function as detailed explicitly  in Ref. \cite{RyuMooreLudwig2012} for the case of symmetry class DIII in $d=3$. - See also the discussion in Ref. \cite{QiWittenZhangAxionTopSuperconductors-PRB2013} obtaining the general
integer from an alternative point of view for symmetry class DIII in $d=3$}.  Yet,  the argument shows that there are non-trivial entries in the corresponding positions in the Table. (The same comments as those made in the preceding paragraph on class AIII also apply  here. Indeed the non-interacting integer classification in symmetry class DIII in $d=3$ spatial dimensions  is  actually known to collapse in the presence of interations to $Z_{16}$\cite{KitaevZ16,FidkowskiChenVishwanathNoAbelianTopOrderSurface3D2013,WangSenthilInteractingFermionic3D-2014,HsiehChoRyu-arXiv2015,Witten-arXiv2015}. A similar collapse has been found
in class CI in $d=3$\cite{WangSenthilInteractingFermionic3D-2014}. We will briefly discuss both cases  in section \ref{LabelSubSectionRecentResults-d-3-PresenceOfInteractions} below. -  A full description of symmetry  classes DIII and CI  in the presence of interactions, and more general, of all Topological Insulators (superconductors)  in the presence of interactions not fully  characterized by perturbative anomalies,  was conjectured\cite{RyuMooreLudwig2012} to be provided by a global anomaly; this
 has recently been confirmed - see 
\cite{RyuNobelProc2015,HsiehChoRyu-arXiv2015,Witten-arXiv2015} for a recent discussion.)

\vskip .2cm

\textcolor{blue}{{\bf (c): Mixed anomalous responses} (AI, BDI, AII, CII).} We consider coupling the members of the group  (AI, BDI, AII, CII)  to a mixed  background, consisting  of both classical gauge and gravitational background fields. We read off from the generating function for mixed anomalous responses 
  (using case $(c)$ of (\ref{DEFGaugeGravitationsMixedAnomaly})) that a $\theta$-term $\Omega_D$ can exist in every space-time
dimension $D=4k+2$ ($k=$ integer). The resulting  mixed Chern-Simons term 
$\Omega^{(0)}_{D-1}$,
%\ref{LabelSubSectionRecentResults-d-3-PresenceOfInteractions} 
from  (\ref{GeneralWtopforChernSimonsTerm})  and first descent relation (\ref{DEFFirstDescentEquation}),
predicts  a Topological Insulator (Superconductor)  can exist
in every space-time dimension $D=4k+1$,  corresponding to spatial dimensions $d=4k$, i.e. to 
spatial dimensions $d=0$ (mod 8) and $d=4$ (mod 8).
The intersections of these dimensions $d$ with the group  (AI, BDI, AII, CII)     of symmetry classes are marked
with blue bold face \textcolor{blue}{${\bf Z}$}  in 
TABLE \ref{fig:periodic table with gravity}. 
Note that this predicts precisely the spatial dimensions of the Topological Insulators  in these classes
with an integer classification. The presence of the corresponding Chern-Simons term, and  of  the integer classification, 
is expected to be robust to adding interactions in these cases. - Moving on to make use of (the mixed version of) the $\theta$-term $\Omega_{D}$  which,  as discussed below
(\ref{GeneralWtopForThetaTerm}),
 predicts  that an anomalous response can exist  in
every even space-time dimension $D=4k+2$, corresponding to  spatial dimensions $d=4k+1$, i.e. to spatial dimensions $d=1$ (mod 8) and $d=5$ (mod 8). These are marked with  a blue \textcolor{blue}{$Z$} (not bold face)
 in TABLE \ref{fig:periodic table with gravity}. Similar comments as those made regarding  classes AIII, DIII and CI above are in order here. The present  argument only shows that in these dimensions  Topological Insulators in classes BDI and CII  can exist which are characterized by an odd  integer in the integer classification and does not demonstrate the existence of the entire integer classification. Yet,  the argument shows that there are non-trivial entries in the corresponding positions in the Table.  (The same comments as those made at the end of the preceding paragraphs regarding symmetry classes AIII, as well as DIII and CI also apply here. A collapse of the integer classification in class BDI to $Z_8$  in the presence of interactions in $d=1$ was found in Ref. \cite{FidkowskiKitaevPRB81-2010-134509} for the superconductor version of class BDI\footnote{We have considered here a version with an even number of Majorana Fermion species, which possesses at conserved $U(1)$ charge. This accesses only the even integers in the non-interacting ${\bf Z}$ classification in that symmetry class}. -  A full description of symmetry  classes of all Topological Insulators (Superconductors)  in the presence of interactions not fully  characterized by perturbative anomalies,  was conjectured\cite{RyuMooreLudwig2012} to be provided by a global anomaly; this
 has recently been confirmed - see 
\cite{RyuNobelProc2015,HsiehChoRyu-arXiv2015,Witten-arXiv2015} for a recent discussion.)

\vskip .2cm

\begin{table}[!hbt]
\begin{center}
\resizebox{0.65\textwidth}{!}{
\begin{tabular}{ccccccccc}\hline
Cartan$\backslash d$ & 
\textcolor{blue}{\bf 0}  & 
\textcolor{blue}{1} & 
\textcolor{red}{\bf 2} & 
\textcolor{red}{3} & 
\textcolor{blue}{\bf 4} & 
\textcolor{blue}{5} & 
\textcolor{red}{\bf 6} & 
\textcolor{red}{7}  \\ \hline\hline\hline
A   & ${\bf Z}$& 0  & ${\bf Z}$& 0 &  ${\bf Z}$& 0  & ${\bf Z}$& 0  \\ \hline
AIII & 0& ${Z}$& 0  & ${Z}$& 0 &  ${Z}$& 0  & ${Z}$\\ \hline\hline
AI  & \textcolor{blue}{${\bf Z}$} & 0 & 0 & 0 & $2$\textcolor{blue}{${\bf Z}$} & 0 & ${\bf {Z}_2}$ & $\mathbb{Z}_2$  \\ \hline
BDI &\textcolor{green}{$\bf {Z}_2$} & \textcolor{blue}{${Z}$} & 0 & 0 & 0 & $2$\textcolor{blue}{${Z}$} & 0 & $\mathbb{Z}_2$  \\ \hline
D   & \textcolor{green}{$\bf {Z}_2$} &\textcolor{green}{$\bf {Z}_2$} & \textcolor{red}{${\bf Z}$}& 0 & 0 & 0 & $2$\textcolor{red}{${\bf Z}$ }& 0  \\ \hline
DIII& 0 &\textcolor{green}{$\bf {Z}_2$} & ${\bf {Z}_2}$ & \textcolor{red}{${Z}$ }& 0 & 0 & 0 &$2$\textcolor{red}{$Z$}  \\ \hline
AII & $2$\textcolor{blue}{${\bf Z}$} & 0 & ${\bf {Z}_2}$ & $\mathbb{Z}_2$ & \textcolor{blue}{${\bf Z}$} & 0 & 0 & 0 \\ \hline
CII & 0 & $2$\textcolor{blue}{${Z}$}  & 0 & $\mathbb{Z}_2$ & $\mathbb{Z}_2$ & \textcolor{blue}{${Z}$} & 0 & 0 \\ \hline
C   & 0  & 0 & $2$\textcolor{red}{${\bf Z}$}  & 0 & $\mathbb{Z}_2$  & $\mathbb{Z}_2$ & \textcolor{red}{${\bf Z}$} & 0  \\ \hline
CI  & 0 & 0  & 0 & $2$\textcolor{red}{${Z}$}  & 0 & $\mathbb{Z}_2$  & ${\bf {Z}_2}$  & \textcolor{red}{${Z}$}  \\ \hline
\end{tabular}
}
\end{center}
\caption{
Topological insulators (superconductors) with an
integer (${Z}$) classification,  predicted from {\bf gauge}, \textcolor{red}{\bf purely \ gravitational}, and 
\textcolor{blue}{\bf mixed}  anomalous responses.
 - Note that  2-fold Bott periodicity  of the two complex symmetry classes (A and AIII), and the 8-fold Bott periodicity of the remaining eight real symmetry classes emerges automatically from the anomaly perspective.  (In the Table, bold face (not bold face)  integer symbols denote the existence of Chern-Simons type ($\theta$-term type) anomalous responses; black stands for  'gauge', red for 'gravitational', and blue for 'mixed'. Green  and black bold face  $Z_2$ symbols denote the existence of global gravitational
anomalies in the corresponding dimensions.)
}
\label{fig:periodic table with gravity}
\end{table}

We end this section, as already mentioned,  with a brief discussion of  global anomalies\cite{RyuMooreLudwig2012}. 
Whereas the so-far discussed "perturbative anomalies"
appear as anomalous responses present already to low order in a perturbative  expansion in  small background fields, global anomalies appear only when \lq large\rq  \  (i.e. of order unity)  gauge or coordinate transformations are considered. Here we consider the case of global gravitational
anomalies, and so we are interested in  the response to large coordinate transformations.
It was found in \cite{AlvarezGaumeWittenNPB1983}
that global gravitational anomalies can exist, under certain conditions, (i) in $D=8k$, (ii) in $D=1+8k$. It was also found\cite{AlvarezGaumeWittenNPB1983}
that  global anomalies
can exist in  (iii) in $D=2+4k$ space-time dimensions,
where $k$ is an integer.
As pointed out in Ref. \cite{RyuMooreLudwig2012}, 
following the general logic discussed above these would decribe anomalous behavior at the boundary of the system indicating the presence of
a Topological Insulator (Superconductor) in one higher dimension, i.e. in space-time dimensions (i) $D=8k+1$ and  (ii) $D=2+8k$; moreover
this can also occur in  (iii) $D=3+4k$. 
These cases  correspond to spatial dimensions (i) $d=0+8k$, (ii) $d=1+8k$, and  (iii) $d=2+8k$ as well as  $d=6+8k$.  In each of those
four spatial dimensions (mod 8) there exist ${\it two}$ $Z_2$ topological insulators. The $Z_2$ cases corresponding to (i) and (ii) are marked in green
in  TABLE  \ref{fig:periodic table with gravity}; those corresponding to (iii) bold-face  black. We have conjectured  in \cite{RyuMooreLudwig2012}
that all  $Z_2$ entries in TABLE \ref{fig:periodic table with gravity}
can be obtained from consideration of these and other global anomalies. 

Moreover, the collapse of the integer classification
in odd spatial dimensions (these are the ones where we have obtained the odd  integer representative from an
anomalous $\theta$-term response in the previous paragraphs) has in some cases already been understood in terms of global
anomalies\cite{HsiehChoRyu-arXiv2015,Witten-arXiv2015}, thus supporting our previous general conjecture made in Ref. \cite{RyuMooreLudwig2012}.

\vskip 2cm

\subsection{Some Results in $d=3$ in the  Presence of Interactions}
\label{LabelSubSectionRecentResults-d-3-PresenceOfInteractions}

We close by briefly mentioning some relatively recent results on the effect of interactions in $d=3$ spatial dimensions\cite{WangSenthilInteractingFermionic3D-2014,WangPotterSenthil-Science2013} that build on the work presented in this review. A summary of the main results is listed in TABLE \ref{fig:RecentResultInteractionsSenthilWang}. In the first
column of this TABLE are listed the symmetries of the Fermionic  systems that were studied in that work. 
($U(1)$ and $SU(2)$  indicates the presence of  charge conservation and $SU(2)$ spin rotation symmetry, and $Z_2^T$ and $Z_2^C$ denote
time-reversal and charge-conjugation symmetries, respectively.)
The Cartan Symmetry Class to
which each system belongs is listed in parenthesis. In the second column is listed first the non-interacting classification of the
corresponding Cartan Symmetry classes  in $d=3$ spatial dimensions\footnote{It is known\cite{FosterLudwig-AIII-PRB2008} that Symmetry Class AIII
can also be realized in Bogoliubov DeGennes Hamiltonians describing superconductors with time-reversal symmetry and conservation
of the $S_z$ component of spin}
  (see e.g. TABLE \ref{fig:periodic table with gravity} of the previous section), followed
by an arrow that points to the classification that ensues when interactions are added to the non-interacting systems. (Note
that all $Z$ classifications reducing to  a smaller classification are only protected by the $\theta$-term anomalous response
discussed in section \ref{LabelSubSubSectionGeneralAnomalousResponsesDescentRelations} which, as mentioned  in that section,
guarantees  only the presence of the $n=$ odd  sector in the classification, in particular  including $n=1$.)
%, but it guarantees the existence of this non-trivial topological sector even in the presence of interactions.) 
Finally, in the  third (and last)
column  is listed the "Complete" classification of  the interacting Fermionic systems in the respective $d=3$ symmetry classes. These include the classifications
of the second column but contain additional topological  phases. These are phases that are not simply obtained by adding interactions to the non-interacting
Topological Insulator (Superconductor), but which can occur in other ways.

Some of the results in specific symmetry classes listed in TABLE  \ref{fig:RecentResultInteractionsSenthilWang} were also obtained independently
for Symmetry Class AII in Ref. \cite{MetlitskiKaneFisherBosTopIns3DWittenEffect2013}, and for Symmetry Class DIII in 
Ref.\cite{FidkowskiChenVishwanathNoAbelianTopOrderSurface3D2013}.

%\vskip .1cm

%\begin{figure}[!hbt]
%%  \includegraphics[width=\columnwidth]{BandStructure.pdf}
%\includegraphics[width=15cm]{SenthilWang-Table.pdf}
 %\caption{Some recent results in $d=3$ spatial dimenions on the effect of interactions. 
%Question marks indicate that the authors have arguments, but
%not a full proof of the respective claims, all of which pertain to  the "Complete" column.
%- Adapted from Ref. \cite{WangSenthilInteractingFermionic3D-2014,WangPotterSenthil-Science2013}.
%}
%\label{fig:RecentResultInteractionsSenthilWang} 
%\end{figure}

\begin{table}[!hbt]
\begin{center}
\resizebox{1.0\textwidth}{!}{
{
\renewcommand{\arraystretch}{1.5}
\begin{tabular}{lr|l|l}\hline
Symmetry...........................................................       &Symmetry Class & Reduction of free-& Complete \\
                                                                                                                   &(Cartan Symbol)&Fermion States      & \\ \hline
$U(1)$ only                                                                                                 &(A)                      & $0 \to 0$              & 0 \\ \hline
$U(1) \rtimes Z_2^T$ where ${\hat {\cal T}}^2 = + {\bf 1}$                  & (AI)                  & $0 \to 0$              & $\left ( Z_2 \right )^2$ \\ \hline
$U(1) \rtimes Z_2^T$ where ${\hat {\cal T}}^2= (-1)^{{\hat F}}$        & (AII)		      & $Z_2 \to Z_2$       & $ \left ( Z_2 \right )^3$ \\ \hline
$U(1) \times Z_2^T$ 							       &(AIII)			&$Z \to Z_8$		& $ Z_8 \times Z_2$     \\ \hline
$U(1) \rtimes \left ( Z_2^T \times Z_2^C\right )$,  where
		${\hat {\cal T}}^2={\hat {\cal C}}^2 = (-1)^{{\hat F}}$   & (CII)			&$Z_2 \to Z_2$	&$\left (  Z_2 \right )^5$ \\ \hline
$Z_2^T$ where ${\hat {\cal T}}^2 = (-1)^{{\hat F}}$			& (DIII)		&$Z \to Z_{16}$	&$Z_{16}$ \ (?) \\ \hline
$SU(2) \times Z_2^T$ where ${\hat {\cal T}}^2 = + {\bf 1}$		&(CI)			&$Z \to Z_4$		&$Z_4 \times Z_2$ \ (?) \\ \hline
%Cartan$\backslash d$ & 
%\textcolor{blue}{\bf 0}  &  \textcolor{blue}{1} &  \textcolor{red}{\bf 2} & 
%\textcolor{red}{3} &  \textcolor{blue}{\bf 4} &  \textcolor{blue}{5} & 
%\textcolor{red}{\bf 6} &  \textcolor{red}{7}  \\ \hline\hline\hline
%A   & ${\bf Z}$& 0  & ${\bf Z}$& 0 &  ${\bf Z}$& 0  & ${\bf Z}$& 0  \\ \hline
%AIII & 0& ${Z}$& 0  & ${Z}$& 0 &  ${Z}$& 0  & ${Z}$\\ \hline\hline
%AI  & \textcolor{blue}{${\bf Z}$} & 0 & 0 & 0 & $2$\textcolor{blue}{${\bf Z}$} & 0 & ${\bf {Z}_2}$ & $\mathbb{Z}_2$  \\ \hline
%BDI &\textcolor{green}{$\bf {Z}_2$} & \textcolor{blue}{${Z}$} & 0 & 0 & 0 & $2$\textcolor{blue}{${Z}$} & 0 & $\mathbb{Z}_2$  \\ \hline
%D   & \textcolor{green}{$\bf {Z}_2$} &\textcolor{green}{$\bf {Z}_2$} & \textcolor{red}{${\bf Z}$}& 0 & 0 & 0 & $2$\textcolor{red}{${\bf Z}$ }& 0  \\ \hline
%DIII& 0 &\textcolor{green}{$\bf {Z}_2$} & ${\bf {Z}_2}$ & \textcolor{red}{${Z}$ }& 0 & 0 & 0 &$2$\textcolor{red}{$Z$}  \\ \hline
%AII & $2$\textcolor{blue}{${\bf Z}$} & 0 & ${\bf {Z}_2}$ & $\mathbb{Z}_2$ & \textcolor{blue}{${\bf Z}$} & 0 & 0 & 0 \\ \hline
%CII & 0 & $2$\textcolor{blue}{${Z}$}  & 0 & $\mathbb{Z}_2$ & $\mathbb{Z}_2$ & \textcolor{blue}{${Z}$} & 0 & 0 \\ \hline
%C   & 0  & 0 & $2$\textcolor{red}{${\bf Z}$}  & 0 & $\mathbb{Z}_2$  & $\mathbb{Z}_2$ & \textcolor{red}{${\bf Z}$} & 0  \\ \hline
%CI  & 0 & 0  & 0 & $2$\textcolor{red}{${Z}$}  & 0 & $\mathbb{Z}_2$  & ${\bf {Z}_2}$  & \textcolor{red}{${Z}$}  \\ \hline
\end{tabular}
}
}
\end{center}
\caption{
Some recent results in $d=3$ spatial dimensions on the effect of interactions\cite{WangSenthilInteractingFermionic3D-2014,WangPotterSenthil-Science2013}. 
Question marks indicate that the authors have arguments, but
not a full proof of the respective claims, all of which pertain to  the "Complete" column.
}
\label{fig:RecentResultInteractionsSenthilWang}
\end{table}

%\newpage

It is worth mentioning some interesting more detailed and specific results on some of the seven Symmetry Classes in TABLE  \ref{fig:RecentResultInteractionsSenthilWang}. This makes contact with some of the more general statements about 
Symmetry Protected
Topological (SPT) phases,  made
in section \ref{sectionEntanglementPerspective}:

\vskip .2cm

\noindent $\bullet$ $U(1)\rtimes Z_2^T$ where ${\hat{\cal  T}}^2=(-1)^{{\hat F}}$ (AII):
\vskip .1cm

\noindent There are only two topologically distinct phases, a topological  $n=1 \in Z_2$
and a non-topological  $n=0\in Z_2$ phase.
A surface of this topological  $d=3$ bulk Fermonic  SPT phase in this class  which respects the
$U(1)\rtimes Z_2^T$  symmetry
must have non-Abelian (Moore-Read) statistics.
But this non-Abelian statistics is such that it cannot be realized in $d=2$ spatial dimensions for the given action of
this symmetry on the surface.

\vskip .4cm

\noindent $\bullet$ $U(1)\times Z_2^T$ (AIII):

\vskip .1cm

\noindent There are eight phases, $n=0, 1, ..., 7 \in Z_8$, only one of which is topologically trivial.
A surface of the $n=1$  bulk Fermonic  SPT phase in this class 
which respects the $U(1)\times Z_2^T$ symmetry
must have non-Abelian (Moore-Read) statistics.
But this non-Abelian statistics is such that it cannot be realized in $d=2$ spatial dimensions for the given action of
this symmetry on the surface.

\vskip .4cm

\noindent $\bullet$ $Z_2^T$  where ${\hat{\cal  T}}^2=(-1)^{{\hat F}}$ (DIII):

\vskip .1cm

\noindent There are sixteen  phases, $n=0, 1, ..., 15 \in Z_{16}$, only one of which [$n  =  0 \ ({\rm mod} \ 16)$]  is topologically trivial.
A surface characterized by $n=$odd of the  bulk Fermonic  SPT phase in this class  
which respects ${\hat {\cal T}}^2=(-1)^{{\hat F}}$ time-reversal symmetry must have non-Abelian statistics
 (with topological order characterized by $SU(2)_3$); the surface carries a realization of time-reversal symmetry 
that is impossible to realize
at a purely $d=2$-dimensional surface. The surface of the topological bulk phases characterized by $n=$even but $n \not =  0 \ ({\rm mod} \ 16)$  possess
Abelian topological order. Furthermore, the $n=8$ surface has only Fermions.

\vskip .4cm

\noindent $\bullet$ $SU(2)\times Z_2^T$  where ${\hat {\cal  T}}^2=+1$ (CI):

\vskip .1cm

\noindent There are four  phases, $n=0, 1, ...,3 \in Z_{4}$, only one of which is topologically trivial.
A surface characterized by $n=1$ of the  bulk Fermonic  SPT phase in this class
remains gapless even with interactions.

\vskip 3cm

\begin{acknowledgments} The author thanks the organizers of the Nobel Symposium for arranging this stimulating event.  He would like to thank  S. Ryu, A. Schnyder, A. Furusaki and  J. Moore for their collaboration in the work that was reviewed here. This work was  supported, in part,  by NSF DMR-1309667.  

\end{acknowledgments}

\clearpage
%\newpage

\appendix

\section{Bogoliubov DeGennes (BdG) Hamiltonians of  Superconductors}
\label{LabelSectionAppendixSuperconductor}

In  a superconductor, the  Hamiltonian (\ref{DEFSecondQuantizedHamiltonianNormalSystem})
is replaced by the Bogoliubov-de-Gennes (BdG) Hamiltonian which governs the dynamics of the (Fermionic) quasiparticles of the superconductor.  Specifically, we replace the ($N$-dimensional) column vector ${\hat \psi}$, introduced below (\ref{DEFSecondQuantizedHamiltonianNormalSystem}),
by the vector (Nambu Spinor)
%\begin{equation}\label{DEFNambuSpinor}
%\left (
%\begin{matrix}
%{\hat \chi}_{a} \\
%{\hat \chi}_{a+N}
%\end{matrix}
%\right )
%\equiv 
% \left (
 %\begin{matrix}
%{\hat \psi}_a \\
%{\hat \psi}_a^\dagger
%\end{matrix}
%\right ), \qquad (a=1, ..., N)
%\end{equation}

\begin{equation} \label{DEFNambuSpinorAlternative}
{\hat \chi} \equiv 
\left (
\begin{matrix}
{\hat \chi}_1 \\
...\\
{\hat \chi}_N \\
{\hat \chi}_{N+1} \\
... \\
{\hat \chi}_{2N}
\end{matrix}
\right )
\equiv
\left (
\begin{matrix}
{\hat \psi}_1 \\
... \\
{\hat \psi}_N \\
{\hat \psi^\dagger_1} \\
... \\
{\hat \psi^\dagger_N}
\end{matrix}
\right )
=
\left (
\begin{matrix}
{\hat \psi} \\
({\hat \psi}^\dagger)^t
\end{matrix}
\right )
\end{equation}
(where ${}^t$ denotes the transpose).
With this notation, the BdG Hamiltonian of a superconductor reads, in complete analogy with (\ref{DEFSecondQuantizedHamiltonianNormalSystem})
\begin{equation}\label{DEFSecondQuantizedHamiltonianBdG}
{\hat H}
=
{1\over 2}\sum_{A,B=1}^{2N}
{\hat \chi}^\dagger_A
H_{A,B}
{\hat \chi}_B
=
{1\over 2}
{\hat \chi}^\dagger H {\hat \chi}
=
{1\over 2}
\left (
\begin{matrix}
{\hat \psi}^\dagger,  & {\hat \psi}^t 
\end{matrix}
\right )
 H
\left (
\begin{matrix}
{\hat \psi} \\
({\hat \psi}^\dagger)^t
\end{matrix}
\right)
\end{equation}
The factor of ${1\over 2}$ arises from the property in  (\ref{AppendixAVectorChiVectorChiDagger}) below.
In view of the last equality of (\ref{DEFSecondQuantizedHamiltonianBdG})  the  first quantized BdG Hamiltonian $H$
can be written in terms of  $N\times N$  blocks
\begin{equation} \label{DEFBdGHamiltonianBlocks}
H = \left (
\begin{matrix}
\Xi & \Delta \\
 \Delta^* & - \Xi^t 
\end{matrix}
\right )
\end{equation}
where $\Xi=\Xi^\dagger$ (by Hermiticity of $H$) and $\Delta = - \Delta^t$ (by Fermi statistics). The second quantized BdG Hamiltonian in
(\ref{DEFSecondQuantizedHamiltonianBdG}) then reads
\begin{equation}\label{SecondQuantizedBdGHamiltonian}
{\hat H}
=
\sum_{a,b=1}^N 
{\hat \psi}_a^\dagger \  \Xi_{a,b} \   {\hat \psi}_b
+ {1\over 2} \sum_{a,b=1}^N 
\left (
 {\hat \psi}^\dagger_a \  \Delta_{a,b} \  {\hat \psi}^\dagger_b
+
{\hat \psi}_a \ \Delta^*_{a,b} \  {\hat \psi}_b
\right )
\end{equation}
Note that while the vectors 
\begin{equation}\label{AppendixAVectorPsiVectorPsiDagger}
{\hat \psi} =
 \left (
\begin{matrix}
{\hat \psi}_1 \\
... \\
{\hat \psi}_N 
\end{matrix}
\right)
\quad {\rm and} \quad
({\hat \psi}^\dagger)^t
=
\left (
\begin{matrix}
{\hat \psi}^\dagger_1 \\
... \\
{\hat \psi}^\dagger_N 
\end{matrix}
\right)
\end{equation}
are linearly independent, this is not the case for the vectors ${\hat \chi}$ and $({\hat \chi}^\dagger)^t$ defined in (\ref{DEFNambuSpinorAlternative}), since due to (\ref{DEFNambuSpinorAlternative})
\begin{equation} \label{AppendixAVectorChiVectorChiDagger}
({\hat \chi}^\dagger)^t = 
\left (
\begin{matrix}
({\hat \psi}^\dagger)^t \\
{\hat \psi}
\end{matrix}
\right )
=
\tau_1 \left (
\begin{matrix}
{\hat \psi} \\
({\hat \psi}^\dagger)^t
\end{matrix}
\right )
= \tau_1 {\hat \chi},
\qquad {\rm or:} \qquad {\hat \chi}^\dagger
={\hat \chi}^t \tau_1
\end{equation}
where
\begin{equation}\label{DEFTauParticleHoleSpace}
\tau_1 = 
\left (
\begin{matrix}
{\bf 0}_N & {\bf 1}_N \\
{\bf 1}_N & 0_N
\end{matrix}
\right )
\end{equation}
is the $\tau_1$ Pauli matrix in \lq particle-hole\rq \  (\lq annihilation-creation operator\rq)  space.  (Here ${\bf 1}_N$ is the $N\times N$ identity matrix, and ${\bf 0}_N$ is the $N\times N$ matrix
with   all matrix elements zero.)
As a consequence of (\ref{AppendixAVectorChiVectorChiDagger}) we can write (\ref{DEFSecondQuantizedHamiltonianBdG}) as
\begin{eqnarray}
&&{\hat H} ={1\over 2} {\hat \chi}^\dagger H {\hat \chi}=
 {1 \over 2} {\hat \chi}^t \tau_1 H \tau_1 ({\hat \chi}^\dagger)^t
={1\over 2}
\sum_{A,B=1}^{2N}
{\hat \chi}_A (\tau_1 H \tau_1)_{A,B} {\hat \chi}^\dagger_B=\\ \label{RewriteBdGHamiltonian}
&&
={1\over 2} \sum_{A,B=1}^{2N} (\tau_1 H \tau_1)_{A,B} (- {\hat \chi}^\dagger_B{\hat \chi}_A + \delta_{A,B})
=-{1\over 2} {\hat \chi}^\dagger (\tau_1 H \tau_1)^t {\hat \chi} + {1\over 2} Tr (\tau_1 H \tau_1)
\end{eqnarray}
Due to (\ref{DEFBdGHamiltonianBlocks}) the trace in the last equality vanishes, and we obtain 
\begin{equation} \label{BdGChargeConjugationSymmetry}
\tau_1H^*\tau_1=\tau_1 H^t \tau_1= -  H. 
\end{equation}
(The first equality follows from Hermiticity.)  Therefore, {\it any} first quantized BdG Hamiltonian $H$ will automatically  (i.e. by construction) 
satisfy the charge conjugation
(particle hole) symmetry property, equation (\ref{DEFChargeConjugationSingleParticleHamiltonian}), with $U_C=\tau_1$.
Charge conjugation \lq symmetry\rq \  is therefore not a symmetry  of the superconducting system but rather a property of {\it any} BdG Hamiltonian.

\section{The Alternative Chiral (Sublattice) Symmetry Operation}
\label{LabelSectionAppendixAlternativeChiralSymmetry}

In this Appendix we discuss the alternative chiral (sublattice) operation defined by
\begin{equation}\label{AppendixDEFChiralSymmetrySecondQuantizedAlternative}
{\hat {{\cal S}'}} \equiv
 {\hat {\cal C}}
\cdot  {\hat {\cal T}}
\end{equation}
(see (\ref{DEFChiralSymmetryPrimedSecondQuantized})), in which the order of factors is exchanged as compared  to the chiral symmetry operation
defined in (\ref{DEFChiralSymetrySecondQuantized}).
One readily  checks using   (\ref{DEFTimeReversalSecondQuantized}),(\ref{DEFChargeConjugationSecondQuantized})
that 
\begin{equation}\label{DEFChiralPrimedSecondQuantized}
{\hat {\cal S}} {\hat \psi}_A {\hat {\cal S}}^{-1} = \sum_B  \  ({U^{* \dagger}_{S'})}_{A,B} \  {\hat \psi}^\dagger_B; \quad \  \ 
{\hat {\cal S}} {\hat \psi}^\dagger_A {\hat {\cal S}}^{-1} = \sum_B  \ {\hat \psi}_B\ {(U^*_{S'})}_{B,A};\quad \ \ 
{\hat {\cal S}} i {\hat {\cal S}}^{-1} = - i  \  {\rm (anti-unitarity)}
\end{equation}
where  $U_{S'}\equiv U_C  U_T^*$.
%{\color{blue} <- check this !!}

Let us now go back to the  original version of chiral symmetry ${ {\mathbf S}} = U_T \cdot U_C^* = U_S$, where ${\mathbf S}^2=(U_S)^2={\bf 1}$,
 as in (\ref{DEFChiralSymmetrySingleParticleHamiltonianAlt}),
(\ref{DEFChargeConjugationHamiltonianAlt}),(\ref{ChiralSquaredSchursLemmaNoPhase}).
% Writing out the latter condition,
%$(U_T U^*_C) (U_T U_C^*) = {\bf 1}$,
%${ {\mathbf S}} = U_T \cdot U_C^* = U_S$, where ${\mathbf S}^2=(U_S)^2={\bf 1}$. 
Writing out the latter condition,
$(U_T U^*_C) (U_T U_C^*) = {\bf 1}$,
and
multiplying both sides  by $U_T^\dagger$ form the left
and by $U_T$ form the right yields $U^*_C U_T U^*_C U_T = {\bf 1}$. Taking the complex conjugate yields
$(U_C U^*_T) (U_C U_T^*) =  {\bf 1}$, and hence
\begin{equation}\label{USPrimedSquareIdentity}
(U_{S'})^2 = {\bf 1}.
\end{equation}
Moreover we  conclude from 
$U_S= U_T U_C^*$, $(U_S)^2 = {\bf 1}$ and $U_{S'}= U_C U_T^*$, $(U_{S'})^2={\bf 1}$ that  $Tr U_S =$
$Tr (U_S)^*=$ $ Tr U_{S'}$.
Therefore, both $U_S$ and $U_{S'}$  have the same spectrum of eigenvalues (they have only eigenvalues $\pm 1$, and the difference of positive and negative eigenvalues is the same for both matrices). Therefore, there must
be a unitary matrix $W$ such that 
\begin{equation}\label{USPrimedUS}
U_{S'}= W U_S W^\dagger
\end{equation}
Hence  we see that if $H$ anticommutes with $U_S$,
\begin{equation}\label{SAntiCommutesWithUS}
U_S H  = - H U_S
\end{equation}
from which we obtain
$W^\dagger U_{S'} W H =$ $- H W^\dagger U_{S'} W$
upon inserting (\ref{USPrimedUS})
which can be rewritten as
\begin{equation}\label{USPrimedHW}
U_{S'} \left (W H W^\dagger \right ) 
=
-\left (W H W^\dagger \right )  U_{S'}.
\end{equation}
Thus $H$ anticommutes with $U_S$ if and only if $W H W^\dagger$ anticommutes with $U_{S'}$ for a suitable choice of the unitary matrix $W$
(specfied in (\ref{USPrimedUS})).

Note that we are interested in the situation where the first quantized Hamiltonian $H$ runs over the set of "symmetry-less" blocks
discussed in section \ref{LabelSubSubSectionUnitarily Realized Symmetries}. Clearly, if $H$ is one of these blocks in this set, so is   
$W H W^\dagger $ which just corresponds to writing $H$ in another basis.
\vskip 1cm

\section{Description Of "Simplified Hamiltonians" in the Chiral Symmetry Classes in TABLE \ref{fig:ListOfSimplifiedHamiltoniansProjectors}  }
\label{LabelSectionAppendixChiralClassesFIGSimplifiedHamiltonians}

Let us first discuss the form of a general Hamiltonian $H$ which satisfies the condition (\ref{DEFChiralSymmetrySingleParticleHamiltonianAlt})
\begin{equation} \label{Appendix-C-ChiralInvarianceHamiltonian}
{\bf S} H {\bf S} = - H
\end{equation}
of invariance under the chiral symmetry ${\bf S}=$ $ U_S$, equation (\ref{DEFChargeConjugationHamiltonianAlt}). We know that ${\bf S}$
is unitary and squares to the identity matrix, equation (\ref{ChiralSquaredSchursLemmaNoPhase}). Therefore, we can chose a basis in which
\begin{equation}\label{AppendixC-S-diagonalized}
{\bf S}= U_S = \left ( 
 \begin{matrix}
{\bf 1}_n & {\bf 0} \\
{\bf 0} & -{\bf 1}_m
\end{matrix}
\right ).
\end{equation}
Now, writing out the condition (\ref{Appendix-C-ChiralInvarianceHamiltonian}) in the basis in which ${\bf S}$ is diagonal as in (\ref{AppendixC-S-diagonalized}),
one immediately finds that $H$ has the following block structure
\begin{equation}\label{Appendix-C-H-in-this-basis}
H =
 \left ( 
 \begin{matrix}
{\bf 0}_n & {\bf h}_{12} \\
{\bf h}_{21} & {0}_m
\end{matrix}
\right ), \qquad {\rm where} \quad {\bf h}_{21}^\dagger={\bf h}_{12}.
\end{equation}
Here $ {\bf h}_{12}$ is a $n\times m$ matrix, and ${\bf h}_{21}$ is a $m \times n$ matrix, and the equality follows from the 
Hermiticity of ${ H}$.

Let us now specialize to the case where $H$ is the "simplified Hamiltonian" ${\cal \bf Q}(\vec k)$ discussed in section \ref{LabelSubSectionClassificationByTopolgyOfTheBulk}. First consider the case of the complex symmetry class A. If ${\cal \bf Q}(\vec k)$
denotes the specialized Hamiltonian in symmetry class A, satisfying in addition the chiral symmetry condition (\ref{Appendix-C-ChiralInvarianceHamiltonian}),
it represents a "simplified Hamiltonian" in symmetry class AIII. In this case, it will
take   the form 
\begin{equation}\label{Appendix-C-Q-special-basis}
{\bf Q}(\vec k)
=
 \left ( 
 \begin{matrix}
{\bf 0}_n & {\bf q}_{12} \\
{\bf q}_{21} & {0}_m
\end{matrix}
\right ),
\qquad \qquad {\rm where} \quad {\bf q}_{21}^\dagger={\bf q}_{12}.
\end{equation}
%The last equality follows because ${\bf Q}(\vec k)$, being a Hamiltonian, must be   Hermitian.
But since ${\cal \bf Q}(\vec k)$ is a "simplified Hamiltonian", it has by definition only eigenvalues $\pm 1$, and thus squares to the identity,
\begin{equation}\label{Appendix-C-Q-squares-to-identity}
\left [ {\cal \bf Q}(\vec k)\right ]^2 = {\bf 1}_{n+m}.
\end{equation}
Writing this condition out upon making use of (\ref{Appendix-C-Q-special-basis})
one immediately obtains
\begin{equation}\label{Appendix-C-Result-Q-k}
{\bf Q}(\vec k)
=
 \left ( 
 \begin{matrix}
{\bf 0}_n & {\bf q} \\
{\bf q}^\dagger & {0}_m
\end{matrix}
\right ),
\qquad \qquad {\rm where} \quad {\bf q}\cdot {\bf q}^\dagger={\bf 1}_n, \quad {\bf q}^\dagger\cdot {\bf q}={\bf 1}_m.
\end{equation}

In summary, we conclude that the "simplified Hamiltonian" ${\bf Q}(\vec k)$ in symmetry class AIII, which is the only  symmetry class possessing chiral symmetry
and no other symmetry (it is hence he simplest symmetry class possessing chiral symmetry),
is completely characterized by the $n\times m$ matrix ${\bf q}$ as above. This is the meaning of the entry AII  in the  second column of TABLE \ref{fig:ListOfSimplifiedHamiltoniansProjectors}. (In that TABLE we have considered the special case where $n=m$, so that
the last condition in (\ref{Appendix-C-Result-Q-k}) implies that the $n\times n$ matrix ${\bf q}$ is unitary.)
All the other entries in  TABLE \ref{fig:ListOfSimplifiedHamiltoniansProjectors} for symmetry classes possessing chiral symmetry (which are classes
BDI, CII, DIII, and CI, as can be read off from the 4th column of FIG. \ref{fig:TenFoldWay}), correspond to symmetry classes which are defined
by the presence of additional conditions
imposed on the unitary matrix  ${\bf q}$ which arise from time-reversal or charge-conjugation symmetries imposed on
${\bf Q}(\vec k)$. This explains the notation used
in TABLE \ref{fig:ListOfSimplifiedHamiltoniansProjectors} for all symmetry classes with chiral symmetry.

\vskip 1cm

\section{Symmetry Properties of Homotopy Groups}
\label{LabelSectionAppendixSymmetryPropertiesHomotopyGroups}

%We summarize in the columns 1 and 2 of the Table below the information contained in the Table of Homotopy Groups, FIG. %\ref{fig:TableOfHomotopyGroupsForAllTenSymmetricSpaces} .

In this Appendix we explain the details of the discussion  in section \ref{LabelSubSectionAgreementBetweenBulkAndBoundaryMethods}, in particular
in the paragraph
preceding (\ref{ComparisonBulkBoundaryCondition}), that shows that
the bulk-based and the boundary-based approach to the classification yields the same result. Focus on the Table  \ref{AppendixRelationshipBetweenHomotopyGroups} below. We first focus on the eight real Cartan Symmetric Spaces.
The 2nd, 3rd and 4th columns  just repoduce  the 1st (Cartan Label),  3rd (target space  $G/H$ of the NLSM describing
Anderson Localization at the boundary)  and 4th column (Classifying Spaces used in the bulk-based classification method) of TABLE
\ref{fig:TableThreeOccurrencesOfTenCartanSpaces}, respectively. Now, from the first column (with heading "$\overline{d}=0$")
of FIG. \ref{fig:TableOfHomotopyGroupsForAllTenSymmetricSpaces} which precisely lists the Homotopy Groups of the NLSM target spaces $G/H$ 
listed in the 3rd column of (\ref{AppendixRelationshipBetweenHomotopyGroups}),
we can read off the Homotopy Groups listed in the 5th column of  (\ref{AppendixRelationshipBetweenHomotopyGroups}). By bringing
this list in the order of the 4th  column of (\ref{AppendixRelationshipBetweenHomotopyGroups}) we can fill in the 6th column of the same table.
By re-ordering the 6th column, we can also find directly the Homotopy Groups in the 7th column of  (\ref{AppendixRelationshipBetweenHomotopyGroups}). Now we are in the position to easily verify the correctness of the statement made in 
equation (\ref{ComparisonBulkBoundaryCondition}) from section \ref{LabelSubSectionAgreementBetweenBulkAndBoundaryMethods}, which
we repeat below for the convenience of the reader:
\begin{equation}\label{ComparisonBulkBoundaryConditionRepeated}
\left \{
\begin{matrix}
{\rm \bf\underbar{\bf Bulk:} \ Topological \ Band \ Theory}&& {\rm \bf \underbar{\bf Boundary:} \ Anderson \ Localization (NLSM)} \\
&&\\
(a):\qquad \qquad \textcolor{blue}{\pi_0\left ( R_{q-d}\right)= {\bf Z_2}} \qquad &{\rm \it  if \ and \ only \ if}&    \left [ \pi_{d-1}\left ( R_{4-q}\right) = \right ] \ \ 
\textcolor{blue}{\pi_0\left ( R_{3-(q-d)}\right ) 
= {\bf Z_2}}\\ 
(b): \qquad  \qquad \textcolor{red}{ \pi_0\left (R_{q-d}\right )={\bf Z}} \  \qquad    &{\rm \it  if \ and \ only \ if}&  
 \   \ \left [  \pi_{d}\left ( R_{4-q}\right) =\right ] \ \ \textcolor{red}{\pi_0 \left( R_{4-(q-d)}\right)
 ={\bf Z}}  \\
&&\\
                                                                                       & {\rm using:} \ \pi_d\left ( R_p\right ) = \pi_0\left (R_{p+d}\right ) &
\end{matrix}
\right \}
\end{equation}
By comparing the 6th and the 5th columns of  (\ref{AppendixRelationshipBetweenHomotopyGroups}) we see that condition $(b)$ is
satisfied (red color), whereas by comparing the 6th and the 7th column  of 
(\ref{AppendixRelationshipBetweenHomotopyGroups} we see that condition $(a)$ is
satisfied (blue color). Therefore, the conditions on the right hand sides (boundary-based classification) of \ref{ComparisonBulkBoundaryConditionRepeated}
are satisfied if and only the conditions on the left hand sides (bulk-based classification) is satisfied. This agreement therefore lies in a particular
("symmetry") properties of the list of Homotopy Groups (e.g. of the list in the 6th column).

\vskip 2cm

\begin{equation}\label{AppendixRelationshipBetweenHomotopyGroups}
%\begin{table}
%\begin{center}
\begin{tabular}{|c||c||c||c||c|c|c|}\hline
%&&&& \\
1st column&2nd column&3rd column&4th column&5th column&6th column&7th column\\ 
                 &    Cartan Label               &NLSM $G/H$&Classifying space&&& \\    \hline     \hline
\quad $(q-d)$ \quad  && \quad  $R_{4-(q-d)}$ \quad & \quad $R_{q-d}$ \quad& \quad \textcolor{red}{ $\pi_0\left (R_{4-(q-d)}\right )$} \quad  &  \quad $\pi_0\left ( R_{q-d}\right )$\quad  &  \quad \textcolor{blue}{ $\pi_0\left ( R_{3-(q-d)}\right)$}\quad  \\ \hline \hline
$0$&AI& $R_4$& $R_0$&\textcolor{red}{${\bf Z}$}&$$\textcolor{red}{$\Leftrightarrow{\bf Z}$} \quad  \quad &$0$ \\ \hline
$1$&BDI& $R_3$& $R_1$& $0$&\qquad  \textcolor{blue}{${\bf Z_2}\Leftrightarrow$}& \textcolor{blue}{ ${\bf Z_2}$} \\ \hline
$2$&D&$R_2$& $R_2$& ${\bf Z_2}$&\qquad\textcolor{blue}{${\bf Z_2}\Leftrightarrow$}&\textcolor{blue}{${\bf Z_2}$} \\ \hline
$3$&DIII&$R_1$&$R_3$& ${\bf Z_2}$& $0$&${\bf Z}$\\ \hline
$4$&AII&$R_0$& $R_4$&\textcolor{red}{${\bf Z}$}&\textcolor{red}{$\Leftrightarrow{\bf Z}$}\quad \quad  &$0$\\ \hline
$5$&CII&$R_7$&$R_5$&$0$&$0$&$0$\\ \hline
$6$&C&$R_6$&$R_6$&$0$&$0$&$0$ \\ \hline
$7$&CI&$R_5$&$R_7$&$0$&$0$&${\bf Z}$ \\ \hline
\end{tabular}
%\end{center}
%\caption{Here goes FigureCaption
%}
%\label{periodic table with gravity}
%\end{table}
\end{equation}

%$$
%q  R_{4-q}  \pi_0\left (R_{4-q}\right )  \pi_0\left ( R_q\right )  \pi_0\left ( R_{3-q}\right) 
%$$

\bibliography{cslproc-asubmit}

\end{document}